\documentclass[rmp,twocolumn,amsmath,amssymb,seceqn]{revtex4}
\usepackage{graphicx}
\usepackage{graphics}
\usepackage{epsfig}
\usepackage{dcolumn}
\usepackage{bm}
\usepackage{makeidx}
\usepackage{subfigure}
\usepackage{color}
\usepackage{longtable}

\makeatletter
\@addtoreset{equation}{section}
\makeatother

\newcommand{\kk}{{\bm{k}}}

\def  \bsig    {\mbox{\boldmath$\sigma $}}

\newcommand{\bfig}{\begin{figure}}
\newcommand{\efig}{\end{figure}}

\newcommand{\be}{\begin{equation}}
\newcommand{\ee}{\end{equation}}
\newcommand{\kf}{{\bf k}}

\newcommand{\rf}{{\bf r}}

\def\sigic{\sigma_{xy}^{H-int}}
\def\sigsj{\sigma_{xy}^{H-sj}}
\def\sigij{\sigma_{xy}^{H-int+sj}}

\newcommand{\bracks}[1]{\ensuremath{\left[ #1 \right]}}

\newcommand{\ad}{\dagger}
\newcommand{\abs}[1]{\left|\,#1\,\right|}

\newcommand{\Ms}{M_{\text{s}}}

\renewcommand{\phi}{\varphi}

\newcommand{\tstt}{\kappa}

\begin{document}

\title{Spin Hall effect}

\author{Jairo~Sinova}
\affiliation{Institut f\"ur Physik, Johannes Gutenberg Universit\"at Mainz, 55128 Mainz, Germany}
\affiliation{Institute of Physics ASCR, v.v.i., Cukrovarnick\'a 10, 162 53
Praha 6, Czech Republic}

\author{Sergio~O.~Valenzuela}
\affiliation{ICN2 - Institut Catala de Nanociencia i Nanotecnologia, Campus UAB, Bellaterra, 08193 Barcelona, Spain}
\affiliation{ICREA - Instituci\'o Catalana de Recerca i Estudis Avan\c{c}ats, 08010 Barcelona, Spain}

\author{J. Wunderlich}
\affiliation{Institute of Physics ASCR, v.v.i., Cukrovarnick\'a 10, 162 53
Praha 6, Czech Republic}
\affiliation{Hitachi Cambridge Laboratory, Cambridge CB3 0HE, United Kingdom}

\author{C.~H.~Back}
\affiliation{Universit{\"a}t Regensburg, Universit\"atstra{\ss}e 31, 93040 Regensburg, Germany}

\author{T. Jungwirth}
\affiliation{Institute of Physics ASCR, v.v.i., Cukrovarnick\'a 10, 162 53
Praha 6, Czech Republic}
\affiliation{School of Physics and Astronomy, University of Nottingham, Nottingham NG7 2RD, United Kingdom}

\date{\today}

\begin{abstract}

{Spin Hall effects are a collection of relativistic spin-orbit coupling phenomena in which electrical currents can generate transverse spin currents and vice versa. Although first observed only a decade ago, these effects are already ubiquitous within spintronics as standard spin-current generators and detectors. Here we review the experimental and theoretical results that have established this sub-field of spintronics. We focus on the results that have converged to give us a clear understanding of the phenomena and how they have evolved from a qualitative to a more quantitative measurement of spin-currents and their associated spin-accumulation.
Within the experimental framework, we review optical, transport, and magnetization-dynamics based measurements and link them to both phenomenological and microscopic theories of the effect.
Within the theoretical framework, we review the basic mechanisms in both the extrinsic and intrinsic regime which are linked to the mechanisms present in their closely related phenomenon in
ferromagnets, the anomalous Hall effect.
We also review the connection to the phenomenological treatment based on spin-diffusion equations applicable to certain regimes, as well as the spin-pumping theory of spin-generation which has proven important in the measurements of the spin Hall angle.
We further connect the spin-current generating spin Hall effect to  the inverse spin galvanic effect, which often accompanies the SHE, in which an electrical current induces a
non-equilibirum spin polarization. These effects share common microscopic origins
and can exhibit similar symmetries when present in ferromagnetic/non-magnetic structures
through their induced current-driven spin torques.
Although we give a short chronological overview  of the evolution of this field, the main body of this review is structured from a pedagogical point of view, focusing on well-established and accepted physics.
In such a young field, there remains much to be understood and explored, hence
we outline from our own perspective some of the future challenges and opportunities of this rapidly evolving area of spintronics.
}
\end{abstract}

\maketitle
\tableofcontents

\section{Introduction}\label{sec:intro}
Spintronics is a field that jointly utilizes  the spin and charge degrees of freedom to control equilibrium and non-equilibrium properties of materials and devices \cite{Wolf2001,Zutic2004,Bader2010}. The generation, manipulation, and detection of spin-currents  is one of the key aspects of the field of spintronics.
Among the several possibilities to create and control spin-currents 
the spin Hall effect  (SHE) has gained its distinct place since its first observation a decade ago \cite{Kato2004d,Wunderlich2004,Wunderlich2005,Day2005}.  
In the direct SHE, an electrical current passing through a material with relativistic spin-orbit coupling can generate 
a transverse pure spin-current polarized perpendicular to the plane defined by the charge and spin-current. Its reciprocal effect,
the so called inverse SHE (ISHE), is the phenomenon in which a pure spin-current through the material generates a transverse charge current.

The SHE borrows its concept from the well established anomalous Hall effect (AHE) where relativistic spin-orbit coupling generates an asymmetric deflection of the charge carriers depending on their spin direction \cite{Nagaosa2010}. The AHE can be detected electrically in a ferromagnet (FM) via a transverse voltage because of the difference in population of majority and minority carriers. The generalization of this effect to a pure spin-current generated by the SHE in a non-magnetic material (NM) was proposed over four decades ago~\cite{Dyakonov1971} based on the  idea of asymmetric Mott scattering \cite{Mott1929}.  This so called extrinsic SHE 
remained unexplored until recent proposals that put forward a similar prediction \cite{Hirsch1999,Zhang2000} as well as the possibility of a strong intrinsic effect \cite{Murakami2003,Sinova2004}. 

The initial challenge for SHE detection was primarily the lack of direct electrical signals; therefore initial experiments detected it by optical means, both in the extrinsic regime \cite{Kato2004d} and the intrinsic regime \cite{Wunderlich2004,Wunderlich2005}. The ISHE was detected soon thereafter \cite{Valenzuela2006,Saitoh2006,Zhao2006}. Early measurements were mostly qualitative. However,  more accurate quantitative measurements of spin Hall angles have been established in later experiments through the aid of FM detectors in static or dynamic magnetization regimes, and a much firmer situation has arisen in the field. 

Adding to this flurry of activity and increased understanding, 
recent experiments in magnetic tunnel junctions have aimed to use spin-currents injected from an adjacent spin Hall NM for spin-transfer torque  (STT) switching of a FM \cite{Miron2011b,Liu2012}. In addition to this SHE induced torque there is also a spin-orbit torque (SOT) \cite{Bernevig2005c,Chernyshov2009}, which is generated via the inverse  spin galvanic  effect (ISGE) \cite{Belkov2008}. In the ISGE, a charge current can generate a non-equilibrium homogeneous spin-polarization via relativistic spin-orbit coupling and it is often a companion effect to the spin-current generating SHE \cite{Kato2004b,Kato2004d,Wunderlich2004,Wunderlich2005}.
These results underscore the relevance of the SHE for applications.

\begin{figure}[h!]
\hspace*{-0cm}\epsfig{width=1\columnwidth,angle=0,file=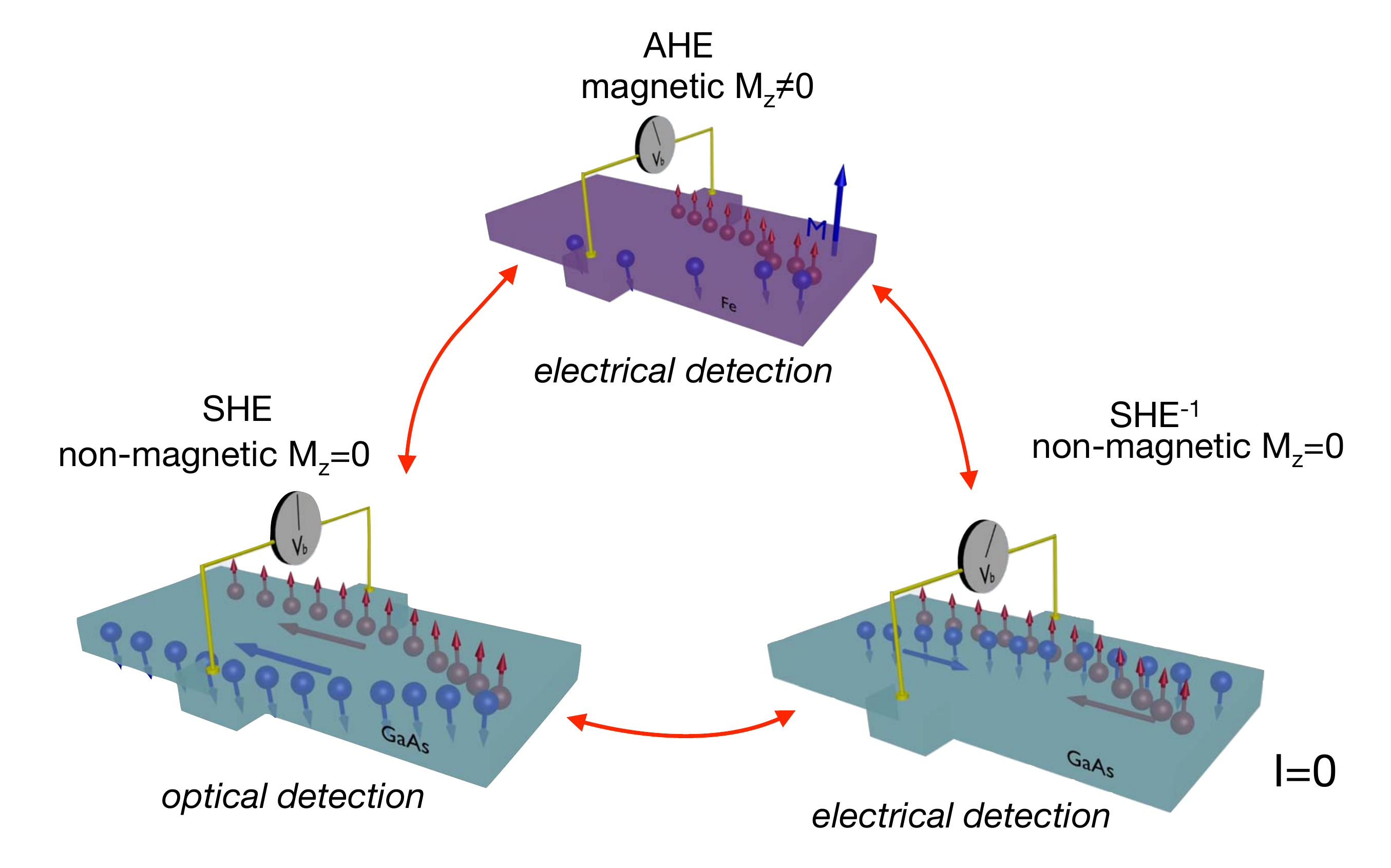}
\caption{An illustration of the connected family of the spin-dependent Hall effects. In the AHE, a charge current generates a polarized transverse charge current. In the SHE an unpolarized charge current generates a transverse pure spin-current. In the ISHE a pure spin-current generates a transverse charge current.}
\label{SHEfamily}
\end{figure}

As already mentioned, the SHE borrows directly from the physics and mechanisms of the AHE and correspondingly much of their descriptions are parallel. The family of these three key spin-dependent Hall effects is illustrated in Fig.~\ref{SHEfamily}. The important caveat is that, unlike the AHE which correlates charge degrees of freedom via relativistic spin-orbit interaction, the SHE and ISHE correlate the charge degree of freedom, a conserved quantity, and the spin degree of freedom, a non-conserved quantity subject to decay and dephasing.

\vspace*{0cm}

The aim of this review is to survey the rapid developments of the SHE field, to give an overview of its current experimental understanding, 
the basic theoretical tools that are being applied to describe it and their current level of success and limitations, the connection to important related phenomena, as well as the potential of the SHE for applications, particularly in the area of magnetization dynamics. 

Given the enormous work that has been done in just a decade, we can only highlight what we have deemed the reports
that have contributed significantly to the field. As with any review we are shackled by our own views and the reader 
interested in this field should complement this reading  with other recent reviews of the subject, such as \cite{Hoffmann2013,Maekawa2012,Valenzuela2012,Jungwirth2012,Hankiewicz2009,Sinova2008,Culcer2009,Vignale2010,Raimondi2012}.

\section{Overview}
\label{over}

In this section we provide an overview that starts from the original seeds of the SHE field and  connects afterwards to the broader context of the phenomenon within 
spintronics. The overview is organized as follows: First we look back to how the  Mott scattering of electron beams in vacuum and the skew scattering of electrons  in FMs germinated into the prediction of the extrinsic SHE in NMs. 
 Second we discuss that in a solid-state system there is in addition an intrinsic spin-deflection, arising from the internal spin-orbit coupling forces in a perfect crystal. This key distinction from electrons in a vacuum 
 makes the spin-dependent  Hall physics in condensed matter systems much richer.
 We also note here the connection of this intrinsic mechanism  to the quantum Hall effects. Third, we summarize 
 studies of spin injection and detection in hybrid FM/NM structures, which were particularly impactful on the research of SHE.
 Here we highlight DC transport as well as AC ferromagnetic resonance (FMR) experiments. 
 Finally, we connect the physics of the SHE, which 
 considers pure spin-currents and non-uniform spin accumulations, to 
 the physics of the spin galvanic effects. The latter effects represents a seemingly distinct family of relativistic phenomena 
 relating to the generation or detection of uniform non-equilibrium spin-polarizations.
  However, as we point out,  the spin Hall and spin galvanic effects can have common features in their microscopic physical origins and both can generate current-induced torques in magnets. These two relativistic effects are now at the forefront of current-induced magnetization dynamics research aimed at future spintronic technologies. 

\subsection{Spin Hall, anomalous Hall, and Mott polarimetry}
\label{mott}
In their original work, \onlinecite{Dyakonov1971}  referred to the 
phenomena of  Mott scattering \cite{Mott1929} and of the AHE  \cite{Hall1881}  to  theoretically predict  the extrinsic SHE.
 In particular, they pointed out the following: (i) Spin-dependent asymmetric deflection is observed in electron beams in  vacuum due to Mott scattering \cite{Mott1929,Mott1932,Shull1943,Gay1992}. (ii) Mott's skew scattering is regarded among the origins of the AHE  of electron carriers in FMs \cite{Karplus1954,Smit1955,Smit1958,Berger1979,Nagaosa2010}. The two points imply that under an applied electrical current, asymmetric spin-dependent deflection should occur  in NMs. Unlike in FMs, NMs in equilibrium have the same number of spin-up and spin-down electrons and no  transverse charge imbalance will occur. Instead, the SHE generates an edge spin-accumulation that has opposite polarization at  opposite edges. 

Let us now explore the Mott scattering seed  of the SHE in more detail.  Mott proposed its scattering experiment \cite{Mott1929,Mott1932} to provide a direct evidence that spin, inferred four years earlier from atomic spectra \cite{Uhlenbeck1925}, is an intrinsic property of a free electron. Mott anchored his proposal in the then recently derived Dirac equation \cite{Dirac1928}. The ensuing quest for the experimental verification of Mott scattering \cite{Shull1943} was among the founding pillars not only for verifying the  electron spin but over the entire relativistic quantum mechanics concept. Since Mott scattering   of electron beams from heavy nuclei in a vacuum chamber can be regarded as the SHE in a non-solid-state environment, the seeds of the SHE date back to the very foundations of the electron spin and relativistic quantum mechanics.

\begin{figure}[h!]
\hspace*{-0cm}\epsfig{width=0.8\columnwidth,angle=0,file=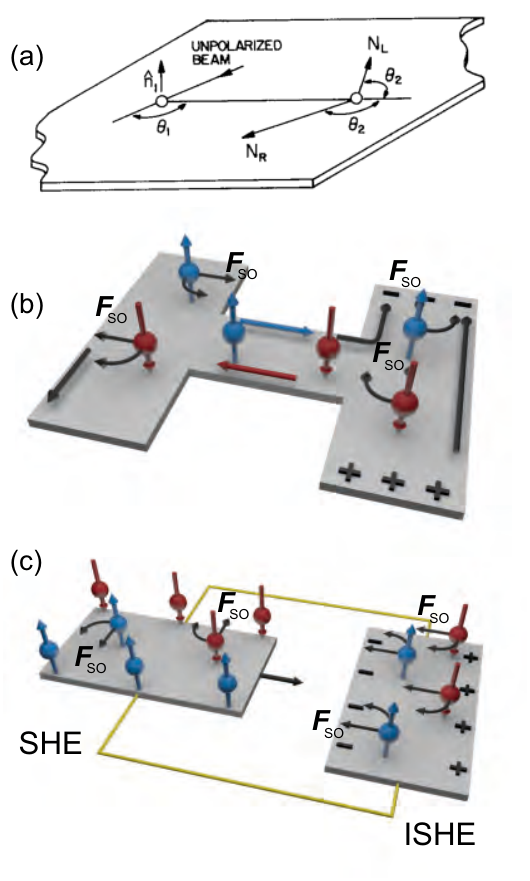}
\vspace*{-0.5cm}
\caption{(a) Schematics of \onlinecite{Mott1929} original double-scattering proposal, (b) SHE/ISHE analogue of Mott double-scattering in \onlinecite{Hankiewicz2004b} H-bar device, (c) SHE (left) and ISHE (right) wired as proposed by \onlinecite{Hirsch1999}. Instead of directly injecting a spin-current generated in the SHE part of the experiment, as suggested by Mott and by Hankiewicz {\em et al.}, Hirsch considered that the pure spin-current is generated from the opposite spin-accumulations at the edges of the SHE part of the "double-scattering" device. $F_{SO}$ represents an effective spin-orbit force that deflects the spins in the SHE/ISHE. Panel (a) from Ref. \onlinecite{Gay1992}.}
\label{fig_exp_MHH}
\end{figure}

Figure~\ref{fig_exp_MHH}(a) shows \onlinecite{Mott1929}  double-scattering experiment proposal. First, an unpolarized beam of electrons is scattered from heavy nuclei in a target. Because of the relativistic spin-orbit coupling, large angle ($\sim 90^\circ$) scattering from the first target produces a polarized beam with the spin-polarization transverse to the scattering plane. Scattering of these polarized electrons from the second target results, again due to the spin-orbit coupling, in a left-right scattering asymmetry that is proportional to the polarization induced by the first scattering.

\vspace*{0cm}

In a complete analogy to the Mott double-scattering effect, but instead of vacuum now considering a solid state system, \onlinecite{Hankiewicz2004b} proposed an H-bar microdevice schematically shown in Fig.~\ref{fig_exp_MHH}(b). An unpolarized electrical current driven through the first leg of the device generates a transverse spin-current due to the relativistic SHE. The spin-current injected into the second leg generates, via the ISHE, an electrical current, or  in an open circuit geometry a voltage across the second leg. This H-bar SHE/ISHE experiment has been realized in a NM semiconductor by \onlinecite{Brune2010}.

In Fig.~~\ref{fig_exp_MHH}(c) we show an earlier variant of the double-scattering experiment proposed by  \onlinecite{Hirsch1999} for observing the SHE/ISHE in a solid state device. Instead of considering the spin-current produced directly by the charge current via the SHE, Hirsch focused on the edges of the SHE sample. Here the transverse spin-current accumulates, forming a non-equilibrium spin polarization of opposite sign at the two opposite edges. In NMs, the non-equilibrium spin polarization corresponds to a splitting of the spin-up and spin-down chemical potentials. When connecting the two edges, the gradient of the spin-dependent chemical potentials will generate a circulating spin-current which is then detected by the ISHE spin-current meter inserted into the closed spin-current circuit. The idea for the experiment was borrowed from the ordinary Hall effect (HE) in which opposite charge accumulates at opposite edges due to the Lorentz force, and the resulting electro-chemical potential gradient  generates a circulating charge current when the two edges are connected in the closed circuit geometry.

Realizing \onlinecite{Hirsch1999} SHE/ISHE device remains a challenge. Similarly to the \onlinecite{Hankiewicz2004b} design directly copying the Mott double-scattering experiment, the wires connecting the SHE and ISHE parts of Hirsch's device have to be shorter than the characteristic spin-conserving length-scale. The spin-orbit coupling required for the SHE/ISHE in the first place, however, tends to make the spin life-time short. The additional complication  is that the spin-orbit coupling also limits, again via the finite spin life-time,   the width of the sample edge with non-zero spin accumulation from which the spin-current is extracted in Hirsch's device proposal.

While difficult to realize experimentally, Hirsch's concept  is stimulating for comprehending the general  key distinctions between charge and spin-current.  Electron charge is a conserved quantity but its spin direction is not conserved. In the charge HE, the difference between electro-chemical potentials at the edges determines the uniform charge current which in steady-state flows through the closed circuit. In the SHE, on the other hand, the spin-current in the connecting wire of Hirsch's device is not uniform and is not determined by the difference between the spin-dependent chemical potentials at the left and right edges. It is determined by the local gradient of the spin-dependent chemical potentials which vanishes, i.e. also the spin-current vanishes, on the length-scale given by the spin life-time. As long as the connecting wire is longer than the characteristic spin-conserving length-scale, there is no difference between a closed and an open spin-current circuit.

Hirsch's concept  also points to  the general applicability of the ISHE as an electrical  spin detector. Even in electrically open circuits, the non-conserving, non-uniform spin-current can still flow. It is then readily separated from the charge current and can be detected by the ISHE. The Mott polarimetry of electron beams in vacuum chambers and AHE polarimetry of charge currents in itinerant magnets can, therefore, be complemented by the ISHE polarimetry of pure spin-currents.

A spin-current in a NM of any origin (not only of the SHE origin) can be detected by the ISHE. Indeed, ISHE  detectors of pure spin-currents became a standard measurement tool. They led to, e.g., the discovery of the spin Seebeck effect \cite{Uchida2008,Uchida2010,Jaworski2010} and helped establishing the emerging field of spin caloritronics \cite{Bauer2012}.

Given the inherent challenges in realizing Hirsch's device it is not surprising that experimentalists initially avoided attempts to perform the SHE/ISHE "double-scattering" experiments and that the first  observations of the SHE \cite{Kato2004d,Wunderlich2004,Wunderlich2005} and ISHE \cite{Valenzuela2006,Saitoh2006,Zhao2006} were made separately. When the \onlinecite{Hankiewicz2004b} H-bar microdevice was eventually realized in experiment by \onlinecite{Brune2010}, both the SHE and ISHE had been already independently established.

\subsection{Intrinsic spin Hall and quantum Hall effects}
\label{he}

\vspace*{0cm}
Remarkably, the H-bar experiment \cite{Brune2010} discussed in the previous subsection was performed in a ballistic transport regime where the picture of Mott scattering, single or double, did not apply. A fundamental physics principle makes the SHE in solid-state systems richer than in the Mott electron beams scattered from spin-orbit coupled targets in vacuum chambers.  For electrons moving in a crystal, a transverse spin-dependent velocity can be generated by the relativistic spin-orbit field of a perfect crystal even in the absence of scattering. The roots of this intrinsic SHE are clearly distinct from the Mott (skew) scattering AHE and from the Mott scattering of  free electron beams.

\begin{figure}[h!]
\hspace*{-0cm}\epsfig{width=1\columnwidth,angle=0,file=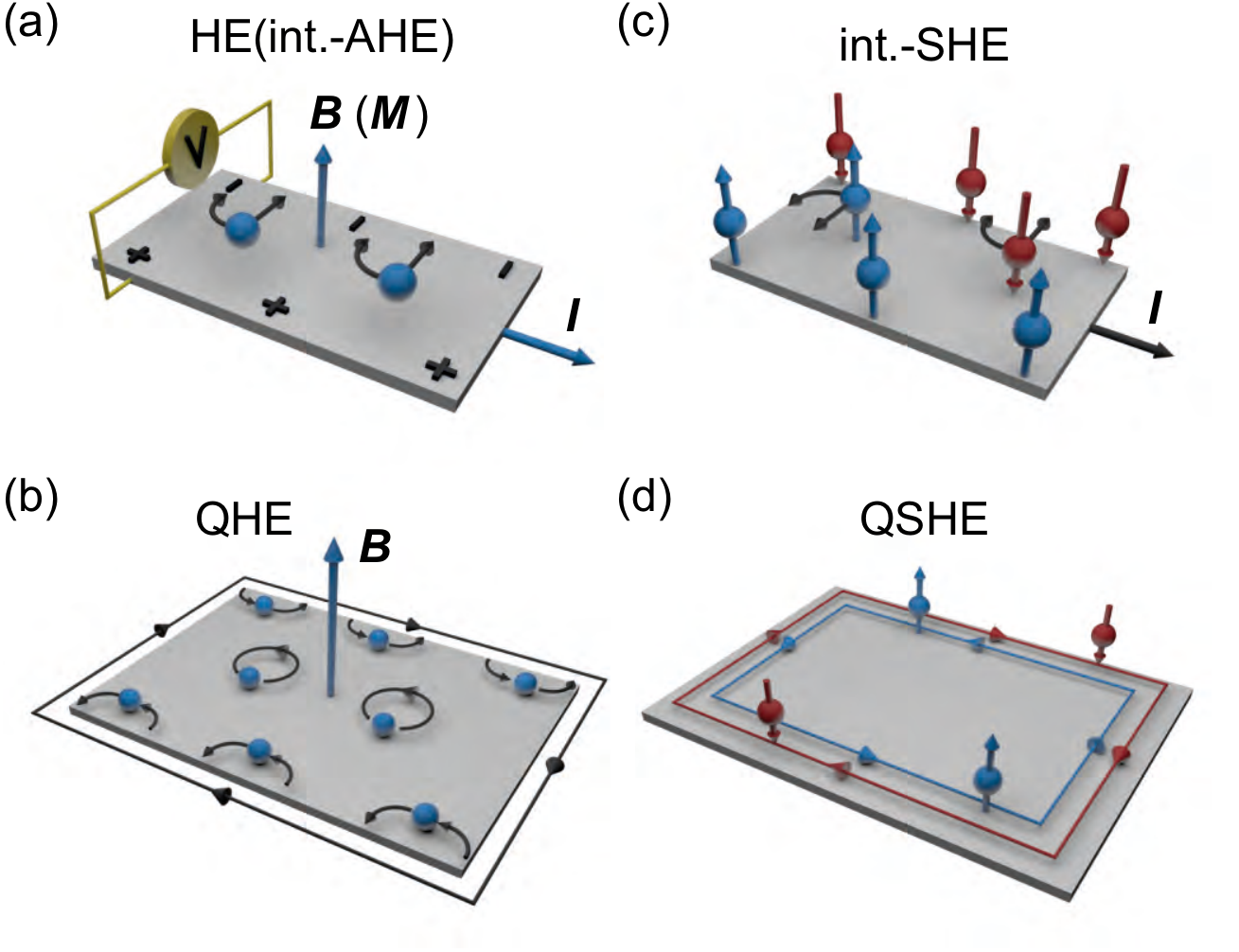}
%
\caption{Schematics of the HE and the AHE (a), the QHE (b), SHE (c),  and the QSHE (d). In the HE and QHE, the carrier deflection
is a reaction to the Lorentz force. In the cases of the intrinsic AHE, SHE and QSHE, the carriers experience an internal spin-orbit force.}
\label{fig_exp_Halls}
\end{figure}

The reactive term responsible for the intrinsic SHE is akin to the  ordinary HE in which the transverse deflection of electrons is a reaction to the Lorentz force of the applied magnetic field acting on the moving carriers (see Fig.~\ref{fig_exp_Halls}(a)).
In strong  magnetic fields, the quantum Hall effect (QHE) becomes a precise, disorder-independent measure of the quantum conductance $e^2/h$,  and the integer multiples of $e^2/h$ observed in the QHE correspond to the number of occupied dissipationless chiral edge states in the conductor (see Fig.~\ref{fig_exp_Halls}(b)).

Besides the externally applied Lorentz force, electrons moving in a crystal can experience an internal spin-orbit force. The effect was first recognized in FMs where it generates the intrinsic AHE (see Fig.~\ref{fig_exp_Halls}(a)) \cite{Karplus1954,Onoda2002,Jungwirth2002}. 
\onlinecite{Murakami2003} and \onlinecite{Sinova2004} predicted  that the same spin-orbit force derived directly from the relativistic band structure of a NM can induce the SHE without involving Mott scattering (see Fig.~\ref{fig_exp_Halls}(c)).  The first experimental observations confirmed that the SHE can indeed have the two distinct origins.  While  \onlinecite{Wunderlich2004,Wunderlich2005} ascribed the circularly polarized luminescence signal from the edge of the p-GaAs sample  to the intrinsic SHE, \onlinecite{Kato2004d} detected an edge Kerr rotation signal in $n$-GaAs due to the extrinsic, skew-scattering SHE.

Following the discovery of the phenomenon, the SHE experiments in semiconductors using optical spin detection  have explored the basic phenomenologies of the extrinsic and intrinsic SHEs \cite{Kato2004d,Wunderlich2004,Wunderlich2005,Nomura2005b,Sih2005a,Sih2006,Stern2006b,Chang2007,Stern2007,Stern2008,Matsuzaka2009}. They also experimentally demonstrated the potential of the SHE as a spin-current source \cite{Sih2006}. A two-color optical excitation technique with perpendicular linear polarizations of the incident laser beams was used to detect the ISHE in a semiconductor \cite{Zhao2006}.  The 
spin-current produced by the laser excitation is transferred due to the ISHE into a transverse electrical current, resulting in a spatially dependent charge accumulation which was detected by the optical transmission signal of a probe laser beam. These all-optical measurements in an intrinsic semiconductor were eventually performed on timescales shorter than the scattering time and have provided a direct demonstration of the intrinsic SHE signal \cite{Werake2011}.

The intrinsic SHE proposal  triggered an intense theoretical debate which is summarized in several review articles \cite{Murakami2005,Sinova2006,Schliemann2006,Engel2006,Sinova2008,Hankiewicz2009,Culcer2009,Vignale2010,Raimondi2012}. Combined with the established physics of the dissipationless QHE, it led to the prediction  and subsequent experimental verification of the quantum spin Hall effect (QSHE) \cite{Murakami2004,Kane2005b,Bernevig2006b,Konig2007,Hasan2010}. In the time-reversal symmetric QSHE, the chiral edge states of the QHE are replaced by pairs of helical spin-edge states (see Fig.~\ref{fig_exp_Halls}(d)). This leads to a $2e^2/h$ quantization of the observed transport signal \cite{Konig2007} and resistance values in nonlocal experiments that can be expressed
as specific integer fractions of the inverse conductance quanta \cite{Roth2009,Buttiker2009}.  The QSHE initiated the new research field of topological insulators \cite{Moore2010a,Hasan2010}.

\subsection{Spin Hall effect and non-magnetic/ferromagnetic hybrid structures}
\label{mag}

Among the early SHE device proposals, \onlinecite{Zhang2000} suggested to electrically detect the edge spin accumulation produced by the SHE using an attached FM probe \cite{Silsbee1980,Johnson1985}. In a broader context, the idea of connecting the SHE with the more mature field which utilized FMs for  injection and detection of spins in NMs fueled numerous studies of fundamental importance for the SHE field. Electrical spin injection from a FM contact and electrical observation of the ISHE on a Hall cross patterned in the NM was demonstrated by  \onlinecite{Valenzuela2006}. 

Metal spin Hall devices provided the demonstration of the electrical measurement of the SHE by an attached FM contact  \cite{Kimura2007}, as proposed originally by \onlinecite{Zhang2000}. They showed that the same NM electrode can generate  the SHE or the ISHE, i.e., can be used as an electrical spin injector or detector \cite{Valenzuela2006,Kimura2007,Vila2007,Seki2008,Mihajlovic2009}.

Compared to metals, semiconductor spin transport devices with FM metal electrodes can suffer from the problem of the resistance mismatch  which hinders efficient spin transport across the interface \cite{Schmidt2000}. The introduction of a highly resistive tunnel barrier between the FM metal electrode and the semiconductor channel solves this problem \cite{Rashba2000,Lou2007} and FM tunnel contacts were successfully used to detect the SHE-induced spin accumulation in a semiconductor \cite{Garlid2010}.  Similarly, an electrical spin-injection from a FM/semiconductor tunnel contact was used to demonstrate, side by side, the electrical spin detection by the ISHE and by the FM detection electrode \cite{Olejnik2012}.

\begin{figure}[h!]
\hspace*{-0cm}\epsfig{width=0.7\columnwidth,angle=0,file=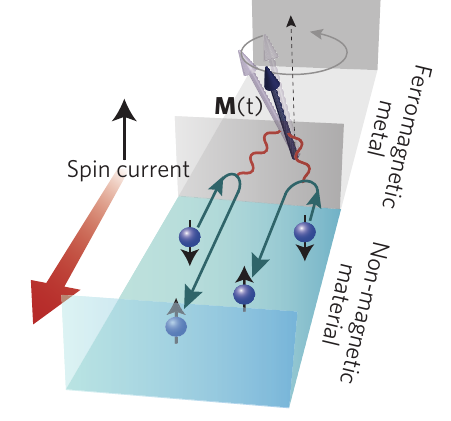}
\caption{Illustration of the SP spin-current generation by magnetization dynamics from a FM into a NM. From Ref.~\onlinecite{Ando2011}.}
\label{fig-Ando2011}
\end{figure}

Using FMs contributed significantly to the basic understanding of the SHE. Apart from the transport measurements, NM/FM hybrid structures also allow to combine the SHE physics with the field of magnetization dynamics. The ISHE and SHE can be investigated using spin-pumping (SP) and other related dynamic methods in structures comprising FMs and NMs, as illustrated in Fig.~\ref{fig-Ando2011} \cite{Saitoh2006,Ando2008b,Ando2009,Mosendz2010,Mosendz2010b,Liu2011,Czeschka2011,Miron2011b,Liu2012,Saitoh2012,Bai2013a,Wei2014}. In return, the SHE was found to provide efficient means for injecting spin-currents into the FM, generating the STT \cite{Ralph2008}, and by this electrically controlling magnetization in FMs with potential applications in spintronic information technologies  \cite{Miron2011,Miron2011b,Liu2012,Emori2013,Ryu2013}. Moreover, the ISHE detection of pure spin-currents did not remain limited to NMs but is now used also in FMs \cite{Miao2013,Azevedo2014} and antiferromagnets \cite{Freimuth2010,Mendes2014}.

In general, when SHE induced torques in the adjacent FM are considered in the description of the dynamic magnetization (the Landau-Lifshitz-Gilbert equation), two types of torques can occur. An (anti)damping-like torque which has the same functional shape as the Gilbert damping term (and thus can manifest itself in an increased or decreased Gilbert damping) and a field-like term which alters the magnetic energy landscape and can be observed as a shift of the resonance line in a FMR experiment. FMR allows the determination of the total internal magnetic field in a sample as well as investigation of dissipation. Thus, in principle FMR like techniques enable determination of field-like and (anti)damping-like contributions of SHE induced torques.

In ISHE experiments using FMR techniques a detailed analysis of field-symmetric and field-antisymmetric contributions of the detected DC output voltage at FMR allows in principle a quantitative determination of the strength and symmetry of the SHE induced torques, as well as the spin Hall angle.

 Note that the torque can induce not only the small-angle FMR precession but the lateral current along a NM/FM interface can drive domain walls at high velocities \cite{Miron2011,Emori2013,Ryu2013} or switch the magnetization in the FM \cite{Miron2011b,Liu2012}.
This may have practical implications for designing domain-wall based memories or for three-terminal magnetic tunnel junction bits with the lateral writing current decoupled from the perpendicular read-out current.

\subsection{Spin Hall effect, spin galvanics, and spin torques}
\label{conv}
From the early experiments with the relativistic torques it was realized that the SHE is not the only possible mechanism responsible for torques induced by the lateral current in the NM/FM bilayers \cite{Manchon2008b}. The  interface breaks the structural inversion symmetry which implies that the SHE-STT can be accompanied by another microscopic mechanism. Its origin is in the so called spin galvanic phenomena that were explored  earlier in inversion-asymmetric NMs \cite{Ivchenko2008}. In the picture discussed in the previous section, the spin-current generated in the NM via the relativistic SHE is absorbed in the FM and induces the  STT. In the competing scenario, a non-equilibrium spin-density  of carriers is generated  in inversion asymmetric systems via the relativistic ISGE \cite{Silov2004,Kato2004b,Ganichev2004b,Wunderlich2004,Wunderlich2005,Ivchenko2008,Belkov2008}. A SOT is then directly  induced  if the carrier spins are exchange coupled to magnetic moments \cite{Bernevig2005c,Manchon2008b,Chernyshov2009,Miron2010}.

\begin{figure}
\centering
\includegraphics[width=8cm]{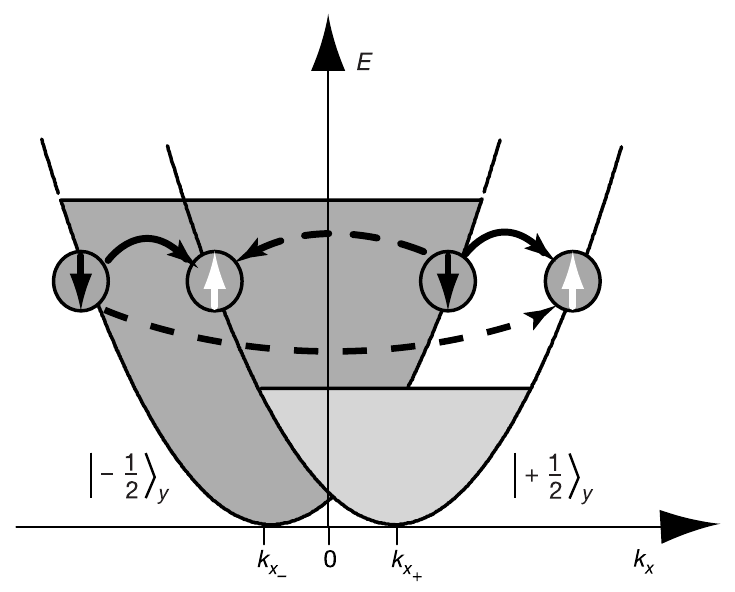}
\caption{Microscopic origin of the spin galvanic current in the presence of $k$-linear terms in the electron Hamiltonian. The $\sigma_yk_x$ term in the Hamiltonian splits the conduction band into two parabolas with the spin $\pm1/2$ in the $y$-direction. If one spin sub-band is
preferentially occupied, for example, by spin injection (the $|-1/2\rangle_y$-states shown in the figure) asymmetric spin-flip scattering results in a current in the $x$-direction. The rate of spin-flip scattering depends on the value of the initial and final $k$-vectors. There are four
distinct spin-flip scattering events possible, indicated by the arrows. The transitions sketched by dashed arrows yield an asymmetric occupation of both sub-bands and hence a current flow. If, instead of the spin-down sub-band, the spin-up sub-band is preferentially occupied the current direction is reversed. From Ref.~\onlinecite{Ganichev2002}.}
\label{SGE}
\end{figure}

From the early observations in NM semiconductors, SHE and ISGE are known as companion phenomena, both  allowing for electrical alignment of spins in the same structure \cite{Kato2004d,Kato2004b,Wunderlich2004,Wunderlich2005}. Hand in hand, SHE and ISGE evolved from subtle academic phenomena to efficient means for electrically reorienting magnets. Understanding the relation between the spin Hall and spin galvanic phenomena is, therefore, important not only from the basic physics perspective but has also practical implications for spintronic devices.

The term spin galvanic effect (SGE) is derived from the analogy to the galvanic (voltaic) cell. Instead of a chemical reaction, however, it is the spin polarization that generates an electrical current (voltage) in the SGE. Inversely, an electrical current generates the spin polarization in the ISGE.

Following theoretical predictions of the phenomena \cite{Ivchenko1978,Ivchenko1989,Aronov1989,Edelstein1990,Malshukov2002}, it was the SGE that was initially observed  in an asymmetrically confined 2D electron gas in a GaAs quantum well \cite{Ganichev2002}. The key signature of the SGE is the electrical current induced by a non-equilibrium, but uniform polarization of electron spins. The microscopic origin of the effect is illustrated in Fig.~\ref{SGE}. In the non-equilibrium steady-state, the spin-up and spin-down sub-bands have different populations, induced in the \onlinecite{Ganichev2002} experiment by a circularly polarized light excitation. Simultaneously, the two sub-bands for spin-up and spin-down electrons are shifted in momentum space due to the inversion asymmetry of the semiconductor structure  which leads to an inherent asymmetry in the spin-flip scattering events between the two sub-bands. This results in the flow of the electrical current.

\begin{figure}[h!]
\hspace*{-0cm}\epsfig{width=1\columnwidth,angle=0,file=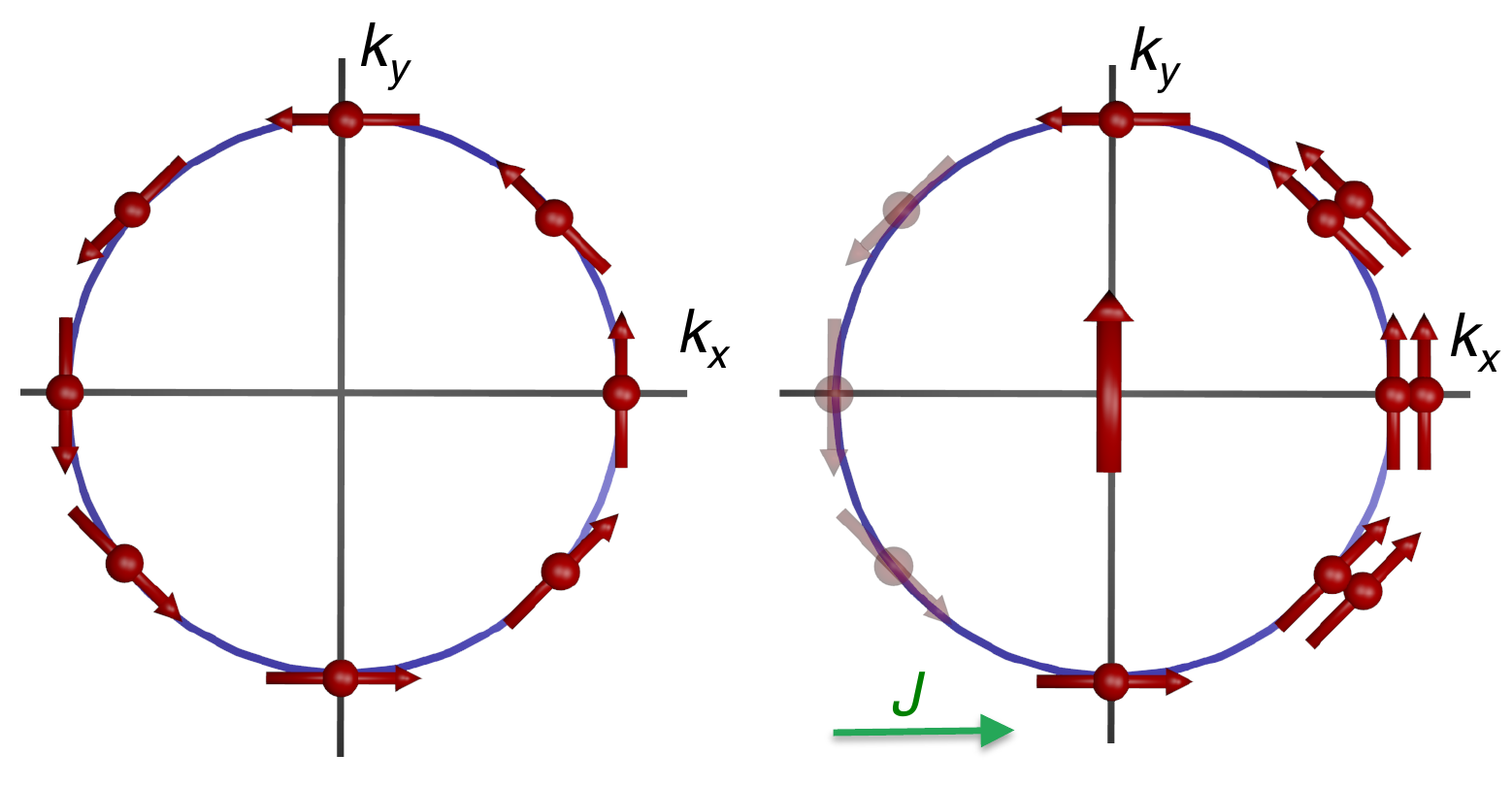}
%
\caption{Left panel: Rashba spin-texture in equilibrium with zero net spin-density. Right panel: Non-equilibrium redistribution of eigenstates in an applied electric field resulting in a non-zero spin-density due to broken inversion symmetry of the spin-texture.
}
\label{fig_ISGE}
\end{figure}

A microscopic picture of the ISGE is illustrated in Fig.~\ref{fig_ISGE}. The uniform non-equilibrium spin-density  occurs as a consequence of an electric-field and scattering induced redistribution of carriers  on the Fermi surface whose texture of spin expectation values has a broken inversion symmetry. For the Rashba spin-orbit coupling, illustrated in Fig.~\ref{fig_ISGE}, the uniform in-plane spin polarization is perpendicular to the applied electrical current.

Initial observations of the ISGE were made in parallel with the initial SHE experiments, in both cases employing the Kerr/Faraday magneto-optical detection methods or circularly polarized luminescence \cite{Silov2004,Kato2004b,Kato2004d,Ganichev2004b,Wunderlich2004,Wunderlich2005,Ivchenko2008,Belkov2008}. \onlinecite{Kato2004b,Kato2004d} observed the SHE and ISGE in the same strained bulk n-InGaAs sample and  \onlinecite{Wunderlich2004,Wunderlich2005} detected the two effects in the same asymmetrically confined 2D hole gas (2DHG) in a AlGaAs/GaAs heterostructure.

Subsequently, it  was predicted \cite{Bernevig2005c} and experimentally verified \cite{Chernyshov2009} that the ISGE can generate relativistic SOTs in a FM semiconductor (Ga,Mn)As with broken inversion symmetry in the strained crystal structure of a thin film sample.
Both the ISGE and SHE based mechanisms have been found to contribute to the relativistic spin torques in the NM/FM bilayers with broken structural inversion symmetry \cite{Manchon2008b,Miron2010,Pi2010,Suzuki2011,Miron2011b,Kim2013,Garello2013,Pai2014}.

As mentioned earlier, the SHE and the Mott scattering of free electron beams can have the same skew scattering origin which is captured by the second-order Born approximation (third order in the scattering potential). Moreover, in condensed matter systems, the SHE can arise from the spin-dependent transverse deflection  induced by the intrinsic spin-orbit coupling in a perfect crystal with no impurities. We have also already mentioned that this intrinsic SHE has its direct counterpart in systems with broken time reversal symmetry in the intrinsic AHE.

\begin{figure}[h!]
\hspace*{-0cm}\epsfig{width=1\columnwidth,angle=0,file=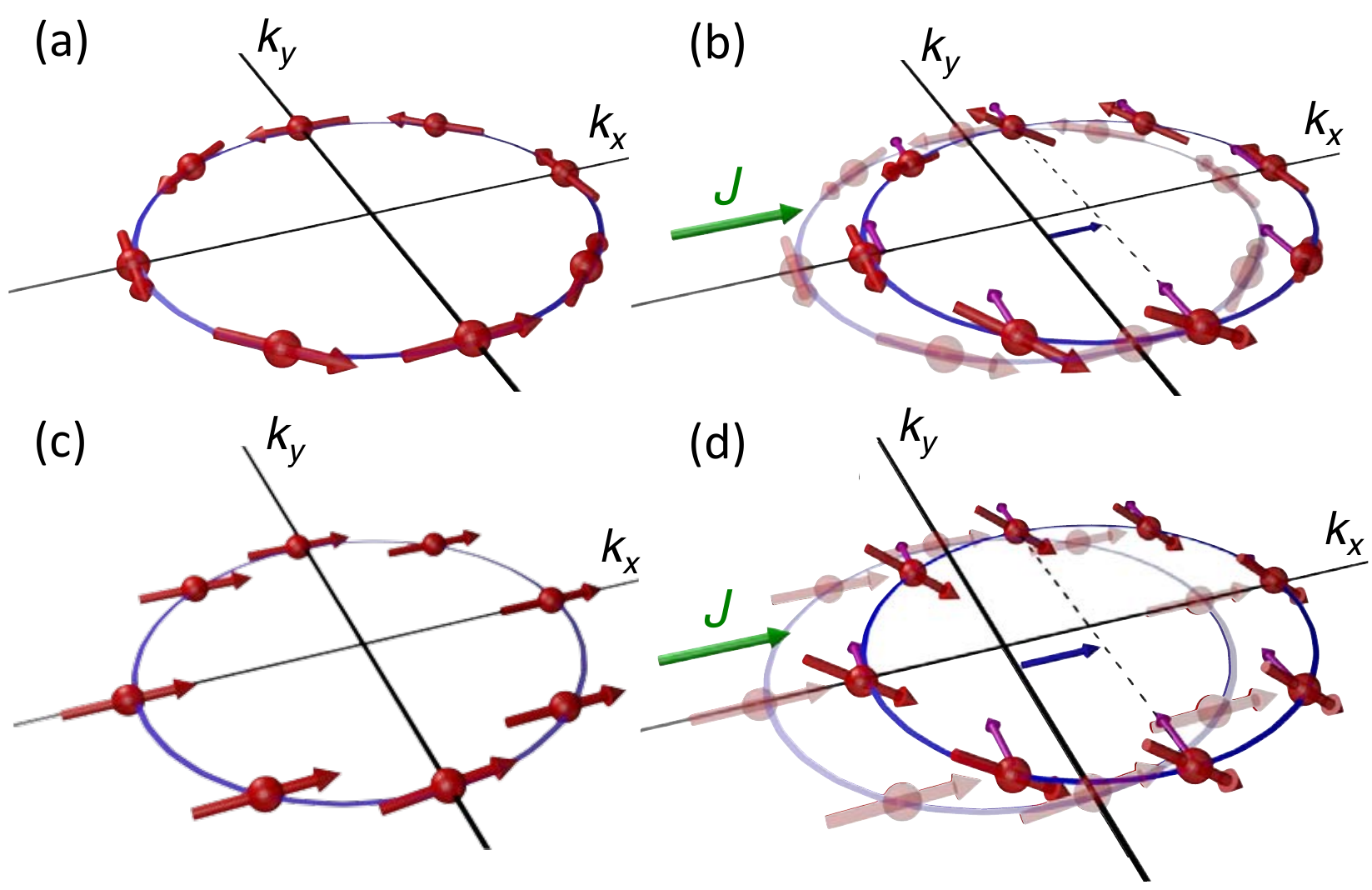}
%
\caption{(a) A model equilibrium spin texture in a 2D Rashba spin-orbit coupled system with spins (red arrows) pointing perpendicular to the momentum.  (b) In the presence of an electrical current along the $x$-direction the
Fermi surface (circle) is displaced along the same direction. When moving
in momentum space, electrons experience an additional spin-orbit field (purple arrows). In reaction to this non-equilibrium current induced field,  spins tilt up for $k_y>0$ and down for $k_y<0$, creating a spin-current in the $y$-direction. (c) A model equilibrium spin texture in a 2D Rashba spin-orbit coupled system with an additional time-reversal symmetry breaking exchange field of a strength much larger than the spin-orbit field. In equilibrium, all spins in this case align approximately with the direction of the exchange field. (d) The same reactive mechanism as in (b) generates a uniform, non-equilibrium out-of-plane spin-polarization. From  Refs.~\onlinecite{Sinova2004,Kurebayashi2014}.
}
\label{fig_exp_Berry}
\end{figure}

The spin galvanic phenomena, on the other hand, are traditionally considered to originate in NMs only from extrinsic origins (seen already in the first-order Born approximation scattering). Nevertheless,  the physics of the SHE, AHE, spin galvanics, and relativistic spin torques can be entangled even when considering the intrinsic effects. In Fig.~\ref{fig_exp_Berry} we illustrate that the same current-induced reactive mechanism that generates the transverse spin-current in the intrinsic SHE can induce a uniform spin polarization, i.e. a signature characteristic of the ISGE,  in systems with broken space and time reversal symmetry. Relativistic SOTs generated by the non-equilibrium uniform spin polarization of this intrinsic origin were identified in the FM semiconductor (Ga,Mn)As \cite{Kurebayashi2014}.

\section{Theory of spin Hall effect}

The SHE is a prime example of a field germinated directly from several key theoretical predictions and one which needed the correct timing to come to its full life. It all began with the seminal prediction 
of the extrinsic SHE by \onlinecite{Dyakonov1971}  based on a phenomenological theory that 
considered the consequences of chiral Mott scattering in a solid-state system. 
This  prediction laid  dormant for almost three decades until \onlinecite{Hirsch1999} and  \onlinecite{Zhang2000} made a similar prediction, but did it at a time that the nascent field of spintronics could fully exploit 
the notion of the SHE.

Shortly after this,  \onlinecite{Murakami2003} and \onlinecite{Sinova2004} predicted  the intrinsic SHE based on linear response microscopic theories of strong spin-orbit coupled materials. It is perhaps at this point that the field of SHE surged forward in a flurry of enormous theoretical activity culminating later on the parallel discoveries of the
extrinsic \cite{Kato2004} and intrinsic \cite{Wunderlich2004,Wunderlich2005} SHE. 

The theories of the SHE have naturally emerged from the  theory of the AHE. 
However, the latter, although remaining  complex and for many decades quite controversial, relies on the very important
pillar of charge conservation  for its development. 
The ever-present key difference between the SHE and the AHE is that spin, unlike charge, is not a conserved quantity in most cases, which makes the examination of experiments and predictions more involved in the SHE.

In the initial predictions of the extrinsic SHE this was dealt with by writing down phenomenological theories based on coupled spin-charge drift-diffusion equations derived from symmetry considerations. This approach is  well justified in the weak spin-orbit coupling regime \cite{Dyakonov1971,Dyakonov2008,Hirsch1999,Zhang2000}. However, within the strong spin-orbit couple regime of, e.g., heavy transition metals, the dominant coherent effects of the intrinsic SHE are more difficult to couple to such phenomenological theories. 

It is within this strong spin-orbit coupling regime that the AHE has made its furthest progress within the last decade by
systematic approaches that aimed at reaching agreement in non-trivial models among linear response theories based on different formalisms that, ultimately, should provide equivalent predictions. This has led to a better established microscopic theory of the AHE \cite{Sinitsyn2007,Kovalev2010,Nagaosa2010} and a full understanding of its mechanisms that have now been linked to the earlier ad-hock introduction of the semiclassically defined mechanisms: intrinsic, skew scattering, and side-jump scattering.  

We spend the first part of this section, Sec.~\ref{mechanisms}, defining and explaining each of those contribution and their origins in the more modern parsing of the spin-dependent Hall transport theory. We will try to clarify in particular the typical misconceptions that sometimes linger in the literature regarding which aspect of the spin-orbit coupling - within the crystal itself or within the disorder potential - contribute to each mechanism. We will borrow in this part extensively from \onlinecite{Nagaosa2010} and we will direct the reader to this previous review for detailed explanations of the different linear response theories and the resolution of some of the historical controversies.

We follow in Sec.~\ref{drift-diffusion} by a description of the phenomenological spin-charge drift-diffusion equations that are often used to fit experiments. 
Because of the challenge of merging the strong spin-orbit coupled microscopic theories and the phenomenological weak spin-orbit coupling theories, one of the more popular models that is used to describe the SHE is  based on a simple Hamiltonian in which the spin-orbit coupling is only present in the disorder potential. 
We discuss such a model in Sec.~\ref{CB-model}. This theory has the benefit of having a single parameter - the strength of the spin-orbit coupling parameter for the disorder - which can be fitted to the spin-diffusion length and from this value estimate the SHE angle \cite{Crepieux2001b,Engel2005,Zhang2000,Maekawa2012}. However, as seen by comparing to experiment, it gives sensible results in the weak spin-orbit coupling regime but misses the coherent effects of the band structure in strongly spin-orbit coupled materials. 

A smaller number of microscopic ab-initio theory studies on impurities indicate also a significant contribution due to skew scattering \cite{Zimmermann2014,Long2014a}.

In Sec.~\ref{spin-pumping-theory} we discuss in detail the theory of SP and how it is utilized to measure the ISHE and the spin Hall angle. 
It introduces the concept of the spin-mixing conductance \cite{Tserkovnyak2002}, 
another parameter borrowed from the weak spin-orbit coupled systems, 
which is at present often used in analyzing magnetization dynamics experiments in connection to the physics of the SHE \cite{Saitoh2006}. 
Besides introducing the basic concepts of SP and 
its connection to the measurements of the ISHE, we
discuss the range of assumptions and limits which are often used in experiments.

Lastly, in Sec.~\ref{sec:theory:Kubo}, we present the formalism primarily used in 
the strong spin-orbit coupled systems. This formalism is  based in the Kubo formula and exploited successfully in transition metals \cite{Tanaka2008,Freimuth2010}. Calculations seem to indicate that the principal contribution - as in the AHE - arises from the intrinsic deflection mechanism. 

\subsection{Mechanisms of the spin Hall effect}
\label{mechanisms}
The family of spin-dependent Hall effects (AHE, SHE, and ISHE) 
originate from three distinct microscopic mechanism that they all share. 
These mechanisms have been first identified in the AHE \cite{Nagaosa2010}.
The 
mechanisms 
originate from coherent  band mixing effects due to the external electric field and the disorder potential. It makes them
more complex than the simpler single-band diagonal transport. 
As with other coherent interference transport phenomena,
 they cannot be satisfactorily explained using traditional semiclassical Boltzmann theory. It is then not surprising that the original proposals of the intrinsic, skew scattering, and side-jump mechanisms  which were based on semiclassical theory considerations brought in both insightful new concepts but also seeds for ensuing controversies in the debate over the quantum-mechanical microscopic origins of the AHE.  
While initially based in the semiclassical theory, there exists now a more modern strincter definition of the mechanisms within microscopic theories.
However, to keep continuity and not create further confusion, this more modern parsing of the different contributions has inherited the 
already established lexicon
(see Ref.~\onlinecite{Nagaosa2010}, Sec.~IV).

The modern parsing of the microscopic mechanisms is based on both experimental and microscopic transport
theory considerations, rather than on the identification of one particular effect from within semiclassical theory. 
The justification here is of course primarily on the AHE, not the SHE, for which the spin-Hall conductivity and its consequences have to 
be ultimately coupled  to the spin-accumulation that it induces and can therefore depend on the method of measurement.
In other words, depending on the measurement 
the spin accumulation may vary, e.g. in non-local transport measurements vs. FMR based measurements.

The link to semiclassically defined processes, as they have been historically attributed, is established through the works on AHE after  
 developing a fully generalized Boltzmann transport theory which takes  inter-band coherence effects into
account  and is fully equivalent to  multi-band microscopic theories. 
The key recent developments  have been 
 in understanding  the link between  semiclassical and microscopic theory of spin-dependent Hall transport.
 
A very natural classification of contributions based on the AHE, which is guided by experiment and by microscopic theory of 
metals, is  to separate them according to their dependence on the Bloch state transport lifetime $\tau$. 
Within the metallic regime, disorder is treated perturbatively and 
higher order terms  vary with a higher power of the quasiparticle scattering rate $\tau^{-1}$.
 As we will discuss, it is relatively easy to identify contributions to the 
anomalous or spin Hall conductivity, $\sigma_{xy}^{H}$, which vary as $\tau^{1}$ and as $\tau^{0}$.  
In experiments of the AHE a similar  separation can sometimes be achieved  by plotting $\sigma_{xy}$ vs.
 the  longitudinal conductivity $\sigma_{xx} \propto \tau$, when $\tau$ is 
varied by  altering disorder or varying temperature.  

We emphasize that several microscopically distinct terms can share the same $\tau$-dependence. We also note that in this parsing of the AHE and SHE contributions it is the dependence on $\tau$ (or $\sigma_{xx}$) 
which defines it, not a particular mechanism linked to a microscopic or semiclassical theory.
The contribution proportional to $\tau^1$ we {\em define} as the {\em skew-scattering} contribution, $\sigma_{xy}^{H-skew}$.
The second contribution  proportional to $\tau^0$  (or independent of $\sigma_{xx}$) 
 we further separate into two terms: {\em intrinsic} and {\em side-jump}.
 
 The first  term arises from the evolution of spin-orbit coupled quasiparticles as they are accelerated by an external electric field, whereas the second 
 term arises from scattering events from impurities that do not include the skew scattering contribution.  
Although the intrinsic and side-jump terms cannot be separated simply 
experimentally by DC measurements, they in principle {\em can} be separated experimentally (as well as theoretically)
by defining the intrinsic term, $\sigma_{xy}^{H-int}$,
as the extrapolation of the ac-Hall conductivity to zero frequency in the limit of $\tau \rightarrow \infty$,
with $1/\tau\rightarrow 0$ faster than $\omega\rightarrow 0$. This then leaves a unique definition for the 
 side-jump term, as $\sigma_{xy}^{H-sj}\equiv \sigma_{xy}^{H}-\sigma_{xy}^{H-skew}-\sigma_{xy}^{H-int}$. 

We further describe these contributions below. We note that the above definitions have 
not relied on linking the terms to semiclassical processes such as side-jump scattering \cite{Berger1979} or skew-scattering from asymmetric 
contributions to the semiclassical scattering rates \cite{Smit1958} identified in earlier theories. 

The ideas explained briefly in this section are substantiated in the recent review \onlinecite{Nagaosa2010}, 
which analyses the tendencies in the AHE data of several  material classes, 
and extensively  discusses the AHE theory. The extensions to the other spin-dependent Hall effects, such 
as SHE and ISHE, require the coupling of this spin-current generating mechanisms to spin-charge drift-diffusion transport equations 
that are appropriate to describe the particular experimental measurement, be it optical or electrical.

\subsubsection{Intrinsic mechanism}\label{int-intro}
Among the three contributions,  
the easiest to evaluate accurately and the one that has dominated most theoretical studies is the intrinsic contribution.  
We have defined the intrinsic 
contribution microscopically as the {\em dc} limit of the 
interband spin Hall conductivity with $1/\tau \rightarrow 0$ faster than the frequency. There is however a 
direct link to  the semiclassical theory   in which
the induced interband coherence is captured by a momentum-space
Berry-phase related contribution to the anomalous velocity.

In the context of the AHE, this contribution was first derived by \onlinecite{Karplus1954} but
its topological nature was not fully appreciated until recently \cite{Onoda2002,Jungwirth2002}. 
The work of  \onlinecite{Jungwirth2002} 
was motivated by the experimental importance of the AHE
in FM semiconductors and also by the analysis of the relationship
between momentum space Berry phases and anomalous transverse velocities in semiclassical transport theory \cite{Xiao2010,Sundaram1999}.  
Its connection to the SHE was described by 
\onlinecite{Murakami2003} and \onlinecite{Sinova2004}.

The intrinsic contribution to the spin Hall conductivity is 
dependent only on the band structure of the perfect crystal, hence
its name.  Pictorially it can be seen to arise from the non-equilibrium electron dynamics of the Bloch electrons as they 
are accelerated in an electric field and undergo spin-precession due to the induced electric field, as illustrated in Sec.~\ref{conv} Fig.~\ref{fig_exp_Berry}
for a Rashba spin-orbit coupled Hamiltonian. Here the 2D Rashba system is described by the 
Hamiltonian:
\begin{equation}
H=\frac{p^2}{2m}-\frac{\lambda}{\hbar}\vec{\sigma}\cdot(\hat{z}\times\vec{p}),
\label{rashham}
\end{equation}
where  $p=\hbar k$ is the 2D electron momentum, $\lambda$ is the Rashba coupling constant, $\vec{\sigma}$ the Pauli matrices, $m$ 
the electron effective mass, and $\hat{z}$ the unit vector perpendicular to the 2D electron gas (2DEG) plane.

For this example the dynamics of an electron spin in the presence of time-dependent 
spin-orbit coupling is described by the Bloch equation \cite{Sinova2004}: 
\begin{equation}
\frac{\hbar d \hat{n}}{dt} = \hat{n} \times \vec{\Delta}(t) + \alpha \; 
\frac{\hbar d \hat{n}}{dt} \times \hat{n} ,
\label{lleq}
\end{equation}
where $\hat n$ is the direction of the spin and $\alpha$ is a damping parameter that we assume is small. 
For the application of Eq.~(\ref{lleq}) we have in mind the $\vec{p}$ dependent 
spin-orbit coupling term in the spin-Hamiltonian,  $- \vec{s} \cdot \vec{\Delta}/\hbar$, where
$\vec{\Delta}=2\lambda/\hbar(\hat{z}\times\vec{p})$. 
For a Rashba effective magnetic field with magnitude $\Delta_1$ that initially points in the
$\hat x_1$ direction then tilts (arbitrarily slowly) slightly toward $\hat x_2$, where 
$\hat x_1$ and $\hat x_2$ are orthogonal in-plane directions, it follows from the linear 
response limit of Eq.~(\ref{lleq}) that 
\begin{eqnarray}
\frac{\hbar d n_2}{dt} &=& n_z \Delta_1 + \alpha \; dn_z/dt \nonumber \\
\frac{\hbar d n_z}{dt} &=& - \Delta_1 n_2 -\alpha \; dn_2/dt +\Delta_2,
\label{linearizedlleq}
\end{eqnarray}
where $\Delta_2=\vec{\Delta}\cdot\hat{x}_2$.   By solving these inhomogeneous coupled 
equations, it follows that 
to leading order in the slow-time dependences $n_2(t) = \Delta_2(t)/\Delta_1$, {\it i.e.}, the
$\hat x_2$-component of the spin rotates to follow the direction of the spin-orbit field, and that 
\begin{equation}
n_z(t) = \frac{1}{\Delta_1^2} \frac{\hbar d \Delta_2}{dt}.
\label{tilt1}
\end{equation}

These dynamics give rise to the spin-current in the $\hat{y}$ direction:
\begin{eqnarray}
j_{s,y} &=& \int_{annulus} \frac{d^2 \vec{p}}{(2 \pi \hbar)^2} \frac{\hbar n_{z,\vec{p}}}{2} \frac{p_y}{m} \nonumber \\
&=& \frac{-e E_x }{16 \pi \lambda m} (p_{F+}-p_{F-}),
\label{spincurrent}
\end{eqnarray}
where $p_{F+}$ and $p_{F-}$ are the Fermi momenta of the majority and minority spin
Rashba bands \cite{Sinova2004}. 

We choose the example based on the Rashba system because it is simple to see pictorially the intrinsic contribution. However,
for this particular simple example, in a large range of Fermi energies the result for the intrinsic spin Hall conductivity
turns out to be $\sigma^{H-int}_{xy}=-(e/\hbar)j_{s,y}/E_x=e^2/8\pi\hbar$. This contribution is eventually cancelled by short-range disorder scattering 
because the induced spin-current is proportional to the spin-dynamics, which should vanish in the steady state \cite{Sinova2006}.

For other spin-orbit coupled Hamiltonians, corresponding to realistic materials system, this cancellation from the 
vertex corrections arising from disorder does not exist. 
The above result, illustrated in a simple semiclassical form, is usually best evaluated directly from
the  Kubo formula for the spin Hall conductivity 
for an ideal lattice \cite{Sinova2004}:  
\begin{eqnarray}
\sigma^{H-int}_{xy}&=&\frac{e^2}{V}\sum_{{\bf k},n\ne n'}
(f_{n',k}-f_{n,k})\nonumber\\&\times&
\frac{{\rm Im}[\langle n' k|
\hat{j}^z_{\rm spin\,\,x}|nk\rangle\langle nk| v_y|n' k\rangle]} 
{(E_{nk}-E_{n'k})(E_{nk}-E_{n'k} -i \eta)}
\label{SH}
\end{eqnarray}
where $n,n'$ are band indices, $\vec{j}^z_{\rm spin}=\frac{\hbar}{4}\{\sigma_z,\vec{v}\}$ is
the spin-current operator, $\omega$ and $\eta$ are set to zero in the DC clean limit, 
and the velocity operators at each $\vec{p}$ are given by 
$\hbar v_i = \hbar {\partial{H(\vec{p})}/{\partial p_i}}$.

What makes this contribution quite unique, particularly in the AHE, is that it is directly linked to the topological properties of the Bloch states.  
Specifically it is proportional to the integration 
over the Fermi sea of the Berry's curvature of each occupied band, or equivalently \cite{Haldane2004} to 
the integral of Berry phases over cuts of Fermi surface segments.  
This same linear response contribution to the AHE and SHE conductivity can be obtained from the 
semiclassical theory of wave-packets dynamics~\cite{Xiao2010,Jungwirth2002,Sundaram1999}.
\begin{figure}
\includegraphics[width=0.9 \columnwidth]{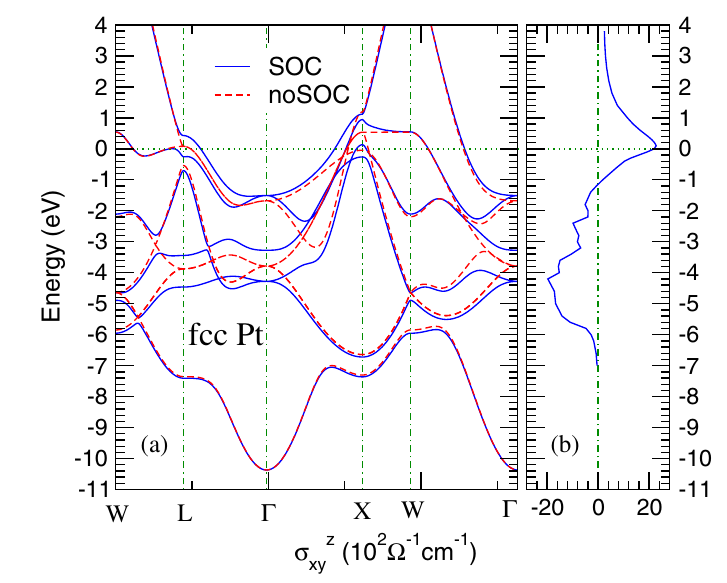}
\includegraphics[width=0.9 \columnwidth]{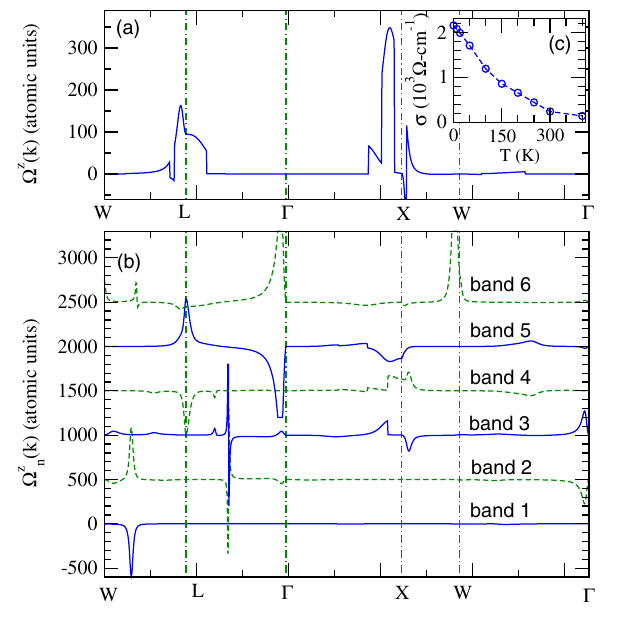}
\caption{Band structure for Pt calculated with (solid lines) and without (dotted lines) spin-orbit coupling. The spin Hall conductivity (b) is shown 
calculated at each energy. In the lower figure the Berry curvature is calculated (total) as well as the one corresponding for each sub band. 
 From Ref.~\onlinecite{Guo2008}.}
\label{Guo2008Fig}
\end{figure}

\begin{figure}
\includegraphics[width=0.9\columnwidth]{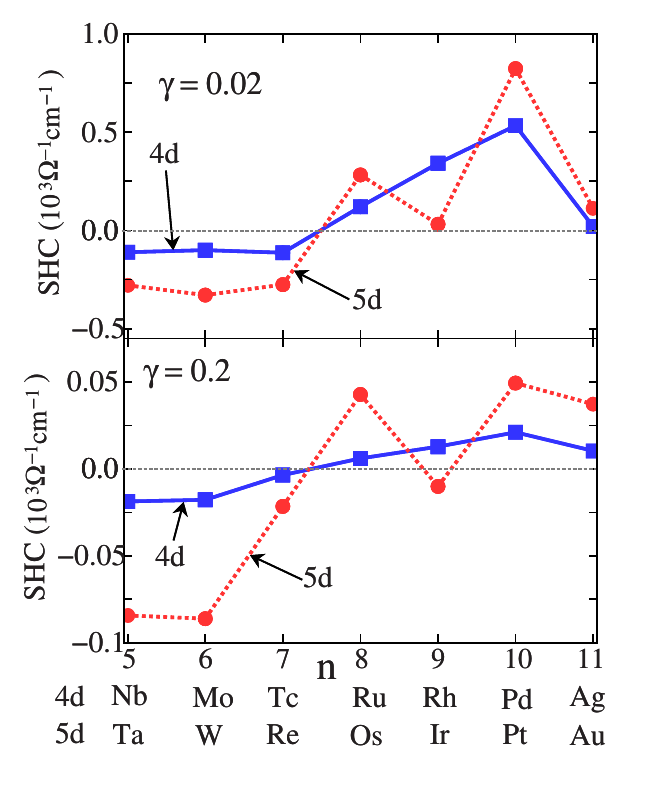}
\caption{Intrinsic spin Hall conductivity for 4d and 5d transition metals. A key feature is the change in sign from Pt to Ta.
 From Ref.~\onlinecite{Tanaka2008}.}
\label{Tanaka2008Fig}
\end{figure}

\begin{figure}
\includegraphics[width=1.0 \columnwidth]{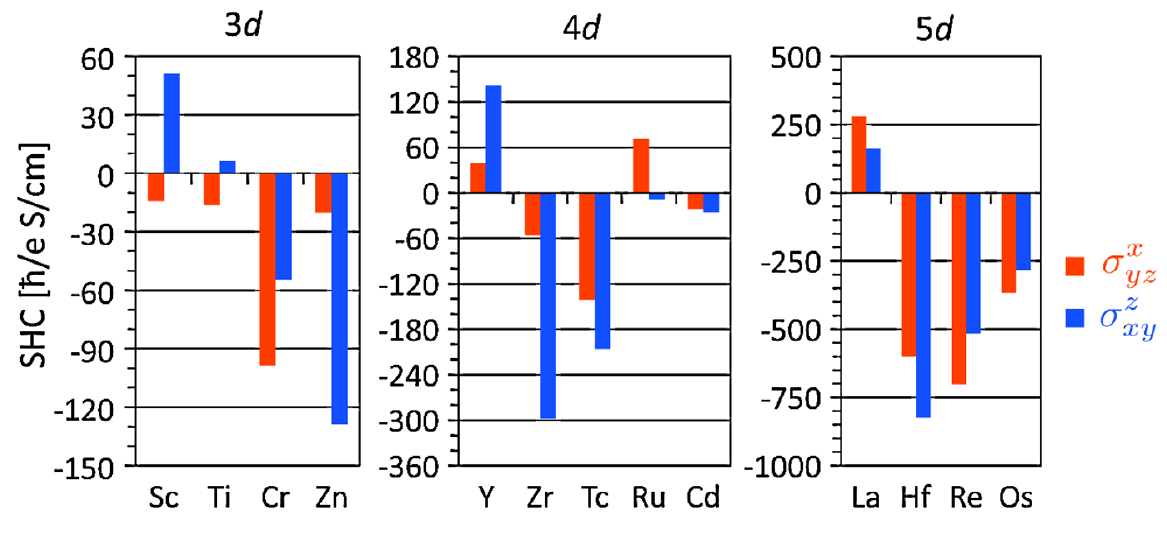}
\caption{Intrinsic spin Hall conductivity  for hcp metals Sc, Ti, Zn, Y, Zr, Tc, Ru, Cd, La, Hf, Re, and Os and for antiferromangetic Cr.
 From Ref.~\onlinecite{Freimuth2010}.}
\label{Freimuth2010Fig}
\end{figure}

One of the motivations for identifying
the intrinsic contribution $\sigma_{xy}^{H-int}$ is that it can be evaluated accurately
even for materials with relatively complex bands using microscopic electronic structure theory techniques.
In many materials which have strongly spin-orbit coupled bands,
the intrinsic contribution seems to dominate the SHE and AHE.
Particularly in metals the calculations have given semi-quantitative predictions of
  the expected spin Hall angles. This is illustrated in the density-functional calculation for Pt \cite{Guo2008}, shown in Fig.~\ref{Guo2008Fig},
  and in the microscopic tight-binding calculations for other 4d and 5d metals \cite{Tanaka2008}, shown in Fig.~\ref{Tanaka2008Fig}.
As it is clear from Fig.~\ref{Guo2008Fig} the largest contributions to the spin Hall conductivity arise, similar to AHE, whenever bands connected via spin-orbit coupling are near each other at the Fermi energy.
The calculated 
spin Hall conductivities are predicted to be large in these transition metals, and, in particular, a sign change is predicted going from Pt to Ta which has been
directly observed in experiments. More recent density-functional calculations on a range of hcp metals and antiferromagnetic Cr,  show a strong anisotropy of the spin Hall conductivity
 \cite{Freimuth2010}, 
as illustrated in Fig.~\ref{Freimuth2010Fig}.

\subsubsection{Skew scattering  mechanism}\label{skew-intro}
\label{skew-section}
The skew scattering contribution to the SHE and the AHE is 
the term which is proportional to the Bloch state transport lifetime $\tau$. It will  therefore  tend to 
dominate in nearly perfect crystals.  
It is  the only contribution to the SHE and AHE which appears in traditional Boltzmann transport theory where
 interband coherence effects  are usually neglected. Skew scattering is due to chiral features 
which appear in the disorder scattering in the presence of spin-orbit coupling.  
This mechanism was first identified in FMs by Smit \cite{Smit1958} and has its origins in the  Mott scattering
in relativistic physics \cite{Mott1929,Mott1932}. 

Typical treatments of semi-classical Boltzmann transport theory found in textbooks 
often appeal to the principle of detailed balance which states that the transition probability 
$W_{n \to m}$ from state $n$ to $m$ is identical to the 
transition probability in the opposite direction ($W_{m \to n}$). 
Although these two transition probabilities are identical in the Fermi's golden-rule approximation, since
$
W_{n \to n'}= { ({ 2 \pi} / {\hbar})} |\langle n |V| n'\rangle|^2 
\delta(E_n - E_{n'}),
$
where $V$ is the perturbation inducing the transition,
detailed balance in this microscopic sense  is not generic.
In the presence of spin-orbit coupling, 
either in the Hamiltonian of the perfect crystal or in the disorder 
Hamiltonian, a transition which is right-handed with respect to the 
magnetization direction has a different  probability than
the corresponding left-handed transition.     
When the transition rates are evaluated perturbatively, asymmetric 
chiral contributions appear at third order. In simple models the asymmetric chiral contribution to the
transition probability is often {\it assumed} to have the form:
\begin{equation}
W^A_{\bm{k} \bm{k}'} = - \tau_A^{-1} (\bm{k} \times \bm{k}')
\cdot \bm{M}_s.
\end{equation}
Inserting  this asymmetry  into the Boltzmann equation 
leads to a current proportional to the longitudinal current driven by $\bm{E}$
and perpendicular to both $\bm{E}$ and $\bm{M}_s$.
This contribution to the Hall conductivity $\sigma_{xy}^{H-skew}$ and 
the conductivity $\sigma_{xx}$ are approximately proportional to the transport lifetime $\tau$ 
and the Hall resistivity, $\rho_{xy}^{H-skew} = \sigma_{xy}^{H-skew}\rho_{xx}^2$, is therefore 
proportional to the longitudinal resistivity $\rho_{xx}$ whenever this contribution dominates.  
 
There are several specific mechanisms for skew scattering (see Sec. IV-B and Sec. V-A. in \onlinecite{Nagaosa2010}).  
Evaluation of the skew scattering contribution to the Hall 
conductivity or resistivity requires that the conventional linearized Boltzmann equation be solved using a 
collision term with accurate transition probabilities, since these will generically include a chiral contribution.  
In practice our ability to accurately estimate the skew scattering contribution to the SHE and AHE of a real material
is limited only by the accuracy of the characterization of its disorder.    

In simple models, the skew scattering contributions to the SHE or AHE are considered 
to arise only from the spin-orbit coupling in the disorder potential. This can only be considered valid when the 
spin-orbit coupling in the bands is not strong enough to split the degenerate spin subbands when
compared to the typical disordered broadening. In systems with strong spin-orbit coupling in the bands, such as 
heavy transition metals, considering the spin-orbit coupling only in the disorder potential would be incorrect. The reason is because a strong contribution to the skew scattering 
also arises from the scattering of the 
spin-orbit coupled quasiparticle from the scalar potential. In fact,  the spin-orbit coupling of the 
disordered potential is typically strongly renormalized by the other nearby subbands as well, and therefore 
the effect of the multi-band character cannot be ignored in  these materials, even when discussing 
skew scattering alone. 

\begin{figure}
\includegraphics[width=1.0 \columnwidth]{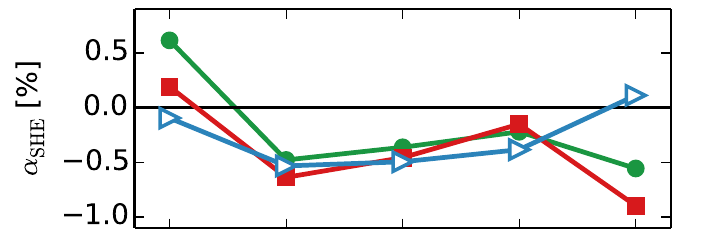}
\caption{Skew scattering spin Hall angle in a Pt host with 1\% level of impurities.
 From Ref.~\onlinecite{Zimmermann2014}.}
\label{Zimmermann2014Fig}
\end{figure}
Recent studies of the skew scattering based on ab-initio studies and Boltzmann equation
in  systems with impurities of Cr, Mn, Fe, Co, and Ni in Pt, Au and Pd hosts have yielded contributions to the spin
Hall angle of a fraction of a per-cent \cite{Zimmermann2014,Long2014a}.  The result related to 1\% doping of impurities to a Pt host is
 shown in Fig.~\ref{Zimmermann2014Fig}.

We end this subsection with a note directed to the reader 
who is more versed in the latest development of the links between the full semiclassical
and the microscopic theory of the SHE and AHE.
 We have been careful above not to define the skew-scattering contribution 
as the sum of {\it all} the contributions arising from the asymmetric scattering 
rate present in the collision term of the Boltzmann transport equation. 
We know from microscopic theory that
this asymmetry also makes an AHE contribution of order $\tau^{0}$ \cite{Sinitsyn2007}.
There exists a contribution from this asymmetry which is  present in the microscopic theory treatment 
associated with the so called ladder diagram
corrections to the conductivity, and therefore of order $\tau^0$. In our parsing of the contributions to the SHE and AHE
we do not associate this contribution with skew-scattering but 
place it under the umbrella of side-jump scattering even though it does 
not physically originate from any side-step type of scattering.

\subsubsection{Side-jump mechanism}\label{sj-intro}

Given the sharp definition we have provided for the intrinsic and skew scattering contributions 
to the SHE and AHE conductivity, the equation
\begin{equation} 
\sigma_{xy}^{H} = \sigma_{xy}^{H-int} + \sigma_{xy}^{H-skew} + \sigma_{xy}^{H-sj} 
\end{equation} 
defines unambiguously the side-jump contribution as the difference between the full SHE/AHE 
conductivity and  the skew and intrinsic contributions.  In using the term side-jump for the remaining contribution,
we are appealing to the historically established taxonomy outlined in the previous section.
Establishing this connection rigorously 
has been the most controversial aspect of the AHE theory and, not surprisingly, some confusion has spilled over to 
the discussion of the SHE.

The basic semiclassical argument for a side-jump contribution 
can be stated straight-forwardly: when considering the scattering of a Gaussian wave-packet from a
spherical impurity with spin-orbit interaction ( $H_{SO} = (1/2 m^2 c^2) (r^{-1} \partial V/ \partial r) S_z L_z$), a wave-packet with incident
wave-vector $\bm{k}$ will undego a displacement transverse to $\bm{k}$ equal to ${ 1\over 6} k \hbar^2/m^2 c^2$. This type of contribution was 
first noticed, but discarded, by Smit \cite{Smit1955,Smit1958} and reintroduced by Berger \cite{Berger1964} who argued 
that it was the key contribution to the AHE.  Most of the earlier developments were based on 
physical arguments of how to incorporate the physics in a semiclassical Boltzmann formalism, 
although it cannot be done systematically and errors have ensued \cite{Nagaosa2010}.

{ A very common misconception} is that the side-jump can be generally computed by taking only into account the spin-orbit coupling interaction
of the disorder scattering potential. This can only be justified in a weak spin-orbit coupled system, e.g. n-doped GaAs, where indeed this is 
likely to be the case. This is the consideration in the  Cr\'epieux-Bruno model \cite{Crepieux2001b} where the spin-orbit coupling is only present in the
disorder potential and which has been subsequently used by others \cite{Engel2005,Maekawa2012}. 
However, when addressing materials with strong spin-orbit coupling it is important to remember that there
is always {\it two} sources of
side-jump scattering:
\begin{enumerate}
\item Extrinsic-side-jump: The contribution arising from the non-spin-orbit coupled part of the wave-packet scatting off the spin-orbit coupled disordered.
\item Intrinsic-side-jump: The contribution arising from the spin-orbit coupled part of the wave-packet formed by the Bloch electrons scattering off the scalar potential alone without spin-orbit coupling.
\end{enumerate}
Both can be important and independent of each other depending on the crystalline environment and the type of scattering impurity. In heavy-element 
materials, such as Pt and Ta, that have become quite important for possible future technological applications, the dominant contribution is likely to be 
the second type of contribution. 
In FMs it has been demonstrated that the second type of contribution, termed here {\it intrinsic-side-jump} to distinguish them clearly, 
can be very large.
Both of these side-jump contributions add to the 
scattering independent mechanisms, i.e., they are independent of $\tau$, which incorporate both side-jump types and the intrinsic mechanism \cite{Weischenberg2011}.

Both types of side-jump and the intrinsic contributions have quite different dependences on more specific system parameters,
particularly in systems with complex band structures (for a detailed review on these delicate issues we recommend 
\onlinecite{Sinitsyn2008}).  
Most of the prior mistakes surrounding the theory of side-jump 
can be traced back to the 
  physical meaning
ascribed to quantities which were  gauge dependent, like the Berry's connection that  
is typically identified as the definition of the side-step upon scattering.
Studies of simple models, e.g. semiconductor conduction bands, also gave
results in which the side-jump contribution seemed to be the same magnitude but opposite in sign 
compared to the intrinsic contribution \cite{Nozieres1973}.  
It is well understood now that 
these 
cancellations are unlikely  in more complex models \cite{Sinitsyn2007,Weischenberg2011}. The prior cancellations can be traced back
to the fact that in these  
very simple band structures the  Berry's curvature of the Bloch electrons is a constant independent of momentum. One is reminded in this case
of the famous quote by Albert Einstein which states that things should be made as simple as possible but not simpler. 

It is only through comparison between different fully microscopic linear response theory calculations, based on 
equivalently valid microscopic formalisms such as Keldysh (non-equilibrium Green's function)
 or  Kubo formalisms, and the systematically developed semi-classical theory 
that the specific contribution due to the side-jump mechanism can be separately identified with 
confidence. Nonsystematic treatments can lead to misconceptions that linger for a long time in 
the community \cite{Maekawa2012}.

Very recently, there has been strong steps forward in the theory of the AHE in developing 
full theories with predictive power to calculate all these contribution in FM materials with a complex band structure \cite{Weischenberg2011,Kovalev2010,Czaja2013,Freimuth2010,Lowitzer2010}.
In the theory of the SHE on the other hand, perhaps because of the complexity of the measurements, the fact that spin decays in a non-trivial way, and 
the lack of practical general theories that can bring one from a weak to a strong spin-orbit coupled regime, such progress still remains to be undertaken fully.

\subsection{Phenomenological Drift-Diffusion Theory}
\label{drift-diffusion}
\onlinecite{Dyakonov1971,Dyakonov1971b}  considered the  phenomenological theory of the SHE by coupling  
 the  usual drift diffusion equation for charge transport to the spin-current drift-diffusion equations. 
Hence, the spin-charge drift-diffusion equations applicable to electrical transport measurements can be written from symmetry considerations as \cite{Dyakonov2008}:
\begin{equation}
{\bf j}^c=e\mu n {\bf E}+D\nabla n+
e\alpha_{SH}\mu ({\bf E}\times {\bf P})+e\alpha_{SH}D(\nabla \times {\bf P}),
\label{jc}
\end{equation}
\begin{equation}
{\bf j}^s_{ij}=-\hbar \mu n E_iP_j+D\frac{\partial P_j}{\partial x_i} 
-\hbar \alpha_{SH}\epsilon_{ijk}
\left( \mu n E_k+D\frac{\partial n}{\partial x_k}\right),
\label{js2}
\end{equation}
where the first two terms of Eqs.~(\ref{jc}) and (\ref{js2}) correspond to the definition of the uncoupled charge and spin-currents. Here $\bf P$ corresponds to the spin polarization, $D$ is the electron diffusion constant, $\mu$ is the {\it spin-independent} electron mobility, $n$ is the electron density, $\bf E$ is the electric field, and $\alpha_{SH}$ is the spin Hall angle defined by the ratio of the spin Hall conductivity to the diagonal charge conductivity.

In Eq.~(\ref{jc}) the third term corresponds to the AHE. The fourth term  describes the ISHE if a charged diffusive current is absent, i.e., in the case  of the pure spin-current in the system. We distinguish this from the situation in which a polarized charge diffusive current, e.g. generated by optical excitation \cite{Bakun1984}, leads to a charge transverse current which we associate here with a regime closer to the AHE. This distinction is made more clear by the fact that the SHE has a precise definition of a pure spin-current being generated by a charge current, and therefore its inverse is associated with a pure spin-current generating a transverse charge current. 
%
In Eq.~(\ref{js2}) the third term represents the SHE from an electric field, while the fourth term represent its diffusion driven counterpart. The equations are written to first order in the spin Hall angle. 

These equations can be directly extended to include thermal gradients and junctions \cite{Johnson1987}. Recently, there has been 
an extension of these to incorporate thermal SHEs within the emerging field of spin caloritronics \cite{Bauer2010,Bauer2012}. The treatment, for the 
most part, remains phenomenological with connections in particular to the Onsager relations between the thermodynamic forces and their corresponding entropy fluxes. 
Within this emerging subfield of spintronics  many theoretical challenges remain, not least a better treatment of scattering coherent effects driven by statistical forces and the ability of going
beyond the simple adiabatic frozen phonon approximations.

\subsection{The Cr\'epieux-Bruno model: extrinsic side-jump and skew scattering model}
\label{CB-model}
We discuss here the \onlinecite{Crepieux2001b} model that incorporates spin-orbit coupling only through the disorder potential, i.e., there is no spin-orbit coupling 
present directly in the Bloch electron bands at the Fermi surface. This model has been applied  to weak spin-orbit coupled materials, such as $n$-GaAs, 
to explain the extrinsic origin of their SHE \cite{Engel2005} and has also been applied to weakly spin-orbit coupled metals \cite{Maekawa2012}. 
The model builds on the influential work of Nozieres and Lewiner \cite{Nozieres1973}, where they studied the AHE in semiconductors with simple band structure and how to account for the effects of the spin-orbit coupling when projecting to an effective two-band model. There are many subtle issues in such treatment already at this simple-model level, but extrapolating some of its results to generalities, e.g. specific cancellations, is dangerous since "the side-jump is no longer given by the simple expression" derived in these works, as the authors themselves state \cite{Nozieres1973}.

The model two-band Hamiltonian is given by
\begin{equation}
H=\frac{\hbar^2 k^2}{2 m^*}+V({\bf r})+\lambda_{e-so} {\bf \sigma}\cdot({\bf k}\times \nabla V)=H_0+W.
\end{equation}
Here $m^*$ is the effective mass of the Bloch electron, $V({\bf r})$ is the disorder potential, ${\bf \sigma}$ are the Pauli metrices, and $\lambda_{e-so}$ is the effective spin-orbit coupling parameter.
 For a free electron, $\lambda_{e-so}=\hbar^2/2m^2c^2$ is an extremely small parameter, but in a solid state environment it is strongly renormalized by nearby bands. For the effective two-band model of conduction electrons, obtained from the $8\times 8$ Kane description of the semiconductor band structure, $\lambda_{e-so}=(P^2/3)[1/E_g^2-1/(E_g+\Delta_{so})]$, with $E_g$ being the gap, $P$ the s-p dipole matrix element, and $\Delta_{so}$ the spin-orbit splitting of the valence band. For $n$-GaAs, for example, this value is $5.3 \,{\rm \AA}^2$. The total scattering potential is $W$.

\begin{figure} \centering
\epsfig{file=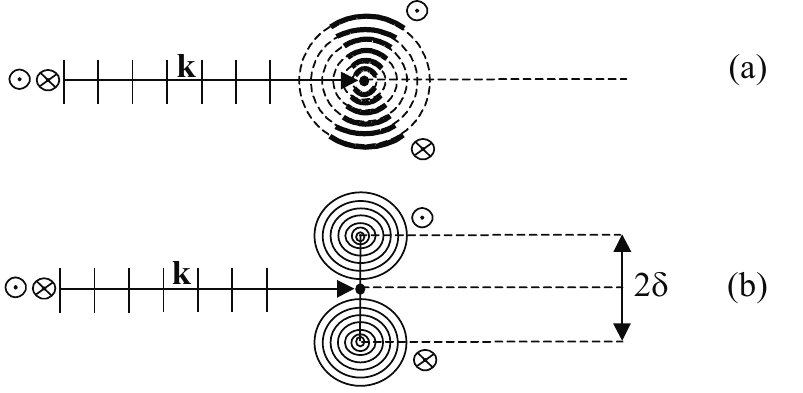}
\vspace{0.25cm} \caption{Schematic
picture of the skew-scattering (a) and the extrinsic side-jump (b) mechanisms
from a quantum point of view ($\odot$ corresponds to spin up and
$\otimes$ to spin down). The bold curves represent the
anisotropic enhancement of the amplitude of the wave-packet due
to spin-orbit coupling. Here the electrons themselves contain no spin-orbit coupling. 
From Ref.~\onlinecite{Crepieux2001b}.}\label{figQuan}
\end{figure}
In this model the velocity operator is modified by the spin-orbit coupled term to read 
\begin{equation}
\hat{\bf v}=\frac{\hat{\bf p}}{m^*}+\frac{\lambda_{e-so}}{\hbar}[{\bf \sigma}\times \nabla V],
\end{equation}
and the scattering amplitude due to the disorder potential from state $|{\bf k},s\rangle$ to $|{\bf k}',s'\rangle$ is given by
\begin{eqnarray}\label{mateleV}
  \langle k',s'|W|k,s\rangle=\tilde{V}_{\bf kk'}\left(\delta_{ss'}
  +i\lambda_{e-so}(\bsig_{s's}\times{\bf k}')\cdot{\bf
  k}\right), \nonumber \\
\end{eqnarray}
where $\tilde{V}_{\bf kk'}$ is the Fourier transform of $V$.
The disorder potential is considered to be short ranged for simplicity (for ionic scattering see \onlinecite{Engel2005}) such that $V({\bf r})=u_{i}\sum_j\delta(\rf-\rf_j)$ and $\tilde{V}_{\bf kk'}=u_i\delta_{\kf,\kf'}$. One can then connect this procedure with the Boltzman equation, which will also yield the spin-diffusion equation \cite{Zhang2000}. Microscopically the scattering from this disorder potential induces a collision integral in the Boltzmann formalism of the form
\begin{equation}
\left(\frac{\partial f_{\kf,s}}{\partial t}\right)_{coll}=\sum_{\kf',s'}[P_{\kf s, \kf' s'}f_{\kf' s'}- P_{\kf' s', \kf s}f_{\kf s}],
\end{equation}
with the transition scattering probabilities available from the T-matrix which yields a symmetric and antisymmetric contribution:
\begin{eqnarray}
P_{\kf' s', \kf s}^{sym}&=&\frac{2\pi}{\hbar}\frac{n_i}{V}u_i^2\left(\delta_{s,s'}+\lambda_{e-so}^2|(\kf'\times\kf)\cdot {\bf \sigma}_{s,s'}\right)\nonumber \\ 
& &\times\delta(\epsilon_{\kf'}-\epsilon_{\kf}),
\end{eqnarray}
\begin{eqnarray}
P_{\kf' s', \kf s}^{ant}&=&-\frac{(2\pi)^2}{\hbar}\lambda_{e-so}\frac{n_i}{V}u_i^3N(0)\delta_{s,s'}(\kf'\times\kf)\cdot {\bf \sigma}_{s,s'})\nonumber \\
& &\times\delta(\epsilon_{\kf'}-\epsilon_{\kf}).
\end{eqnarray}
Here $n_i$ is the density of impurities and $N(0)$ is the density of states at the Fermi level.
Within the framework of the semiclassical Bolzmann equation one then writes \cite{Zhang2000,Maekawa2012}:
\begin{equation}
{\bf v}_\kf \cdot {\bf \nabla} f_{\kf s} + \frac{e {\bf E}}{\hbar}\cdot \nabla_\kf f_{\kf s}=-\frac{\delta f_{\kf s}}{\tau_{tr}}-\frac{f^0_{\kf s}-f^0_{\kf' s}}{\tau_{sf}(\theta)},
\end{equation}
where 
\begin{equation}
\tau_{tr}^{-1}=\sum_{\kf',s'}P_{\kf s, \kf' s'}^{sym}=\frac{1}{\tau_{tr}^0}(1+2k_F^4 \lambda_{e-so}^2/3),
\end{equation}
and
\begin{equation}
\tau_{sf}^{-1}=\sum_{\kf'}P_{\kf 1, \kf' -1}^{sym}=\frac{k_F^4\lambda_{e-so}^2}{3\tau_{tr}^0}(1+\cos^2(\theta)).
\end{equation}
Here $\tau_{tr}$ is the transport lifetime, $\tau_{tr}^0$ is the transport lifetime when neglecting the spin-orbit coupling part of the disorder potential, and $\tau_{sf}$ is the spin-flip time.
Further expanding the equilibrium and non-equilibrium distribution function yields  the spin diffusion equation 
\begin{equation}
\nabla^2(\mu_1-\mu_{-1})=\frac{1}{\lambda_{sd}^2}(\mu_1-\mu_{-1}),
\end{equation}
with $\lambda_{sd}^2=D\tau_{sf}/2$, $D=(1/3)\tau_{tr}v_F^2$, and $\mu_s$ representing the spin-dependent chemical potentials ($s=\pm1$ is the spin index). Averaging over the scattering angle one obtains that the ratio of spin-flip time and transport time {\it for this 
particular model} is approximately
\begin{equation}
\frac{\tau_{tr}}{\tau_{sf}}\approx \frac{1}{2} k_F^4\lambda_{e-so}^2.
\end{equation}

\begin{table}
\scriptsize
\begin{tabular}{lllllll}
\hline
            &  $\lambda_{sd}$     &    $k_F l$             &  $k_F^2{\lambda}_{\rm e-so}$   & $\alpha_{SH}$        &  $\left|\frac{\alpha^{sj}_{SH}}{\alpha_{SH}}\right|$     & Ref. \\
            &  (nm)            &        &                          &  (\%)                &                &      \\
\hline
Al (4.2K)   &  $455(15)$    & 73    & 0.0079                   & $0.032 \pm 0.006$    &   0.67     & 1 NL\\
Al (4.2K)   &  $705 \pm 30$    & 118     & 0.0083                   & $0.016 \pm 0.004$    & 0.88      & 1NL \\
Au (295K)   &  $86 \pm 10$     & 371      & 0.3                      &  11.3                & 0.014      &  2  NL\\
Au (295K)   &  $35 \pm 3^*$    & 253     & 0.52                     &  $0.35\pm 0.03$      &  1.17  & 3   SP \\
CuIr (10K)  &  $5-30$          &                           &                          &  $2.1\pm 0.6$        &        & 4  NL   \\
Mo (10K)    &  10              & 36.8       & 0.32                     &  -0.20               &       8.7       & 5  NL  \\
Mo (10K)    &  10              & 8.11       & 0.07                     &  -0.075              &         23     & 5  NL \\
Mo (10K)    &  $8.6 \pm 1.3$   & 34.1    & 0.34                     &  $-(0.8 \pm 0.18)$   &    2.5          & 6    NL \\
Mo (295K)   &  $35 \pm 3^* $   & 56.7      & 0.14                     &  $-(0.05\pm 0.01)$   & 9.9   & 3  SP \\
Nb (10K)    &  $5.9 \pm 0.3$   & 11.3        & 0.14                     &  $-(0.87 \pm 0.20)$  &          2.9       & 6 NL   \\
Pd (295K)   &  $9^*$           & 24.0       & 0.23                     &  $1.0$               & 1.9   & 7 SP   \\
Pd (10K)    &  $13 \pm 2$      & 26.8    & 0.18                     &  $1.2\pm 0.4$        &    1.1    & 6   NL \\
Pd (295K)   &  $15 \pm 4^*$    & 48.6        & 0.28                     &  $0.64\pm 0.10$      & 1.8   & 3  SP \\
Pt (295K)   &                  & 77.9       & 0.74                     &  0.37                &       5.1       & 8  NL \\
Pt (5K)     &  14              & 97.3      & 0.61                     &  0.44                &       2.9       & 9    NL \\
Pt (295K)   &  10              & 67.6       & 0.58                     &  0.9                 &        1.9      & 9  NL \\
Pt (10K)    &  $11 \pm 2$      & 98.5       & 0.77                     &  $2.1\pm 0.5$        &         0.74     & 5   NL \\
Pt (295K)   &  $7^*$           & 77.8       & 0.97                     &  $8.0$               & 0.31 & 10   SP \\
Pt (295K)   &  $3-6$           & 60.8     & 0.88-1.75                &  $7.6^{+8.4}_{-2.0}$ & 0.57  & 11 SP  \\
Pt (295K)   &  $10 \pm 2^*$    & 29.2       & 0.25                     &  $1.3\pm 0.2$        &  1.31 & 3 SP  \\
Ta (10K)    &  $2.7 \pm 0.4$   & 3.90      & 0.17                     &  $-(0.37 \pm 0.11)$  &    24  & 6  NL  \\
\hline
\end{tabular}
\caption{Experimental spin Hall angles and effective spin-orbit coupling parameters, $k_F^2{\lambda}_{\rm e-so}$.
The values marked by ($^*$) are not measured but taken from the literature. The Fermi momenta are taken to be
$k_F=1.75\times10^8$ cm$^{-1}$ (Al),  
$1.21\times10^8$ cm$^{-1}$ (Au), 
$1.18\times10^8$ cm$^{-1}$ (Nb),
and  $1.0\times10^8$ cm$^{-1}$ (Mo,Pd,Ta,Pt). Here $k_F l= (3\pi/2)\sigma/k_F(h/e^2)$.
References: 1) \cite{Valenzuela2006,Valenzuela2007}, 2) \cite{Seki2008}, 3) \cite{Mosendz2010b}, 4) \cite{Niimi2011}, 5)\cite{Morota2009},
6) cite{Morota2011}, 7) \cite{Ando2010d}, 8) \cite{Kimura2007}, 9) \cite{Vila2007}, 10) \cite{Ando2008d}, 11) \cite{Liu2011}}
\label{table_SH_values}
\end{table}

This is one of the key aspects that has made this model appealing. The model provides a means to obtain its effective spin-orbit parameter $\lambda_{e-so}$ by measuring the spin-diffusion length, independently of the spin Hall angle. At this point, it should to be emphasized that the model is applicable to the weak spin-orbit coupling regime, i.e., when $\tau_{tr}/\tau_{sf}\ll1$.

From either a microscopic or Boltzmann like analysis the result for this model for the side-jump contribution to the spin Hall angle is \cite{Crepieux2001b,Engel2005}:
\begin{equation}
\alpha_{SH}^{sj}\equiv \frac{\sigma_{xy}^{H-sj}}{\sigma_{xx}}= -\frac{2 \lambda_{e-so}m^*}{\tau_{tr}}=-\frac{2 k_F^2 \lambda_{e-so}}{k_F l}.
\label{C-B-sj}
\end{equation} 
Here $l=\tau_{tr}k_F/m^*$ is the mean free path.
The skew scatting contribution is given by
\begin{equation}
\alpha_{SH}^{sk}=\frac{4\pi}{3}k_F^2 \lambda_{e-so} N(0)u_i .
\label{C-B-skew}
\end{equation}
In the SHE experiments in $n$-GaAs, the spin Hall angle is 
dominated by the skew scattering contribution vs. the side-jump contribution by a factor of 2, $\sigma_{xy}^{H-skew}/\sigma_{xy}^{H-sj} \sim -1.7/0.8$ \cite{Engel2005}.

When this simplified model is used for metals it yields a mixture of results and inconsistencies. 
This can be seen in 
Table~\ref{table_SH_values} where we show for a series of metals the experimental SHE angles and the independently experimentally inferred effective spin-orbit coupling parameters  $k_F^2{\lambda}_{\rm e-so}$. Skew scattering is not possible to estimate from the model expression~(\ref{C-B-skew})  without knowing the specific value for $u_i$. However, we know from the AHE that skew scattering has only been seen to dominate for 
extremely conductive metals so it is neglected in the discussion of Table~\ref{table_SH_values}. When comparing the side-jump contribution estimates to the measured 
values of $\alpha_{SH}$ the results vary extensively. In some cases, like Ta, the theoretical side-jump contribution is $24\times$ 
larger than the measured $\alpha_{SH}$. The failure of Eq.~(\ref{C-B-sj}), derived assuming the weak spin-orbit coupling regime, is not surprising here since Ta is a 5d heavy transition metal. In others, like in Pt, Eq.~(\ref{C-B-sj}) gives a value that is smaller than 
the measured $\alpha_{SH}$, in some cases approaching the experimental value rather closely. However, even in this case ascribing the measured spin Hall angle to the side-jump contribution of Eq.~(\ref{C-B-sj}) is questionable because the independently inferred parameter $k_F^2{\lambda}_{\rm e-so}$ is close to 1. This is inconsistent with $\tau_{tr}/\tau_{sf}\ll1$, i.e., with the weak spin-orbit coupling regime assumed in the model.
In the heavy element materials the intrinsic SHE estimates have had much more success. Hence, the simple model expression  Eq.~(\ref{C-B-sj}), although illustrative
and appealing, should perhaps only be consider as such, not as a quantitative predictive theory of the SHE.


\subsection{Theory of inverse spin Hall effect induced by spin-pumping}
\label{spin-pumping-theory}

When measuring the ISHE it is necessary to generate a spin-current that flows into the NM whose spin Hall angle is being measured. In the non-local transport schemes this is achieved indirectly by spin-diffusion into the NM. This can restrict both the types of materials and reliability of the measurements since there are several interfaces involved and the distances of the devices have to be kept shorter than the spin-diffusion length in the area where the spin-current is diffusing.  

A key alternative to generating spin-currents is to exploit SP in a FM/NM bilayer system. This phenomenon was observed experimentally in early 2000's \cite{Mizukami2001,Urban2001,Mizukami2002}. The experiments showed an enhanced Gilbert damping in FMR associated with the loss of angular momentum by a spin-current flowing from the FM to the NM which served as a spin sink. 

\begin{figure}[h!]
\hspace*{-0cm}\epsfig{width=0.8\columnwidth,angle=0,file=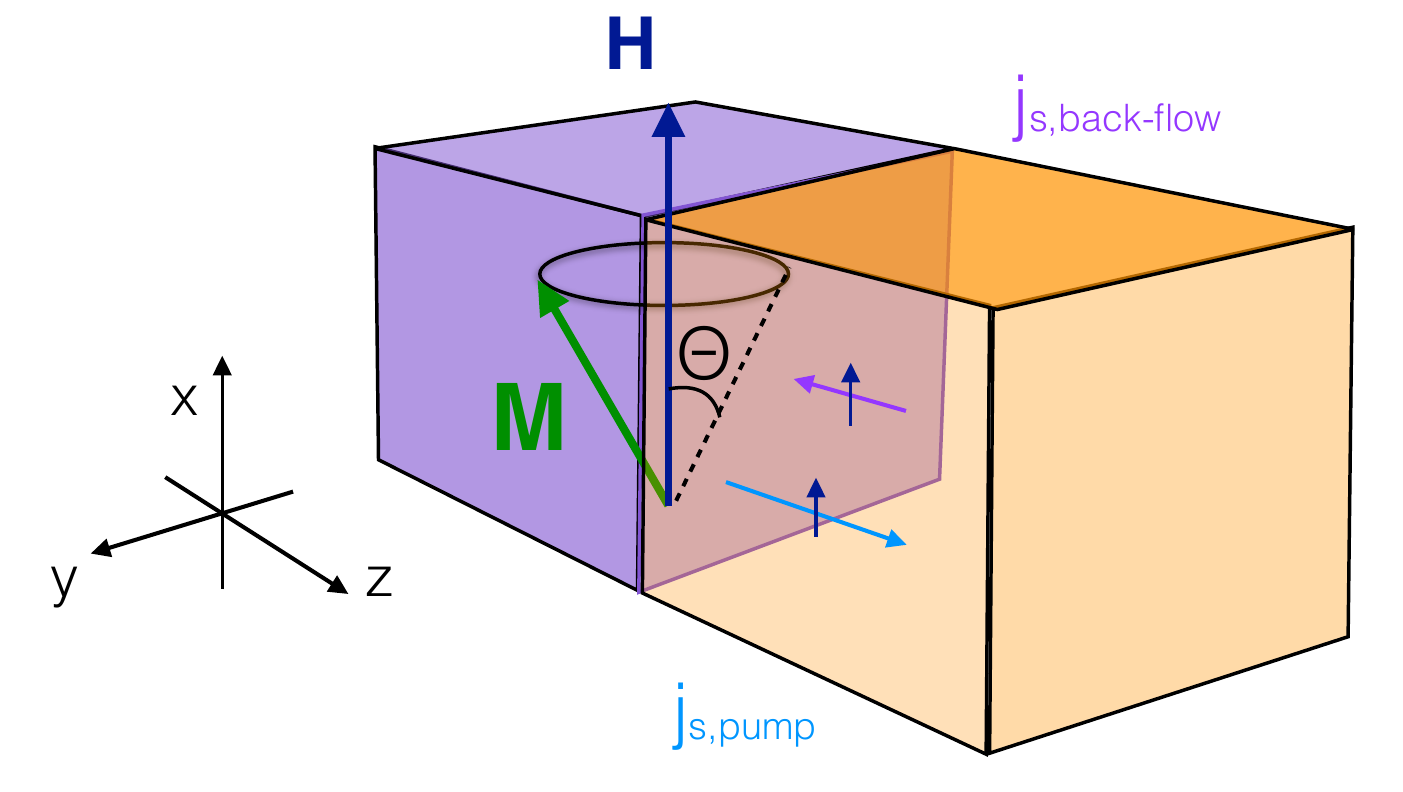}
\caption{Schematic of SP comprising of a spin-pump current flowing from the FM to the NM and a back flow current that 
depends on the thickness of the NM.} 
\label{fig_Bauer_spin_pumping}
\end{figure}

The associated SP theory based on the scattering formalism was developed by Tserkovnyak et al. \cite{Tserkovnyak2002,Tserkovnyak2002b,Tserkovnyak2005}. It extends the theory of adiabatic quantum pumping \cite{M.ButtikerH.Thomas1993,Brouwer1998} by incorporating the spin degrees of freedom.  It can be shown that the precessing magnetization in FM generates a time-dependent spin-current  at the FM/NM interface that flows into the NM given by,
\begin{equation}
\label{ispump}
j_\mathrm{s,pump} \; \vec{\sigma}(t) = \frac{\hbar}{4 \pi} A_\mathrm{r} \; \hat{\vec{m}} \times \frac{\mathrm{d} \hat{\vec{m}}}{\mathrm{d} t}.
\end{equation}
Here $\hat{\vec{m}}(t)$ is the unit vector of the magnetization, $\vec{\sigma}$ is the unit vector of the spin-current polarization, $j_\mathrm{s,pump}$ its magnitude, and $A_\mathrm{r}$ is defined as the SP conductance of the particular sample. The spin-current generated at the interface which propagates into the NM  decays on a length scale connected to the effective spin diffusion length $\lambda_\mathrm{sd}$ of NM. A sketch of the physics is shown in Fig.~\ref{fig_Bauer_spin_pumping}. Note that in systems with strong spin-orbit coupling this length scale can be difficult to define since it can be as short as several atomic layers. Also, proximity effects as well as roughness at the interface with the NM can blur the sharpness of such an interface.  

The scattering-matrix theory introduces the concept of a complex spin-mixing conductance at the interface based on spin-conserving channels and no spin-loss at the interface. Theoretical ab-initio calculations, and the  phase randomization at the scattering interface, seem to indicate that only the real part of the mixing conductance dominates the physics and that in the diffusive regime this will be  approximately the Sharvin conductance given by the number of conducting channels. In this approximation \cite{Tserkovnyak2002b}:
\begin{equation}
A_\mathrm{r} \approx {\rm Re}[g^{\uparrow\downarrow}]=\frac{k_F^2}{4\pi}\approx\frac{1}{4\pi}(3\pi^2 n)^{2/3},
\end{equation}
where $k_F$ and $n$ are the Fermi wavevector and electron density in the NM, respectively.
Direct ab-initio calculations of the mixing conductance \cite{Zwierzycki2005,Carva2007,Xia2002} have verified that for a FM/NM interface with moderate spin-orbit coupling the spin-mixing conductance is of this order of magnitude.

As the magnetization rotates, the spin-current injected from the FM to the NM is time dependent but 
the AC spin-current when averaged over time has a non-zero  DC component, hence the notion of pumping, which is given by:
\begin{equation}
\label{ispump}
j_\mathrm{s,dc}  = \frac{\hbar  \omega}{4 \pi} A_\mathrm{r} sin^2\Theta. 
\end{equation}
Here $\omega$ is the driving radio frequency (RF) and $\Theta$ is the cone angle of precession (see Fig.~\ref{fig_Bauer_spin_pumping}). 
Under the assumption of the NM being a perfect spin-sink the SP conductance will be the spin-mixing conductance. However, in the  systems where the NM has a finite thickness of the order of the spin-diffusion length, the induced spin accumulation in the NM due to the pumped spin-current from the FM will create a spin accumulation which in turn will generate a spin-current back flow. The spin accumulation in the NM within the spin-diffusive regime, 
$\mu_s\equiv \mu_\uparrow-\mu_\downarrow$, is governed by the equation:
\begin{equation}
\frac{d {\bf \mu}_s}{dt}=D\partial_z^2 {\bf \mu}_s -\frac{{\bf \mu}_s}{\tau_{sf}},
\end{equation}
with the boundary conditions:
\begin{equation}
z=0\,\,:\,\,\, \partial_z {\mathbf \mu}_s=- \frac{4 e^2 \rho}{\hbar} j_\mathrm{s,0}  \nonumber
\end{equation}
\begin{equation}
z=t_{NM}\,\,:\,\,\, \partial_z {\mathbf \mu}_s=0\,\,,
\end{equation}
where $\rho$ is the NM resistance and $t_\mathrm{NM}$ is the thickness of the NM. 
In the NM, the spin-current decays away from the FM/NM interface due to the combination of spin diffusion and spin flip scattering. The z-dependent spin-current density $j_\mathrm{s}(z)$ in the NM \cite{Mosendz2010,Azevedo2011} with the above boundary conditions reads:
\begin{equation}
j_\mathrm{s}(z) = -\frac{\hbar}{4e^2 \rho}\partial_z \mu_s(z)= j_\mathrm{s,0} \; \frac{\sinh \left( \frac{t_\mathrm{NM}-z}{\lambda_\mathrm{sd}}\right) }{\sinh \left( \frac{t_\mathrm{NM}}{\lambda_\mathrm{sd}} \right) }.
\end{equation} 
The back-flow current density $j_{s,back}$ at the interface can be taken into account by the relation $j_\mathrm{s,back}(0)\approx 2{\rm Re}[g^{\uparrow\downarrow}]
\mu_s(0)$.  This allows the following expression to be solved for the total spin-current crossing the interface: 
\begin{equation}
\label{is0}
j_\mathrm{s,0} \; \vec{\sigma}(t)= (j_\mathrm{s,pump}-j_\mathrm{s,back}) \vec{\sigma}(t) = \frac{\hbar}{4 \pi} \tilde{A}_\mathrm{r} \; \hat{\vec{m}} \times \frac{\mathrm{d} \hat{\vec{m}}}{\mathrm{d} t}.
\end{equation}
The result is that the effective spin-mixing conductance gets reduced due to a back flow factor given by \cite{Tserkovnyak2002b}:
\begin{equation}
\beta \equiv \frac{\tau_{sf} \delta_{sd}/h}{\tanh{(t_{NM}/\lambda_{sd})}},
\end{equation}
where $\delta_{sd}$ is an effective spin-flip scattering energy obtained by the inverse of the product of 
the volume defined by the scattering cross section and spin-diffusion length and the density of states. 
This then gives the result \cite{Tserkovnyak2002b}:
\begin{align}
\tilde{A}_\mathrm{r} & \approx g^{\uparrow \downarrow} \frac{1}{1+\beta g^{\uparrow \downarrow}} \\
\label{naehat} \tilde{A}_\mathrm{r} & \approx g^{\uparrow \downarrow} \frac{1}{1+\frac{1}{4 \sqrt{\frac{\epsilon}{3}} \tanh\left( \frac{t_\mathrm{NM}}{\lambda_\mathrm{sd}} \right) }} \approx g^{\uparrow\downarrow}_\mathrm{eff},
\end{align}
where $g^{\uparrow \downarrow}$ is now the real part of the spin mixing conductance. The last approximation assumes a weak spin-orbit coupling limit. More explicitly it  assumes $\epsilon=\tau_{tr}/\tau_{sf}\ll 1$.
Hence, the larger $\epsilon$, the more efficient the injected spin-current is relaxed in the NM and the smaller is the amount of backflow~\cite{Tserkovnyak2002b}. However, one has to be aware of the limitations of the approximation since in the strongly spin-orbit coupled systems many of these assumptions fail.

The detection of the net spin-current flowing into the NM can be done electrically via the ISHE as was demonstrated by \onlinecite{Saitoh2006}. By measuring the Hall voltage induced by the spin-current one can infer the spin Hall angle of the material:
\begin{equation}
\label{jcacsp}
\vec{j}_\mathrm{c} = \alpha_\mathrm{SH} \; \frac{2e}{\hbar} \; \vec{j}_\mathrm{s} \times \vec{\sigma}(t).
\end{equation}
Here, the vector of the spin-current density $\vec{j}_\mathrm{s}$ points perpendicular to the NM/FM interface. Note that the vector of the spin-current polarization $\vec{\sigma}(t)$ is a time varying quantity.
In the geometry sketched in Fig.~\ref{fig_Bauer_spin_pumping} the propagation direction of the spin-current is along $z$ and its polarization is along $x$-axis. 
For the detection of a DC voltage along the y-direction one has to consider the charge current
$j_\text{c} \hat{y} = \alpha_\text{SH} \frac{2e}{\hbar} (1/t_{NM}) \int_0^{t_{NM}} j_\text{s}(z) \hat{z}\times \hat{x}$ with magnitude~\cite{Azevedo2011}:\\
\begin{equation}
j_\text{c}  = \alpha_\text{SH} \frac{2e}{\hbar} j_\text{s,0} \frac{\lambda_\text{sd}}{t_\text{NM}} \tanh \left (\frac{t_\text{NM}}{2\lambda_\text{sd}} \right ).
\label{jcsp}
\end{equation}
To convert this charge current density into the actual measured voltage one has to consider the details of the measurement geometry and the resistivity of the bilayer which will be discussed in Sec.~\ref{she-mag}. 
In addition, as described as well in Sec.~\ref{she-mag}, the AC-component can be directly measured. An extension of the above theory to incorporate the AC-component has been done by  \onlinecite{Jiao2013}, with the result that back-flow is important to distinguish between the measured voltages for both the AC and DC configurations.

We conclude this section with a discussion regarding the assumptions of the SP theory. In the above derivations, whenever $\epsilon \sim 1$, the approximations do not hold anymore since for the given boundary conditions and for the use of the spin diffusion equation (and the spin-resolved spin-mixing conductance) one assumes $\epsilon  \ll 1$ ~\cite{Tserkovnyak2002b}. However, for $\epsilon > 0.1 $ most of the spin scattering occurs right at the interface and consequently the films are almost perfect sinks. 
Hence, in this case there is no dependence on the thickness of the film. Since in such films the interface plays the prominent role and scattering occurs at or near the interface, many issues regarding proximity effects, the induced spin-accumulation, and the spin Hall angle inferred from the measurements should be taken as phenomenological parameters rather than direct connections to a quantitative value of the bulk spin Hall angles. 

\subsection{Kubo formalism}
\label{sec:theory:Kubo}
In this section we review  the 
Kubo formalism employed in the calculations of the intrinsic SHE which incorporate the effects of disorder at its simplest level  through finite quasiparticle lifetime.
 It provides a
 fully quantum mechanical formally exact expression for the spin and anomalous conductivity in linear response theory \cite{Mahan2000}.
 Here we
 emphasize the key issues in studying the SHE within this formalism and how it relates to the semiclassical formalism
 described in the previous sections.
 
For the purpose of studying the SHE and AHE  it is best to reformulate the current-current Kubo formula for the conductivity
in the form of the  Bastin formula (see appendix A in \onlinecite{Crepieux2001b})
which can be manipulated into the more familiar form for the conductivity of the
Kubo-Streda formula for the zero-temperature Hall conductivity,
$\sigma_{xy}^{H}=\sigma_{xy}^{I(a)}+\sigma_{xy}^{I(b)}+\sigma_{xy}^{II}$, where
\begin{equation}
\sigma_{xy}^{I(a)}=\frac{e^2}{2\pi V} {\rm Tr} \langle {\{\hat{s}_z,\hat{v}\}}_x G^R(\epsilon_F) \hat{v}_y G^A(\epsilon_F)\rangle_c,
\label{sigmaIa}
\end{equation}
\begin{equation}
\sigma_{xy}^{I(b)}=-\frac{e^2}{4\pi V} {\rm Tr} \langle {\{\hat{s}_z,\hat{v}\}}_x G^R(\epsilon_F) \hat{v}_y 
G^R(\epsilon_F)
+c.c.\rangle_c,
\label{sigmaIb}
\end{equation}
\begin{eqnarray}
\sigma_{xy}^{II}&=&
\frac{e^2}{4\pi V} \int_{-\infty}^{+\infty} d\epsilon f(\epsilon) 
{\rm Tr}[{\{\hat{s}_z,\hat{v}\}}_x G^R(\epsilon) v_y \frac{G^R(\epsilon)}{d\epsilon} 
\nonumber\\&&
-{\{\hat{s}_z,\hat{v}\}}_x \frac{G^R(\epsilon)}{d\epsilon}v_yG^R(\epsilon)+c.c.].
\label{sigmaII}
\end{eqnarray}
Here the subscript $c$ indicates a disorder configuration average. 
The last contribution, $\sigma^{II}_{xy}$, was originally derived by Streda in the context of the 
QHE \cite{Streda1982}. 
In these equations
$G^{R/A}(\epsilon_F)=(\epsilon_F-H\pm i\delta)^{-1}$ are the retarded and advanced Green's functions evaluated
at the Fermi energy of the total Hamiltonian. 

Looking more closely at $\sigma_{xy}^{II}$ we notice that every term depends on  products of retarded Green's functions only,
or on products of advanced Green's functions only. It can be shown
that only the disorder free part of $\sigma_{xy}^{II}$ is important in the weak disorder limit, i.e.,
this contribution is zeroth order in the parameter $1/k_F l$. The only effect of disorder
on this contribution (for metals) is to broaden the Green's functions (see below) through the introduction of a finite
lifetime \cite{Sinitsyn2007}.
By a similar argument, 
$\sigma_{xy}^{Ib}$ is of order $1/k_F l$ and can be neglected in the weak scattering limit \cite{Mahan2000}.
Thus, important disorder effects beyond simple quasiparticle lifetime broadening
are contained only in $\sigma_{xy}^{Ia}$.  For these reasons, it is standard within the Kubo formalism to 
neglect $\sigma_{xy}^{Ib}$ and evaluate the $\sigma_{xy}^{II}$ contribution with a simple lifetime broadening approximation to the
Green's function.

In this formalism the effect of disorder on the disorder-configuration averaged Green's function is 
captured by the use of the T-matrix, defined by the integral equation $T=W+W G_0 T$, where $W=\sum_i V_0\delta(r-r_i)$
is a delta-scatterers potential
and $G_0$ are the Green's function of the pure lattice. 
From this one obtains 
\begin{equation}
\bar{G}=G_0+G_0 T G_0=G_0+ G_0 \Sigma \bar{G}.
\label{G}
\end{equation}
  Upon disorder averaging we obtain 
\begin{equation}
\Sigma=\langle W\rangle_c+\langle W G_0 W\rangle_c
+\langle W G_0 W G_0 W\rangle_c +...
\end{equation}
To linear order in  the impurity concentration, $n_i$, this translates to
\begin{equation}
\Sigma(z,\kk)= n_i V_{\kk,\kk}+\frac{n_i}{V}\sum_\kk V_{\kk,\kk'} G_0(\kk',z) V_{\kk',\kk}+\cdots,
\end{equation}
with $V_{\kk,\kk'}=V(\kk-\kk')$ being the Fourier transform of the single impurity
potential, which in the case of delta scatterers is simply $V_0$.
Note that  $\bar{G}$ and $G_0$ are diagonal in momentum but, due to the presence of spin-orbit coupling,  non-diagonal in 
spin-index in the Pauli spin-basis. 
 
One effect of disorder on the spin and anomalous Hall conductivity is taken into account by inserting the 
disorder averaged Green's function, $\bar{G}^{R/A}$, directly into the expressions (\ref{sigmaIa}) and (\ref{sigmaII}) for  $\sigma_{xy}^{Ia}$ and $\sigma_{xy}^{II}$, respectively.
This step captures the intrinsic contribution to the SHE and AHE and the effect of disorder on it, which is generally weak 
in metallic systems.  

The so-called ladder diagram vertex corrections 
 contribute to the AHE and SHE at the same order in
$1/k_Fl$ as the intrinsic contribution. It is useful to define a ladder-diagram 
corrected velocity vertex $\tilde{v}_\alpha(\epsilon_F)\equiv v_\alpha+\delta\tilde{v}_\alpha(\epsilon_F)$, where 
\begin{equation}
\delta\tilde{v}_\alpha(\epsilon_F)=\frac{n_i V_0^2}{V}\sum_\kk \bar{G}^R(\epsilon_F) (v_\alpha+\delta\tilde{v}_\alpha(\epsilon_F)) \bar{G}^A(\epsilon_F).
\label{vertex_corr}
\end{equation}
Note again that $\tilde{v}_\alpha(\epsilon_F)$ and $ v_\alpha=\partial\hat{H_0}/{\partial \hbar k_\alpha}$ are matrices in the 
spin-orbit coupled band basis.
The skew scattering contributions are obtained by evaluating, without doing an infinite partial sum as in the case of the
ladder diagrams, third order processes in the disorder scattering.

As  may seem obvious from the above machinery, calculating the intrinsic contribution is not very difficult, while calculating the full effects of the disorder
in a systematic way (beyond calculating a few diagrams)
is  challenging for any disorder model beyond the simple delta-scattering model. 

An important recent development  has taken place within the theory 
of the AHE which we hope will have a direct analogy to the spin Hall conductivity. Assuming uncorrelated Gaussian noise disorder, i.e., ignoring any 
skew scattering contribution, all the scattering independent contributions - side-jump and intrinsic - can be formulated in terms of the band structure of the crystal alone 
\cite{Weischenberg2011,Kovalev2010}.

For a short-range scattering disorder model, e.g. scalar delta-correlated Gaussian  disorder, the 
 starting point for this theory of the scattering-independent side-jump is the 
retarded Green's function in equilibrium and the Hamiltionian $H$ of a general multiband 
noninteracting system. 
The first step is to expand the 
self-energy of the system $\Sigma_{eq}$ in powers of potential $V(\rf)$, which 
describes scattering off impurities.
 Inserting the 
expression for the self-energy within these simple disorder models into the appropriate 
equations for the current densities derived following the Kubo-St\v{r}eda formalism mentioned above, 
rotating into eigenstate representation and keeping only the leading order terms 
in the limit of vanishing disorder parameter $V_0$, 
 the scattering-independent  part of the AHE conductivity may be written as 
$\sigma_{xy}^{H-(0)}=\sigma_{xy}^{H-int}+\sigma_{xy}^{H-sj}$, where
\begin{equation}
\sigma^{H-int}_{ij}=\frac{2e^2}{\hbar}\int\frac{d^3k}{(2\pi)^3}{\rm Im}
{\sum\limits_{n\neq m} (f_n-f_m)\frac{
  v_{nm,i}(\kf)
  v_{mn,j}(\kf)}{(\omega_n-\omega_m)^2}}
  \label{ic}
\end{equation}
is the intrinsic contribution. In this expression the band 
indices $n$ and $m$ run from $1$ to $N$, $v_{nm,i}$ are the matrix elements of the 
velocity operator $\hat v_{i}=\partial_{\hbar k_i}\hat H$ and 
$\omega_{n}(\kf)=\varepsilon_{n}(\kf)/\hbar.$ The scattering-independent side-jump contribution to AHE conductivity 
 for inversion-symmetric systems reads:
\begin{equation} 
\begin{aligned}
&\sigma^{H-sj}_{ij}=\frac{e^2}{\hbar}\sum\limits_{n=1}^N
 \int\frac{d^3k}{(2\pi)^3}{\rm Re}\,{\rm Tr}\bigg\{
 \delta(\varepsilon_F-\varepsilon_{n})\frac{ \gamma_c}{[ \gamma_c]_{nn}}\times\\
 &\quad\times\bracks{ S_n  A_{k_i}( {1}-  S_n)
 \frac{\partial\varepsilon_n}{\partial k_j}
 -
 S_\eta A_{k_j}( {1}-  S_n)
 \frac{\partial\varepsilon_n}{\partial k_i}}\bigg\}.
\label{sjc}
\end{aligned}
\end{equation}
Here the imaginary part of the self-energy ${\rm Im}\Sigma_{eq}
=-\hbar V_0\gamma$ is taken to be
in the eigenstate representation, i.e.  $\gamma_c=U^\dagger \gamma U$, with
\begin{equation}
{\gamma}=\frac 12 \sum\limits_{n=1}^N\int\frac{d^3 k}{(2\pi)^{2}}
\,U S_n U^\ad\,\delta(\omega_F-\omega_n),
\end{equation}
$U$ as the unitary matrix that diagonalizes the Hamiltonian at point $\kf$,
\begin{equation}
[U^\ad H(\kf) U]_{nm}=\varepsilon_n(\kf)\delta_{nm},
\end{equation}
$S_n$ is a $N\times N$ matrix that is diagonal in the band indices, 
$[S_n]_{ij}=\delta_{ij}\delta_{in}$, and  the so-called Berry connection matrix is 
given by $ A_\kf=i U^\ad\partial_\kf  U $.  Not included in 
Eq.~(\ref{sjc}) are the vertex corrections,  which vanish for an inversion-symmetric 
system in the Gaussian disorder model. Because this side-jump contribution 
 in  the short-range disorder model  is solely determined by the electronic structure 
of the pristine crystal, it is thus directly accessible by \textit{ab initio} methods. 

\begin{figure}[t!]
\begin{center}
\subfiguretopcaptrue
\subfigure[{Ni [001] $\sigma^{{sj}}$}]{
\includegraphics[scale=0.26]{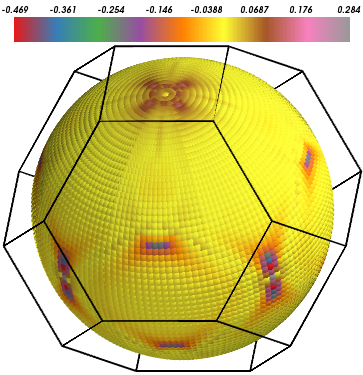}}
\subfigure[{Ni [110] $\sigma^{{sj}}$}]{
\includegraphics[scale=0.26]{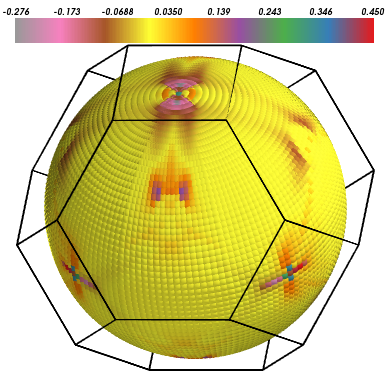}}
\subfiguretopcapfalse
\subfigure[{Ni [001] $\sigma^{{int}}$}]{
\includegraphics[scale=0.26]{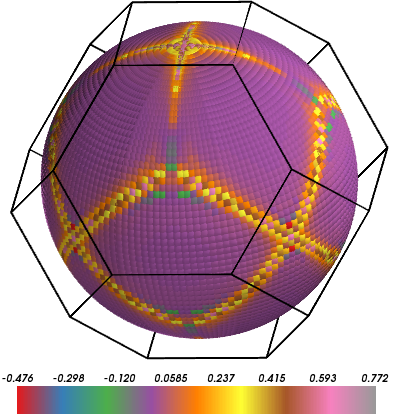}}
\subfigure[{Fe [001] $\sigma^{{sj}}$}]{
\includegraphics[scale=0.26]{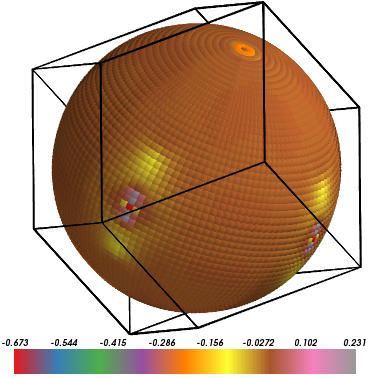}}
\caption{\label{fig:sj} (Color online) Angular distribution of the a) Side-jump contribution for Ni [001], b) Side-jump contribution for Ni [110], 
c) Intrinsic contribution for Ni [001], and d) Side-jump contributionfor Fe [001] on a sphere in the 
Brillouin zone. The color code of each surface point corresponds to 
the sum of all contributions along the path from the origin to the particular surface point. From Ref.~\onlinecite{Weischenberg2011}.}
\label{AHEfig}
\end{center}
\end{figure}

Table~\ref{AHEtable} shows a comparison of the improvement in predictive power of the AHE theory when including the side jump term. Fig.~\ref{AHEfig} shows the non-trivial angular dependence, within the Fermi surface, of the side jump and intrinsic contributions. This is reminiscent of the spin-hot spots observed previously in the theory of spin-dephasing, and emphasizes the importance of anisotropies induced by the band structure itself.  

\begin{table}
\caption{\label{feco}
AHE conductivities for bcc Fe and hcp Co in S/cm for selected high-symmetry 
orientations of the magnetization. $\sigic$, $\sigsj$ and $\sigij$ stand for intrinsic contribution, 
side-jump contribution and their
sum, respectively. The experimental values are for the scattering-independent conductivity.}
\begin{ruledtabular}
\begin{tabular}{llll|llr}
 {\bf Fe} & [001]  &  [111]   &  [110]   & {\bf Co}   &   [001]  & [100]   \\ \hline
 $\sigic$ &  767   &   842    &   810    &  $\sigic$  &    477   & 100    \\
 $\sigsj$ &  111   &   178    &   141    &  $\sigsj$  &    217   & $-$45  \\
 $\sigij$ &  878   &  1020    &  951    &  $\sigij$  &    694   &  55     \\ \hline                    
 Exp.     & 1032 &  &  &  Exp. &  813  &   150   
\end{tabular}
\end{ruledtabular}
\label{AHEtable}
\end{table}

\section{Experimental studies of spin Hall effect} \label{sec:experiments}

Several experimental schemes to detect the SHE were outlined already by \onlinecite{Dyakonov1971} in their seminal theory work.
They proposed to use paramagnetic resonance for detecting the edge spin polarization, to measure the nuclear magnetization resulting from the Overhauser effect, to exploit the gyrotropy, i.e., the difference in the propagation of electro-magnetic waves with opposite helicities through the spin polarized edges, or, in semiconductors,  to detect circular polarization of the luminescence excited by an unpolarised light. 

Variants of the two latter schemes, namely the Kerr magneto-optical microscopy  and circularly polarized electroluminescence from the sample edges, were indeed employed in the pioneering SHE experiments \cite{Kato2004,Wunderlich2004,Wunderlich2005}. These were, however, performed more than thirty years after the \onlinecite{Dyakonov1971} original proposal. Within these three decades, the interest in the phenomenon was scarce. The experimental SHE research only picked up momentum  after the theoretical work by \onlinecite{Hirsch1999} who rediscovered the phenomenology of the extrinsic SHE, and after the prediction of the intrinsic SHE \cite{Murakami2003,Sinova2004}. 

The renewed  theoretical interest occured in the midst of an extraordinary growth of the nascent field of spintronics \cite{Zutic2004}, which had already found important applications in the hard disk drive industry and promised  revolutionary concepts for memory and logic devices. In this setting, the theoretical SHE proposals not only ignited an extensive theoretical debate for their inherent fundamental interest but also attracted significant attention due to the potential of spin Hall phenomena as new spin injection and detection tools. The proposals started to materialize shortly thereafter with  the observations of the SHE in $n$-doped semiconductors \cite{Kato2004} and in the 2DHG \cite{Wunderlich2004,Wunderlich2005}, and of the ISHE in metallic systems \cite{Valenzuela2006,Valenzuela2007,Saitoh2006}.

In this section we review the experimental studies of the spin Hall phenomena.  In Sec.~\ref{early} we summarize AHE experiments  in non-ferromagnetic materials that were performed within the three decades separating the first theoretical proposal and the experimental observations of the SHE. The rest of the section is devoted to modern experiments divided according to the techniques used to generate, detect, and manipulate the SHE and ISHE in experimental samples (Sec.~\ref{opt_tools}-\ref{she-mag}). The overall understanding of the experiments is still incomplete regarding some materials and structures, in particular when trying to quantify the magnitude of the SHE. Therefore, in those cases, we attempt to provide an overview of the current status of the field while stressing the strengths and weaknesses of the different techniques and methods employed.

Apart from the basic research interest in this relativistic quantum-mechanical phenomenon, Sec.~\ref{she-mag} provides an illustration of the application potential of the SHE in spintronic devices. This prompted detailed studies of the SHE efficiency for the charge-spin conversion in a variety of materials. Measurements of the corresponding spin-Hall angles are summarized in the Sec.~\ref{she_angles}.

\subsection{Early experiments of anomalous Hall effect in paramagnets}
\label{early}

\onlinecite{Chazalviel1972} 
reported a pioneering work on a spin dependent Hall effect in non-magnetic semiconductors. They detected the AHE in InSb and $n$-doped Ge at low temperatures ($<25$ K), where the spin polarization was created by the application of a magnetic field and the spin-dependent Hall effect was separated from the larger ordinary HE by magnetic resonance of the conduction electrons \cite{Chazalviel1972,Chazalviel1975}. The magnitude of the measured anomalous Hall angles was of the order of $10^{-4}$ for InSb, and of $10^{-5}$ for Ge, while its sign was observed to change depending on the degree of carrier compensation (InSb) and temperature (Ge). The change in sign was associated with competing contributions from the side-jump and skew scattering mechanisms. The former was expected to be favored in low mobility samples, which was confirmed in the experiment. 

In early 1980's, Fert and collaborators studied diluted magnetic alloys based on non-magnetic hosts, such as Au and Cu, and magnetic impurities such as Mn, Fe, or Cr \cite{Fert1981}. They found that CuMn showed negligible skew scattering effects, but that the exchange scattering by polarized Mn impurities created a spin polarized current. They also noted that the addition of non-magnetic impurities to CuMn gave rise to skew scattering of the polarized current by the unpolarised impurities. By analyzing variations of the Hall coefficient, they were able to extract the Hall angle for the non-magnetic impurities. They found that they varied from $-1.4 \times 10^{-2}$ for Lu to $-2.6 \times 10^{-2}$ for Ir.

In another type of AHE measurements, a circularly polarized beam at the normal incidence to the surface of a bulk semiconductor was used to excite spin-polarized photo-electrons \cite{Bakun1984,Miah2007}. These electrons diffused in the vertical direction from the surface and after aligning their spins  along an axis parallel to the surface by an applied magnetic field (via Hanle precession), an electrical voltage was detected in the transverse in-plane direction \cite{Bakun1984}. Alternatively, the vertically spin-polarized electrons can be accelerated in the in-plane direction by an applied electrical bias yielding also a transverse in-plane voltage \cite{Miah2007}. Since in these experiments the source spin-current is accompanied by a diffusive or drift charge current the geometry corresponds  to the AHE.
\subsection{Optical tools in spin Hall experiments}
\label{opt_tools}

\subsubsection{Optical detection of the spin Hall effect}
\label{opt_she}

Following  \onlinecite{Dyakonov1971} prediction of the extrinsic SHE, the proposed concepts by \onlinecite{Hirsch1999} and  \onlinecite{Zhang2000} for its experimental detection relied on electrical measurements. The concepts considered SHE channels in NMs, in the latter case attached to a FM detection electrode. 

However, the  experimental discovery of the SHE was prompted also by theory works which approached the SHE origin and detection from a different angle. Inspired by studies of the intrinsic nature of the closely related AHE in FMs, \cite{Onoda2002,Jungwirth2002}, \onlinecite{Murakami2003} and  \onlinecite{Sinova2004} predicted  that  a spin-dependent transverse deflection of electrons in non-magnetic systems can originate directly from the relativistic band structure of the conductor without involving the Mott scattering. Unlike \onlinecite{Hirsch1999} and \onlinecite{Zhang2000} who considered  the extrinsic, scattering induced SHE  and electrical detection schemes designed for metals, the intrinsic SHE proposals focused on semiconductors and suggested to utilize the optical activity of these materials for detecting the SHE. As in the original work by \onlinecite{Dyakonov1971}, circularly polarized electro-luminescence was suggested in Ref.~\cite{Murakami2003}, while spatially resolved magneto-optical Faraday or Kerr effects were discussed in  Refs.~\cite{Murakami2003,Sinova2004}. These methods were indeed used  in the first measurements of the phenomenon. \onlinecite{Kato2004d} employed a magneto-optical Kerr microscope to scan the spin polarization across the channel while \onlinecite{Wunderlich2004,Wunderlich2005} used co-planar $p-n$ diodes to detect circularly polarized electro-luminescence at opposite edges of the spin Hall channel. \onlinecite{Wunderlich2004,Wunderlich2005} ascribed the observed signal to the intrinsic SHE while \onlinecite{Kato2004d} to the extrinsic SHE.

\onlinecite{Kato2004d} performed the experiments in $n$-GaAs and $n$-In$_{0.07}$Ga$_{0.93}$As films grown by molecular beam epitaxy on (001) semi-insulating GaAs substrates. The films were doped with Si with $n = 3 \times 10^{16}$ cm$^{-3}$ in order to obtain long spin relaxation lifetimes of $\tau_s \sim 10$ ns,  
which result in spin diffusion lengths $\lambda_{sd} = \sqrt{D\tau_s}\sim 10$ $\mu$m. The unstrained GaAs sample consisted of 2 $\mu$m of $n$-GaAs grown on 2 $\mu$m of undoped Al$_{0.4}$Ga$_{0.6}$As, whereas the strained InGaAs sample had 0.5 $\mu$m of $n$-In$_{0.07}$Ga$_{0.93}$As and 0.1 $\mu$m of undoped GaAs.  Static Kerr rotation measurements were performed at 30~K with a pulsed Ti:sapphire laser tuned to the absorption edge of the semiconductor with normal incidence to the sample. In this technique, the laser beam is linearly polarized and the polarization axis of the reflected beam is determined. The rotation angle is proportional to the net magnetization along the beam direction.

\begin{figure}[h!]
\hspace*{-0cm}\epsfig{width=1\columnwidth,angle=0,file=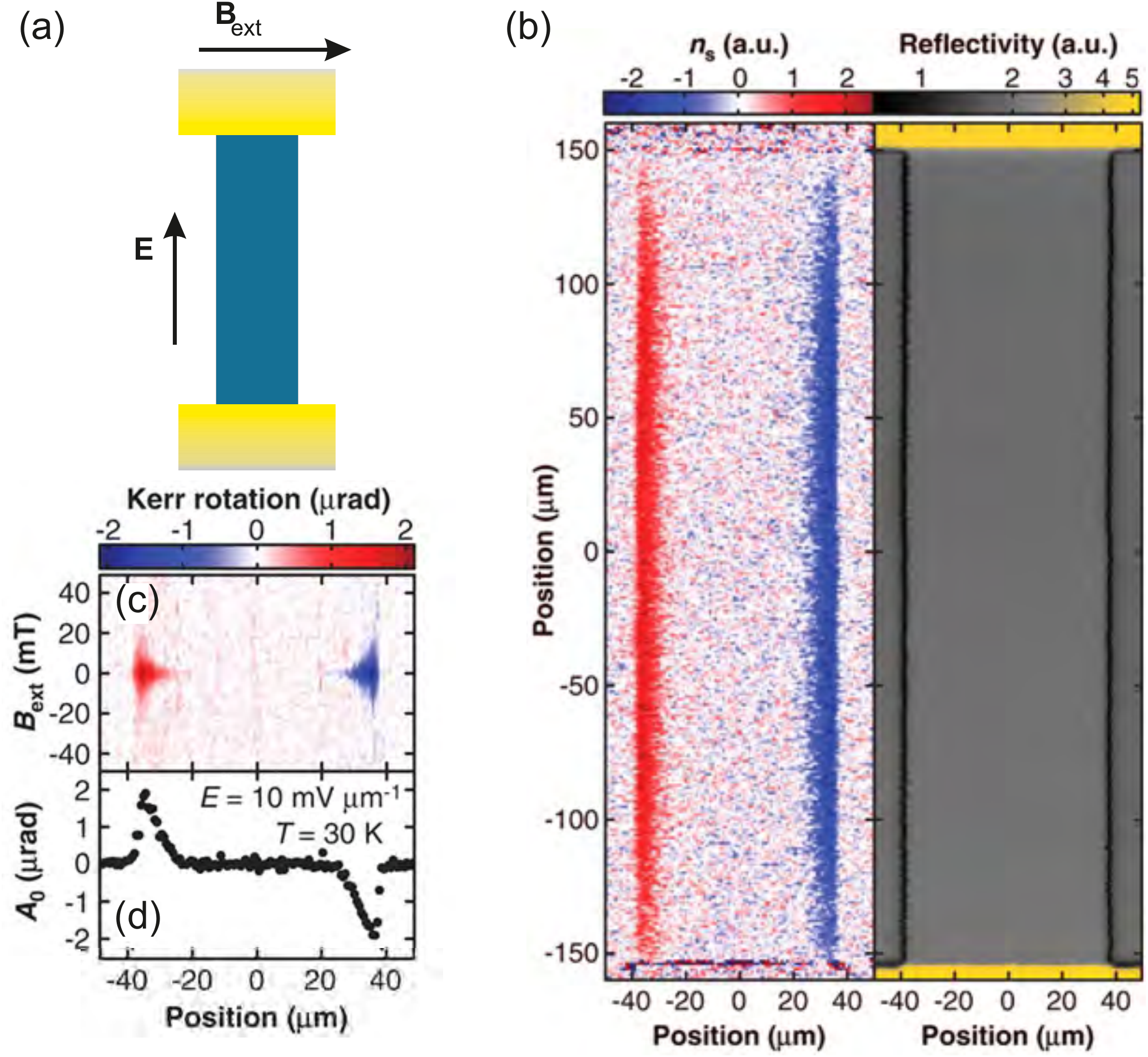}
%
\caption{Observation of the SHE by the magneto-optical Kerr microscope. (a) Schematics of the GaAs sample. (b) Two-dimensional images of the
spin density $n_s$ (left) and reflectivity $R$ (right) for an unstrained GaAs sample measured at temperature 30~K and applied driving electric field $E$=10~mV$\mu$m$^{-1}$. (c) Kerr rotation as a function of $x$ and external magnetic field $B_{ext}$ for $E$=10~mV$\mu$m$^{-1}$. (d) Spatial dependence of the peak Kerr rotation $A_0$ across the GaAs channel. From Ref.~\onlinecite{Kato2004d}.}
\label{fig_2_SHE_MOKE}
\end{figure}

Figure~\ref{fig_2_SHE_MOKE}(a) shows a schematic of the experimental geometry. The epilayers were patterned into $300 \times 77$ $\mu$m$^2$ (GaAs) and $n$-InGaAs $300 \times 33$ $\mu$m$^2$ (InGaAs) channels. An electric field was applied along the channel while a magnetic field $B$ could be applied perpendicular to it in the film plane. Figure~\ref{fig_2_SHE_MOKE}(b) shows a two dimensional scan of the GaAs sample, which demonstrates the existence of spin accumulation close to the edges. The amplitude of the measured edge spin polarizations reaches $\sim 0.1$\%. The polarization has opposite sign at the two edges and decreases rapidly with the distance from the edge as expected for the SHE. This is clearly seen in the one dimensional profile in Fig.~\ref{fig_2_SHE_MOKE}(c). Further experiments demonstrated the effect of spin (Hanle) precession, and associated the suppression of the observed signal to the applied magnetic field, as predicted by \onlinecite{Dyakonov1971b} and \onlinecite{Hirsch1999}.

\onlinecite{Zhang2000} showed by solving the spin-dependent drift-diffusion equations for a finite width channel, that the spin diffusion length $\lambda_{sd}$ defines the length scale of the edge spin accumulation. By fitting to the spin-drift diffusion equation, \onlinecite{Kato2004d}  extracted the transverse spin-current and the spin Hall resistivity $\rho^H$. This analysis, which assumes well-resolved spin-up and spin-down transport channels \cite{Zhang2000,Hirsch1999}, is valid in the weak spin-orbit limit, which is verified by noting that $\Delta_{so}\tau/\hbar \sim 10^{-3} \ll 1$, where $\Delta_{so}$ is the spin-orbit coupling energy and $\tau$ the momentum scattering time. The measured value of $\rho^H \sim 2$ $\Omega$m is consistent with that obtained from modeling based on scattering by screened and short-range impurities \cite{Engel2006,Tse2006}. Noting that the charge resistivity $\rho \sim 4 \times 10^{-6}$ $\Omega$m, this corresponds to a spin Hall angle of $\sim 10^{-4}$. In the weak spin-orbit coupling regime the spin-orbit splitting of the quasiparticle bands is smeared out by disorder which favors the extrinsic SHE interpretation of the measured signal. The absence of the intrinsic SHE was confirmed by measurements in the strained InGaAs sample which showed no dependence of the SHE signal on the strain induced anisotropies of the spin-orbit coupled band structure.

\vspace*{0cm}

Experiments in 2DHG devices (Fig.~\ref{fig_3_SHE_EL}) were carried out in the strong spin-orbit coupling limit, $\Delta_{so}\tau/\hbar \sim 4$, which favors the intrinsic mechanism \cite{Wunderlich2005,Nomura2005b}. The device comprised coplanar $p-n$ junction light emitting diodes (LEDs) that were patterned in (Al,Ga)As/GaAs heterostructures grown by molecular beam epitaxy and using modulation donor (Si) and acceptor (Be) doping in (Al,Ga)As barrier materials. The heterostructure consisted of an $n$-doped AlGaAs/GaAs heterojunction, followed by the growth of 90 nm of intrinsic GaAs and a $p$-doped AlGaAs/GaAs heterojunction. The coplanar $p-n$ junctions were created by removing the $p$-doped layer of the wafer and thus creating a hole channel, with a carrier density $2 \times 10^{12}$ cm$^{-2}$. The 2DEG at the bottom heterojunction was almost depleted. The removal of $p$-doped surface layer populated the 2DEG, forming the $n$-side of the coplanar $p-n$ junction.

A current $I_p$ was applied to drive the electroluminescence at the edge of the channel due to recombination near $p-n$ junctions. The detection of spin-polarization in the 2DHG was done by measuring the circular polarization of the emitted light, shown in Figs.~\ref{fig_3_SHE_EL}(d),(e). The magnitude of the signal reached $\sim 1 \%$ at 4 K. Consistent with the SHE phenomenology, the experiments demonstrated that the spin accumulation was opposite at opposite sides of the channel and that it reversed sign following current reversal.

Calculations of the SHE conductivity showed that the SHE originating from the spin-orbit coupled quasiparticle bands of the 2DHG is only weakly affected by disorder for the parameters of the studied system \cite{Wunderlich2005}. A quantitative microscopic description of the measured edge spin accumulation signal was developed and further experimentally tested  by  
\onlinecite{Nomura2005b}. The theory analysis pointed out that the length scale of the edge spin accumulation is defined in the strong spin-orbit coupling regime by the spin-orbit precession length $L_{so}=v_F\tau_{so}$, where $\tau_{so}=\hbar\pi/\Delta_{so}$ is the precession time of the spin in the internal spin-orbit field and $v_F$ is the Fermi velocity.  With increasing strength of the spin-orbit coupling, the edge spin accumulation region narrows down and, simultaneously, the amplitude of the spin polarization increases. For the experimental parameters of the studied 2DHG, $L_{so}\sim 10$~nm and the calculated amplitude of the edge spin polarization was 8\%, in good agreement with the 1\% polarization of the measured electro-luminescence signal which was averaged over a $\sim 100$~nm sensitivity range of the co-planar light emitting diode. A comparison between measurements in devices with 1.5 and 10 $\mu$m wide channels confirmed the expectation that the SHE signal is independent of the channel width.

\begin{figure}[h!]
\hspace*{-0cm}\epsfig{width=1\columnwidth,angle=0,file=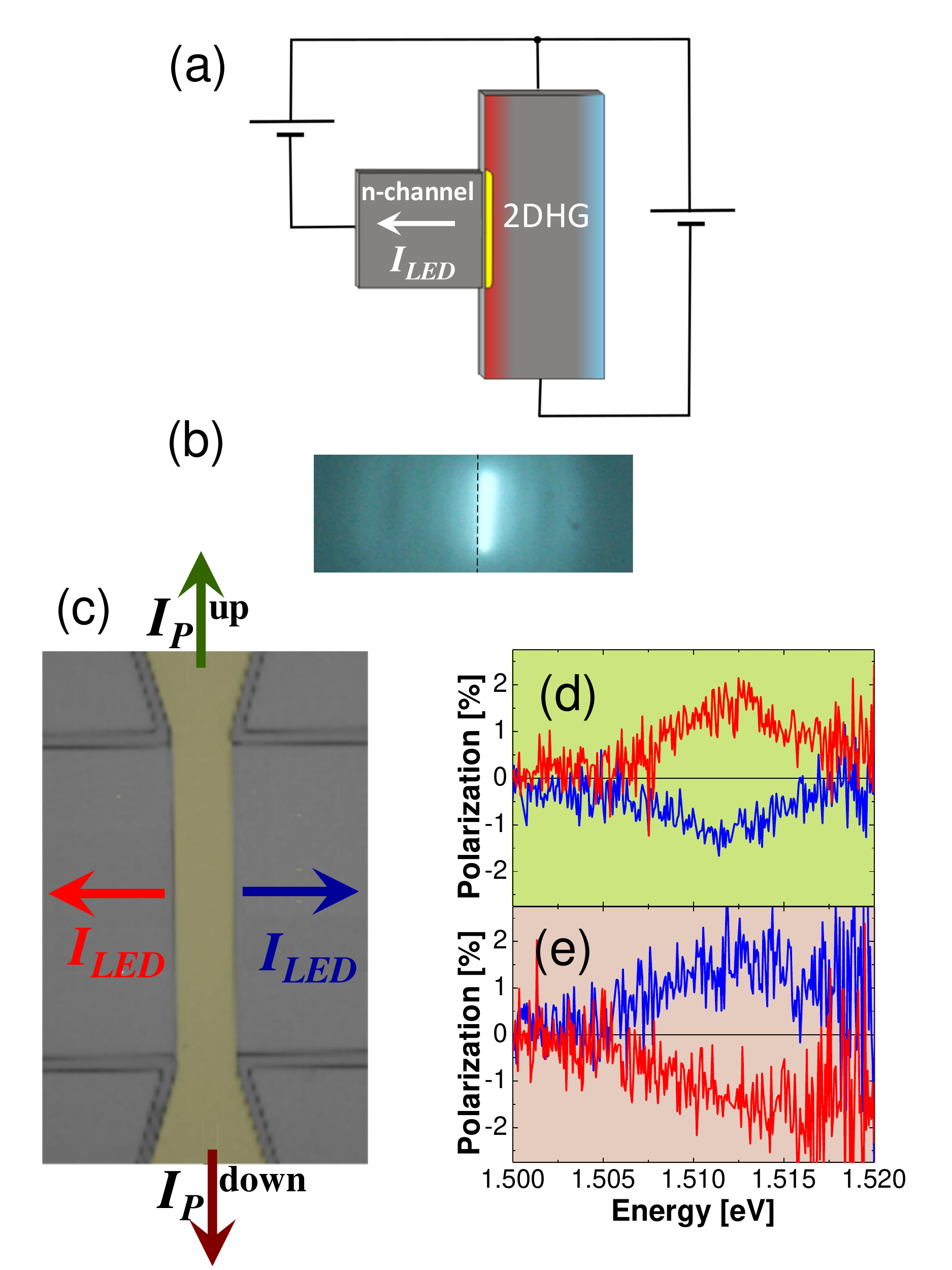}
%
\caption{Observation of the SHE by the circularly polarized electro-luminescence of co-planar $p-n$ diodes. (a) Schematic configuration of the lateral $p-n$ junction to detect spin accumulation. (b) Light emission from the $p-n$ junction recorded by a charged-couple device camera. (c) Electron microscope image of the microdevice with symmetrically placed $p-n$ diodes at both edges of the 2DHG channel. (d),(e) Emitted light polarization of recombined light in the $p-n$ junction for the current flow indicated in (c) at 4~K. From Ref.~\onlinecite{Nomura2005b}.}
\label{fig_3_SHE_EL}
\end{figure}

Subsequent magneto-optical measurements of the SHE in the $n$-GaAs 3D epilayers have experimentally demonstrated that the SHE-induced spin accumulation is due to a transverse  spin-current which can drive spin polarization tens of microns into a region in which there is minimal electric field \cite{Sih2006}. The work proved experimentally that the SHE can be used as a source of spin-current generated in a NM.

A systematic doping dependence of the SHE angle was studied in $n$-GaAs 3D epilayers with electron densities $n=1.8\times10^{16}-3.3\times10^{17}$~cm$^{-3}$  and the results were found consistent with theory predictions for the extrinsic SHE \cite{Matsuzaka2009}. The measured SHE angles of $\sim5\times10^{-4}-5\times10^{-3}$  increase with increasing doping with a tendency to saturate at the high doping end of the studied set of samples at a value corresponding to $\sim 1\%$ edge spin polarization. It was concluded from this systematic analysis that the spin accumulation is reduced by an enhanced spin relaxation due to the Dyakonov-Perel  mechanism, while the spin-current induced by the SHE is enhanced with increasing $n$ \cite{Matsuzaka2009}. The SHE was observed also  in other semiconductor systems including  $n$-ZnSe 3D epilayers \cite{Stern2006b}, and InGaN/GaN superlattices \cite{Chang2007}.

\subsubsection{Optical generation of the inverse spin Hall effect}
\label{opt_ishe}

A traditional way of generating spin-polarized photo-carriers in semiconductors is by absorption of circularly polarized light \cite{Meier1984}. Because of the optical selection rules, the out-of-plane spin polarization of photo-carriers is determined in this technique by the sense and degree of the circular polarization of vertically incident light.  This technique was used to observe the AHE in semiconductors \cite{Bakun1984,Miah2007}, which we have already discussed in Sec.~\ref{early}, and eventually led also to the detection of the ISHE generated by the pure spin current.

\onlinecite{Ando2010} reported an experiment in a NM/semiconductor hybrid structure in which they demonstrated the conversion of circularly polarized light absorbed in a semiconductor to an electrical signal in the attached NM ISHE sensor. The photo-induced ISHE was observed in a Pt/GaAs hybrid structure. In the GaAs layer, circularly polarized light generates spin-polarized carriers, inducing a pure spin-current into the Pt layer through the interface. This pure spin-current is  converted into an electrical voltage due to the ISHE in Pt. Systematic changes of the ISHE signal were observed upon changing the direction and ellipticity of the circularly polarized light, consistent with the expected phenomenology of the photo-induced ISHE. The observed phenomenon allows the direct conversion of circular-polarization information into the electrical voltage and can be used as a spin photodetector.

Using a similar detector configuration, \onlinecite{Kampfrath2013} demonstrated the control of the transmission of terahertz spin-current pulses. The samples consisted of Fe/Au and Fe/Ru heterostructures. The absorption of a femtosecond laser pulse in the Fe layer generates a non-equilibrium electron distribution and associated spin-current, dominated by the majority-spin $sp$-like electrons, that flows into the Au(Ru) non-magnetic layer. The transport dynamics is different in the Fe/Au and Fe/Ru heterostructures because of the much larger electron mobility of Au; the flow of the non-equilibrium electrons occur much more slowly in Ru than in Au, and are accompanied by significantly more spin accumulation. The non-magnetic layer can thus be used to either trap or transmit electrons, and thus engineer ultrafast spin pulses, which change in temporal shape and delay. The detection of the spin-current pulses used by \onlinecite{Kampfrath2013} relied on the ISHE. 

While in the static experiments by \onlinecite{Ando2010} the resulting charge current is measured as a voltage, \onlinecite{Kampfrath2013} detected the electromagnetic pulse emitted by the charge current burst by electro-optical sampling using a GaP crystal. The feasibility of the experiment demonstrated the operation of the ISHE as a spin-current detector 
up to frequencies as high as 20 THz.

\begin{figure}[h!]
\hspace*{-0cm}\epsfig{width=.8\columnwidth,angle=0,file=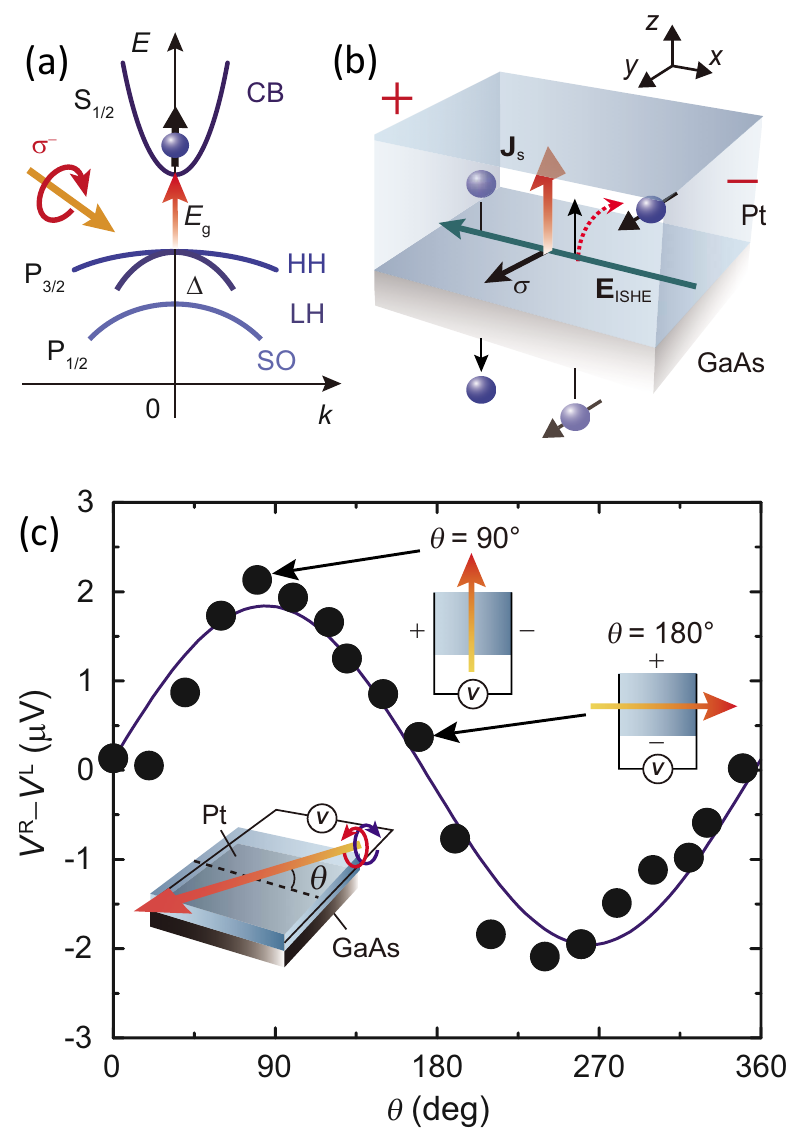}
%
\caption{(a) A schematic illustration of the band structure of GaAs and spin-polarized electrons generated by the absorption of circularly polarized light. (b) A schematic illustration of the ISHE induced by photoexcited pure spin-currents in the Pt/GaAs system. (c) The illumination angle $\theta$ dependence of $V^R-V^L$ measured for the Pt/GaAs hybrid structure. $\theta$ is the in-plane angle between the incident direction of the illumination and the direction across the electrodes attached to the edges of the Pt layer as shown in the inset. $V^R-V^L$ is the difference in the electromotive force for illumination with right and left circularly polarized light. The filled circles are the experimental data. The solid curve shows a fitting result using a function proportional to $\sin\theta$. From Ref~\onlinecite{Ando2010}.}
\label{fig_3_SHE_EL}
\end{figure}

\begin{figure}[h!]
\hspace*{-0cm}\epsfig{width=1\columnwidth,angle=0,file=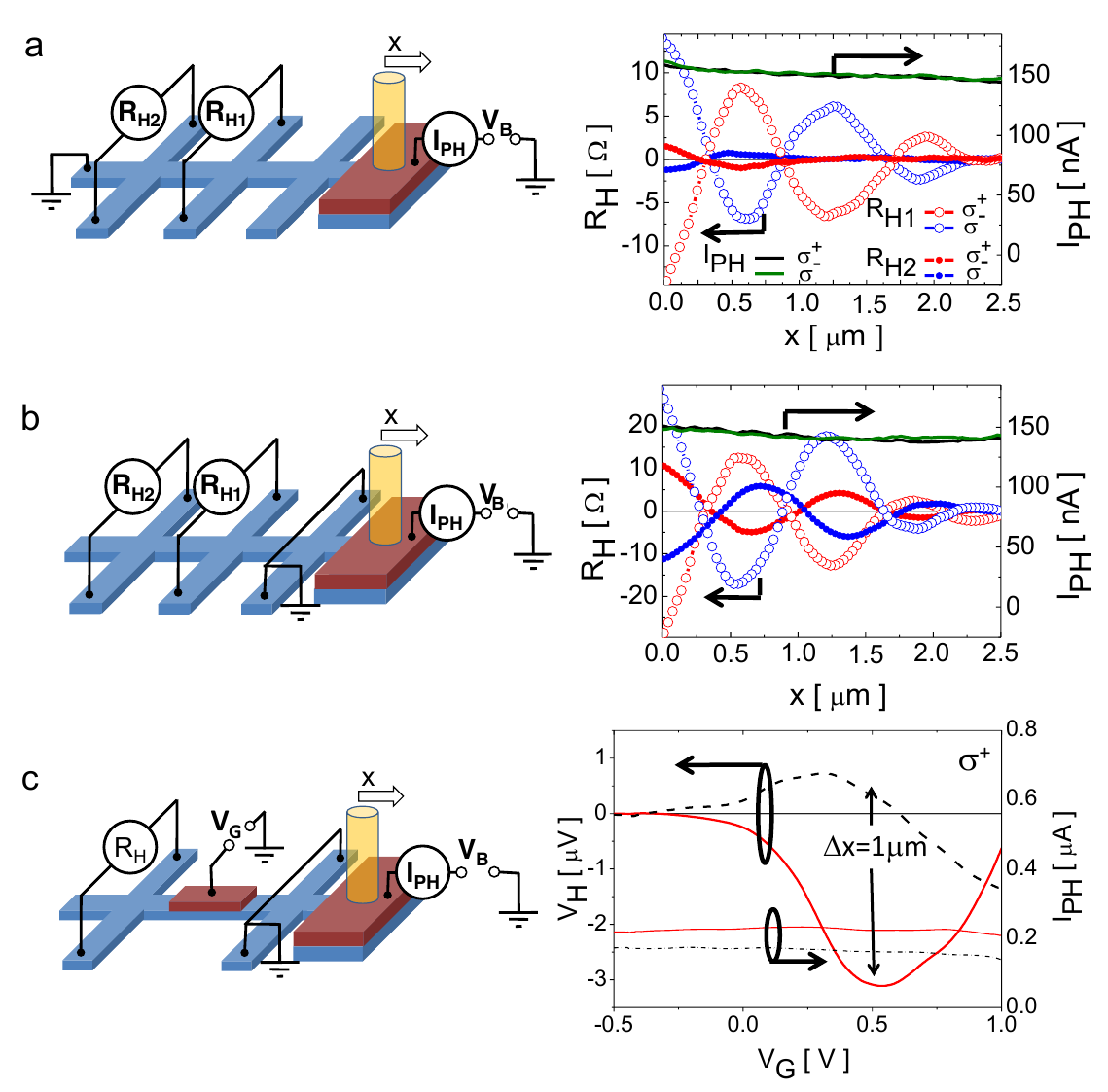}
\caption{ISHE based  transistor. (a) Schematics of the spin injection Hall effect measurement setup with optically injected spin-polarized electrical current propagating through the Hall bar and corresponding experimental Hall effect signals at crosses H1 and H2. The Hall resistances, $R_H=V_H/I_{PH}$, for the two opposite helicities of the incident light are plotted as a function of the focused light spot position, i.e., of the position of the injection point. The optical current $I_{PH}$ is independent of the helicity of the incident light and varies only weakly with the light spot position.  (b) Same as (a) for the ISHE measurement geometry in which electrical current is closed before the first detecting Hall cross H1. (c) Schematics of the setup of the spin Hall transistor and experimental Hall signals as a function of the gate voltage at a Hall cross placed behind the gate electrode for two light spot positions with a relative shift of 1~$\mu$m. From Ref.~\onlinecite{Wunderlich2010}.}
\label{fig_5_SHE_transistor}
\end{figure}

\onlinecite{Wunderlich2009,Wunderlich2010}, using the same type of lateral $p-n$ diodes as in Refs.~\cite{Wunderlich2005,Nomura2005b}, 
exploited optical spin injection by a circularly polarized laser beam to observe the ISHE 
and to fabricate experimental opto-spintronic and spin-transistor devices. 
In the SHE measurements in Refs.~\cite{Wunderlich2005,Nomura2005b}, the $p-n$ junctions were fabricated along the edges of the 2DHG channel and under forward bias could  sense the spin state of recombining electrons and holes through polarized electro-luminescence. In Refs.~\cite{Wunderlich2009,Wunderlich2010}, on the other hand, the spin Hall channel was fabricated in the etched part of the epilayer with the 2DEG, the channel was oriented perpendicular to the $p-n$ junction, and the diode was under zero or reverse bias, operating as a photocell as shown in Fig.~\ref{fig_5_SHE_transistor}. The optical activity of the lateral diode confined to a submicron depletion region, combined with a focused ($\sim 1$~$\mu$m) laser beam, allowed for a well localized injection  of spin-polarized photo-electrons into the planar 2DEG channel.

The Hall signals were detected electrically  on multiple Hall-crosses patterned along the channel. Two regimes of operation of the device are distinguished: One corresponds to an AHE regime, in which the reverse-bias charge current is drained behind the Hall crosses at the opposite end of the channel from the $p-n$ junction injection point (Fig.~\ref{fig_5_SHE_transistor}(a)).  The other regime corresponds to the ISHE measurement since in this case the charge current is drained before the Hall crosses, allowing only the pure spin-current to diffuse further in the channel (Fig.~\ref{fig_5_SHE_transistor}(b)). In both cases the measured transverse electrical signals were consistent with the phenomenology of the spin-dependent Hall phenomena \cite{Wunderlich2009,Wunderlich2010}. The sign of the voltage was opposite for opposite helicities of the incident light, i.e., opposite spin-polarizations of injected photo-electrons. Moreover, the amplitude of the electrical signals was found  to depend linearly on the degree of circular polarization of the light, rendering the device an electrical polarimeter \cite{Wunderlich2009}. The electrical signals were observable over a wide temperature range with spin Hall angles of $10^{-3}-10^{-2}$. The measured 2DEG was in the weak spin-orbit coupling regime, $\Delta_{so}\tau/\hbar\sim 10^{-1}$, and the measured data were consistent with the extrinsic mechanism \cite{Wunderlich2009}.

\subsubsection{All-optical generation and detection}
\label{opt_det}

The SHE and the ISHE were also observed using two-color optical coherence control techniques in intrinsic GaAs at 80 K with polarized 70 fs, 715  and 1430 nm pulses \cite{Zhao2006}. When the pulses were orthogonally polarized, a pure spin source current was generated that yielded a transverse Hall pure charge current via the ISHE. When the pulses were parallel polarized, a pure charge source current was generated that yielded a pure spin-current via the SHE. By varying the relative phase or polarization of the incident pulses, the type, magnitude, and direction of both the source and transverse currents were tuned without applying electric or magnetic fields. In contrast to the previous steady-state experiments, where drift currents are generated by electric fields, the injected currents are ballistic currents with electrons traveling initially at $\sim 1000$ km/s. 

\begin{figure}[h!]
\hspace*{-0cm}\epsfig{width=1\columnwidth,angle=0,file=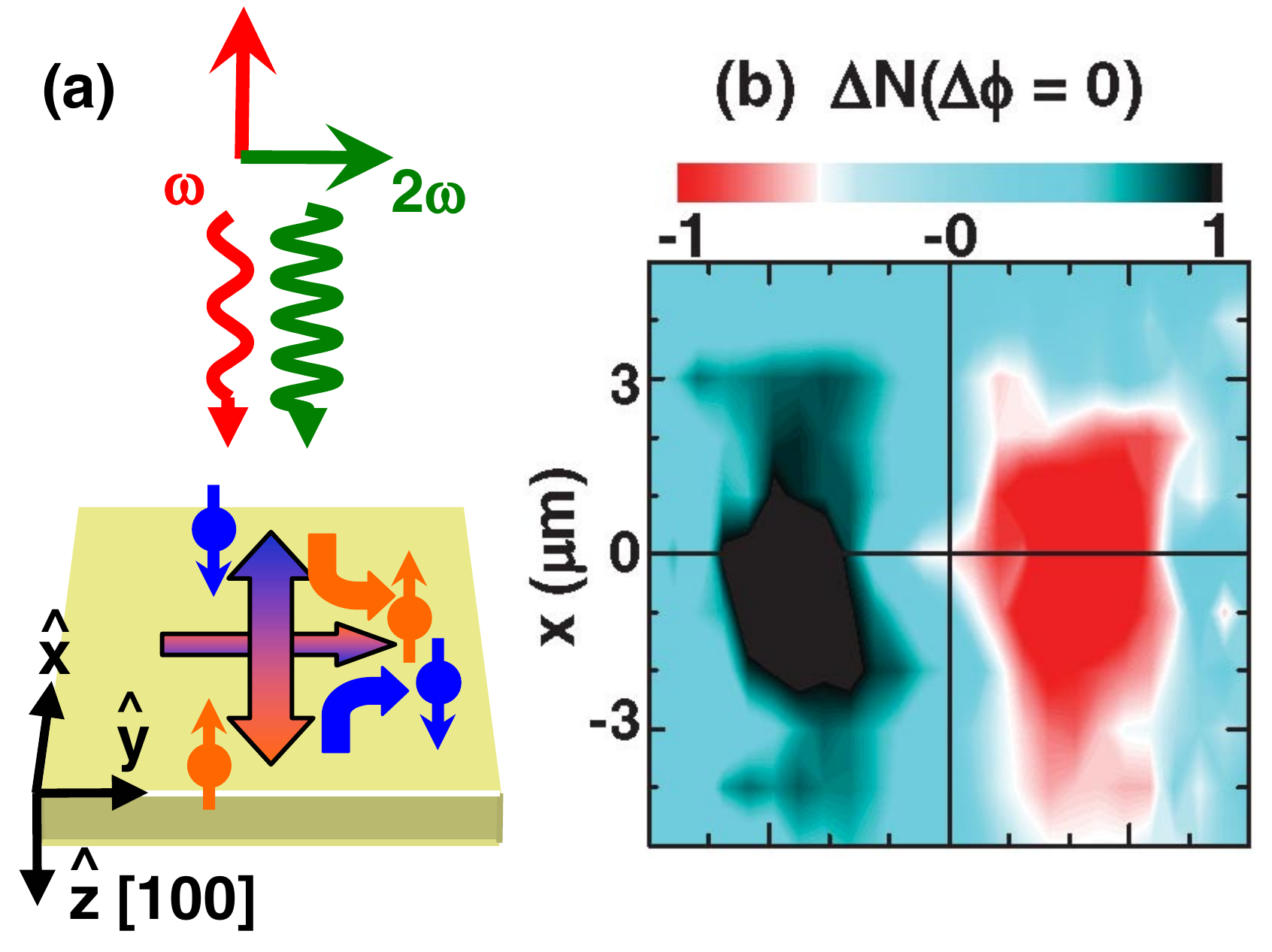}
\caption{Observation of the ISHE using the two-color optical pump-and-probe technique. (a) Illustration of orthogonally polarized $\omega$ and $2\omega$ pulses producing a pure spin-current (double headed
straight arrow) along $\omega$ beam polarization direction ($\hat{x}$). The charge current due to the ISHE (curved arrows) along $\hat{y}$ leads to electron accumulation near one edge of the illuminated region. (b) Measured charge accumulation due to the ISHE. From Ref.~\onlinecite{Zhao2006}.}
\label{fig_4_ISHE_SCs}
\end{figure}

The generation of spin and charge currents results from the quantum interference between absorption pathways for one- and two-photon absorption connecting the same initial and final states as illustrated in Fig.~\ref{fig_4_ISHE_SCs}(a). For a spin-current, a coherent pulse centered at frequency $\omega$ with phase $\phi_{\omega}$ is normally incident along $\hat{z}$ and linearly polarized along $\hat{x}$-direction that can be arbitrary with sespect to crystal axes since the effects are not strongly sensitive to crystal orientation. A co-propagating $2\omega$ pulse with phase $\phi_{2\omega}$  is linearly polarized along the orthogonal $\hat{y}$-direction. Excited spin-up electrons  are polarized along $\hat{z}$ and move preferentially in one direction along $\hat{x}$, while spin-down electrons move in the opposite direction. Together they generate a spin-current proportional to $\cos(\Delta\phi)$ where $\Delta\phi=2\phi_{\omega}-\phi_{2\omega}$. The spin-current is dominated by electrons as holes lose their spin in $<100$ fs. Due to the ISHE, a charge current is generated (Fig.~\ref{fig_4_ISHE_SCs}(a)) that has the same cosine dependence as the spin-current source. Consistent with the ISHE phenomenology, the excess charge on one side and the deficit on the other side of the sample, shown in  Fig.~\ref{fig_4_ISHE_SCs}(b), was observed along the direction perpendicular to the driving spin-current.

The ballistic nature of transport in these experiments 
was fully exploited in  fs time-resolved measurements  \cite{Werake2011}. They allowed to infer the momentum scattering time $\tau\approx 0.45$~ps and with a much shorter time delay of the probe pulses to observe in real time the transverse charge current. The measurements showed that the charge current was generated well before the first scattering event, providing a direct demonstration of the intrinsic ISHE.


\subsubsection{Electrical manipulation}
\label{opt_el_manip}

A distinct feature of the ISHE experiments in the 2DEG is the observed spin precession due to internal Rashba and Dresselhaus spin-orbit fields  \cite{Wunderlich2009,Wunderlich2010}. Since the spin diffusion length scales approximately \cite{Wunderlich2010} as $\sim L_{so}^2/w$, it was possible to observe a few spin precessions in channels of a width $w=1$~$\mu$m for $L_{so}\sim 1$~$\mu$m of the studied 2DEG. The corresponding oscillations of the spin Hall voltages were consistently observed by measuring at different Hall crosses along the channel or by shifting the laser spot, i.e., the spin injection point (Fig.~\ref{fig_5_SHE_transistor}). The lateral ISHE channels also allow to place top gate electrodes in between the Hall crosses as shown in Fig.~\ref{fig_5_SHE_transistor}(c). (The gates are formed by unetched regions of the  wafer). The strength of the Rashba and Dresselhaus spin-orbit fields and, therefore, also the spin precession can be manipulated electrically in the device shown in  Fig.~\ref{fig_5_SHE_transistor}(c). To demonstrate an AND logic functionality, two gates were fabricated on top of the channel and the Hall electrical signal was measured at a cross placed behind both gates. Intermediate gate voltages on both gates represented the input value 1 and gave the largest electrical ISHE signal, representing the output value 1. When a large reverse  gate voltage was applied to any of the two gates, representing input 0, the electrical ISHE signal disappeared, i.e., the output was 0.

A different approach to achieve the control of spin-currents is by directly modifying the spin-orbit coupling strength on a given material, which in turn determines the spin Hall angle. The electronic band structure and impurity states are weakly dependent on an external electric field and therefore cannot be used to change the spin-orbit strength. However,  \onlinecite{Okamoto2014} noted that the electric field can induce a carrier redistribution within a band or multiple bands. Therefore, if the electrons generating the SHE can be controlled by populating different areas (valleys) of the electronic structure, the spin-orbit interaction (and the spin Hall angle) can be tuned directly within a single sample. \onlinecite{Okamoto2014} reported such a tuning in bulk GaAs at room temperature by means of an electrical intervalley transition induced in the conduction band. The spin Hall angle was determined by measuring an electromotive force driven by photoexcited spin-polarized electrons drifting through $n$-GaAs Hall bars. By controlling electron populations in the $\Gamma $ and L valleys with an applied electric field (part of the $p$-character in the L valley provides a larger effective spin-orbit interaction), the angle was changed by a factor of 40, from 0.0005 to 0.02 for moderate electric fields beyond 100 kVm$^{-1}$. Thus the highest spin Hall angle achieved is comparable to that of Pt.



\subsection{Transport experiments}
\label{el_she}

\onlinecite{Hirsch1999} and \onlinecite{Zhang2000}, discussed specific concepts for the experimental detection of the SHE and ISHE using DC transport techniques. \onlinecite{Hirsch1999} proposed a device that consists of a metallic slab in which spin accumulation is generated by an electrical current via the SHE, as described in Sec.~\ref{over} (see Fig.~\ref{fig_exp_MHH}). A transverse strip connects the edges of the slab, allowing the spin-current to flow through it. Due to the ISHE, a voltage is generated that can be measured with a voltmeter. In an alternative approach, \onlinecite{Zhang2000} proposed to detect the spin accumulation electrically using a FM probe. The concept borrows from  techniques for spin injection and detection in NM implemented in nonlocal spin devices \cite{Silsbee1980,Johnson1985}.

Shortly after the optical SHE detection in semiconductors \cite{Kato2004,Wunderlich2004,Wunderlich2005}, \onlinecite{Valenzuela2006} reported an observation of the voltage generated by the ISHE. Instead of generating the spin-current by the SHE, which would render a second order voltage in the spin Hall angle, they used electrical spin injection from a FM in combination with a Hall cross patterned in the ISHE paramagnet. Simultaneously, \onlinecite{Saitoh2006} observed the voltage generated by the ISHE in a set-up where the spin-injection from the FM to the NM was achieved using the  SP techniques.  \onlinecite{Kimura2007} combined the concept of the spin Hall cross and the proposal by \onlinecite{Zhang2000} to detect both the SHE and the ISHE in the same device. It took a few more years to demonstrate the idea of Hirsch of simultaneously exploiting both the SHE and the ISHE in an electrical device. Eventually, \onlinecite{Brune2010} performed the experiment in a ballistic H-bar semiconductor device \cite{Hankiewicz2004b}.  The transport SHE and ISHE experiments are described in detail below in Sec.~\ref{js}-\ref{nometals_sec}.

More recently, the SHE was also detected via the manipulation of magnetization in FMs \cite{Miron2011,Liu2011}. Spin-currents generated by the SHE were shown  to be sufficiently large to induce magnetization dynamics, drive domain walls, or switch magnetization in the FM, demonstrating the potential of the SHE for applications \cite{Miron2011,Emori2013,Ryu2013,Miron2011b,Liu2012}. These SHE experiments together with the ISHE measurements via SP  are discussed in Sec. \ref{she-mag}.

\subsubsection{Concepts of nonlocal spin transport. Electrical injection and detection}
\label{js}

\begin{figure}[t]	
  \centerline{\includegraphics[width=8.5cm]
  {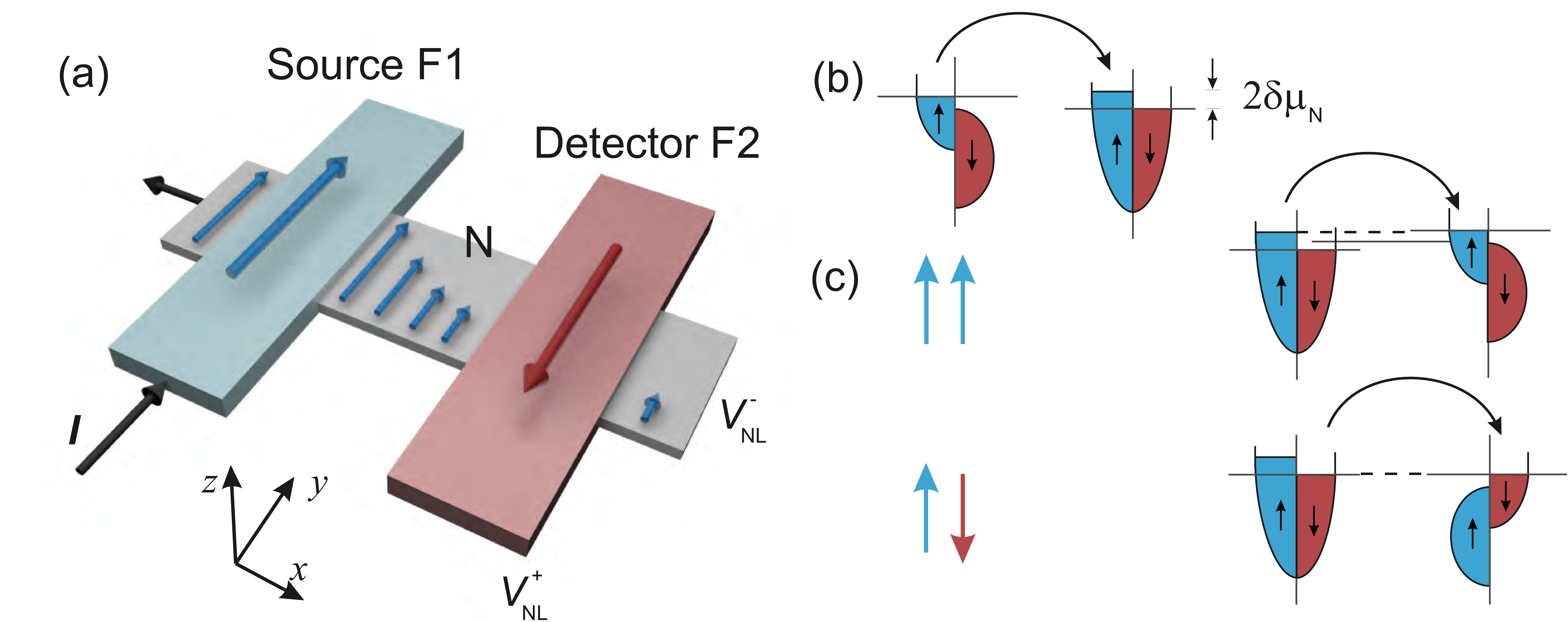}}
  \caption{
(a) Nonlocal spin detection and spin accumulation (a) Schematic illustrations of the device layout. An injected current $I$ on the source (F1) generates spin accumulation in the NM (N) which is quantified by the detector (F2) voltage $V_{\rm NL}$. (b) Schematic representation of the spin splitting in the electrochemical potential induced by spin injection. The splitting decays over characteristic lengths  $\lambda_{sf}$  over the N side. (c) Detector behavior for an idealized Stoner FM with a full spin subband for the parallel magnetization orientation (top) and for the antiparallel magnetization orientation (bottom).
   }	
 \label{FigNL1}		
\end{figure}

\onlinecite{Johnson1985} reported the injection and detection of nonequilibrium spins using a device that consisted of a NM, N, with two attached FM  electrodes (F1, F2), illusrated in Fig.~\ref{FigNL1}. In this device, spin-polarized electrons are injected from F1 into N by applying a current $I$ from F1 that results in spin accumulation in N. The population of, say, spin-up electrons in N increases by shifting the electrochemical potential by $\delta\mu N$, while the population of spin-down electrons decreases by a similar shift of $-\delta\mu N$. Overall, this corresponds to a spin-accumulation splitting of $2\delta\mu N$. The spin accumulation diffuses away from the injection point and reaches the F2 detector, which measures its local magnitude.

As first suggested by \onlinecite{Silsbee1980}, the spin accumulation in N can be probed by the voltage $V_{\rm NL}$, which is induced at F2. Silsbee noted that the density polarization in N, or equivalently the nonequilibrium magnetization, acts as the source of spin electromotive force that produces $V_{\rm NL}$. The magnitude of $V_{\rm NL}$ is associated with $\delta\mu N$, while its sign is determined by the relative magnetization orientation of F1 and F2.

Because the current is applied to the left on N, there is no charge current towards the right, where the detector F2 lies (Fig. \ref{FigNL1}(a)). For this reason, the spin detection is said to be implemented nonlocally, where no charge current circulates by the detection point, and thus $V_{\rm NL}$ is sensitive to the spin degree-of-freedom only. Accordingly, nonlocal measurements eliminate the presence of spurious effects associated to charge transport, such as anisotropic magnetoresistance (AMR) or the ordinary HE that could mask subtle signals related to spin injection. Typically, nonlocal devices exhibit a small output background allowing sensitive spin-detection experiments. This approach has been widely used in recent years to characterize the spin transport in metals, semimetals, semiconductors, superconductors, carbon nanotubes and graphene. It has also been used to study the spin transfer properties of FM/NM material interfaces.

\subsubsection{Nonlocal detection of inverse spin Hall effect with lateral spin-current}
\label{nonlocal_sec}

\begin{figure}[h]	
  \centerline{\includegraphics[width=8cm]
  {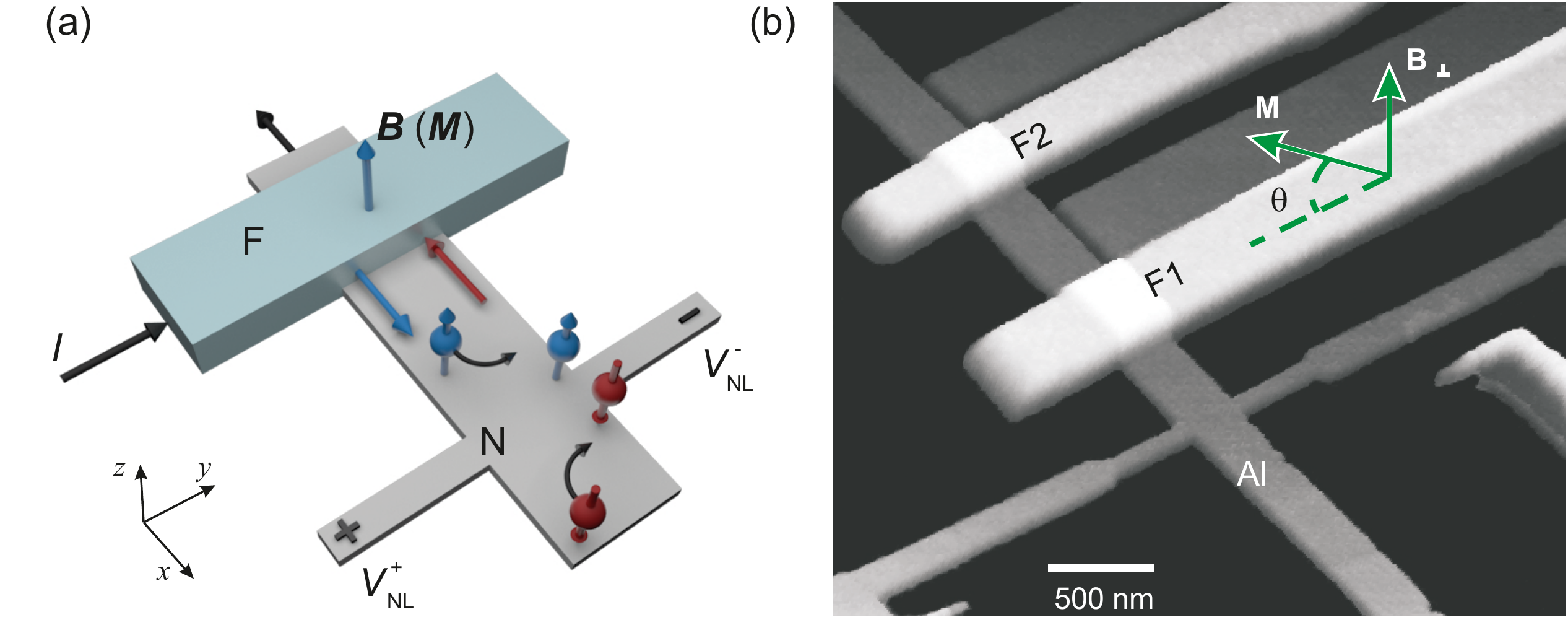}}
  \caption{(a)
Spin-current induced Hall effect or inverse spin Hall effect (ISHE). Schematic representation of an actual device where the pure spin-current is generated by spin injection through a FM (F) with out-of-plane magnetization.
(a) Device fabricated with CoFe electrodes (light color) and an Al channel (dark color). Adapted from Ref. \onlinecite{Valenzuela2006}.
 \label{FigNL2}
 }		
\end{figure}

\onlinecite{Valenzuela2006,Valenzuela2007} adapted the nonlocal detection techniques to study the ISHE. Their device is schematically shown in Fig.~\ref{FigNL2}(a). By using a FM electrode F, a spin-polarized current is injected in a nonmagnetic strip. It propagates to both sides away from the injection point and decays with the spin diffusion length, $\lambda_{sd}$. A laterally induced voltage $V_{SH}$, which results from the conversion of the injected spin-current into charge imbalance owing to the ISHE, is then measured using a Hall-cross structure. The magnitude of $V_{SH}$ is determined by the anomalous Hall operator, $\sigma_{SH} ~\mathbf{\hat{\sigma}} \times \mathbf{E}^{\sigma}$, where $\sigma_{SH}$ denotes the spin Hall conductivity, $\sigma$ is the spin index, and $\mathbf{E}^{\sigma}$ is an effective spin-dependent ``electric" field, which follows from the spin-dependent electrochemical potential $\mu^{\sigma}$ along the NM al strip, i.e. $\mathbf{E}^{\sigma}(\mathbf{r})=-\nabla \mu^{\sigma}(\mathbf{r})$.

In the device of Fig.~\ref{FigNL2}(b), the injector electrode F1 is made of CoFe, while the strip material is Al with thickness $t_{Al}$. The full device is fabricated without breaking vacuum using electron beam evaporation and shadow evaporation techniques. An Al$_2$O$_3$ tunnel barrier is used for spin-current injection. The purpose of the barrier is two-fold. First, it enhances the polarization of the injected electrons and, second, it assures a uniform current injection. The latter is essential because it suppresses the flow of charge current towards the Hall cross, preserving the nonlocal character of the measurements and eliminating the previously mentioned spurious effects.

The FM electrode is magnetized in-plane at zero magnetic field due to shape anisotropy and thus an out-of-plane magnetic field $B_{\perp}$ is used to generate a perpendicularly polarized spin-current at the Hall-cross (Fig.~\ref{FigNL2}(b)). Spin imbalance in the Al film occurs with a defined spin direction given by the magnetization orientation of the F1 electrode. Consequently, $V_{SH}$ is expected to vary when $B_{\perp}$ is applied and the magnetization $\mathbf{M}$ of the electrode is tilted out of the substrate plane. Defining $\theta$ as the angle between $\mathbf{M}$ and the electrode axis, it follows from the cross product in the anomalous Hall operator that $V_{SH}$ is proportional to $\sin \theta$, correlating with the component of $\mathbf{M}$ normal to the substrate (Fig.~\ref{FigNL2}(b)).

\begin{figure}[h!]
\hspace*{-0.5cm}\epsfig{width=1\columnwidth,angle=0,file=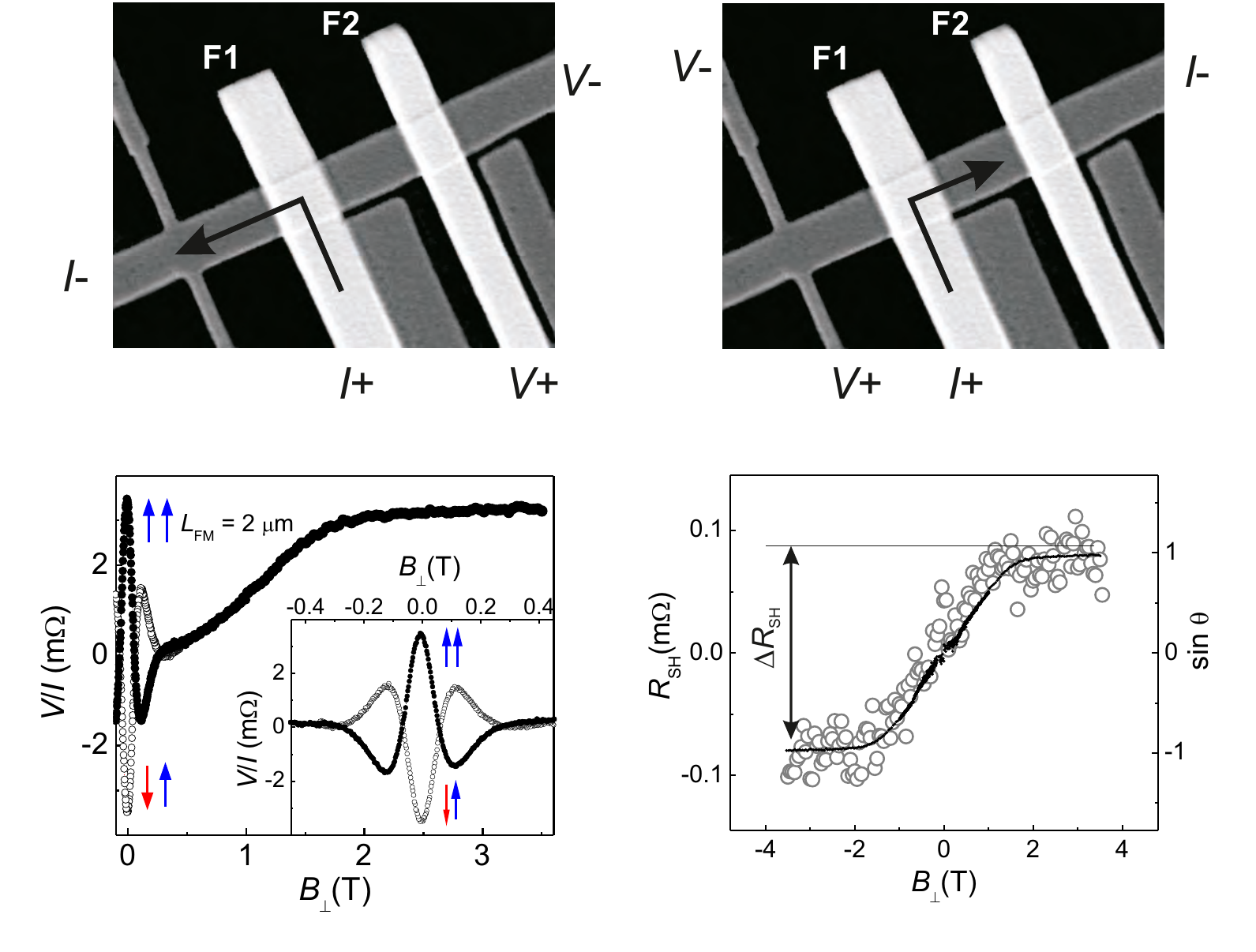}
\caption{Observation of the ISHE (right) in a metal device with an electrical spin injection from a FM, compared with the spin detection by the non-local spin valve effect (left). The light color FM electrodes in the micrographs are made of a CoFe alloy. The dark color Hall cross is made of Al. From Ref.~\onlinecite{Valenzuela2006}.}
\label{FigNL3}
\end{figure}

The device layout in Fig.~\ref{FigNL2}(b) is more sophisticated than the schematics in Fig.~\ref{FigNL2}(a), where only F1 is required. The second electrode (F2) together with F1 and the N strip form a spin injection/detection device (Fig.~\ref{FigNL3}(a)) for the purpose of calibration. Calibration procedures are necessary to demonstrate consistency with standard nonlocal methods. Explicitly, this device can be utilized to measure the spin accumulation in the NM and then determine its associated spin diffusion length $\lambda_{sd}$, the spin polarization of the injected electrons $P$, and the magnetization orientation of the FM electrodes $\theta$ in the presence of an external magnetic field (perpendicular to the substrate). For this purpose, batches of samples are commonly used where the distance between the two FMs, $L_{F}$, is modified and the spin precession signal acquired (Fig.~\ref{FigNL3}(b)). The distance of F1 relative to the Hall cross, $L_{SH}$, is also modified in order to test the consistency of the spin diffusion results. Subsequent measurements in the configuration of Fig.~\ref{FigNL3}(c), performed in Al of different $t_{Al}$, and thus different $\lambda_{sd}$, yielded $\sigma_{SH} \sim 20-40$ ($\Omega$cm$)^{-1}$ and $\alpha_{SH} \sim 1-3 \times 10^{-4}$, which compares well with theoretical estimates based on extrinsic mechanisms \cite{Shchelushkin2005}.

\onlinecite{Olejnik2012} used the same geometry to detect the ISHE in $n$-GaAs using epitaxial ultrathin-Fe/GaAs injection contacts with strong in-plane magnetic anisotropy. Hybrid semiconductor/metal-FM structures suffered for a long time from the resistance mismatch problem \cite{Schmidt2000}. Since the spin transport relies on different conductivities for spin-up and spin-down electrons and is governed by the least conductive part of the device, the effects are weak in devices in which the non-magnetic semiconductor with equal spin-up and spin-down conductivities dominates the resistance of the device \cite{Rashba2000}. The introduction of a highly resistive tunnel barrier between the FM metal electrode and the semiconductor channel solved the problem \cite{Rashba2000,Lou2007}.

\begin{figure}[h]	
  \centerline{\includegraphics[width=1\columnwidth]
  {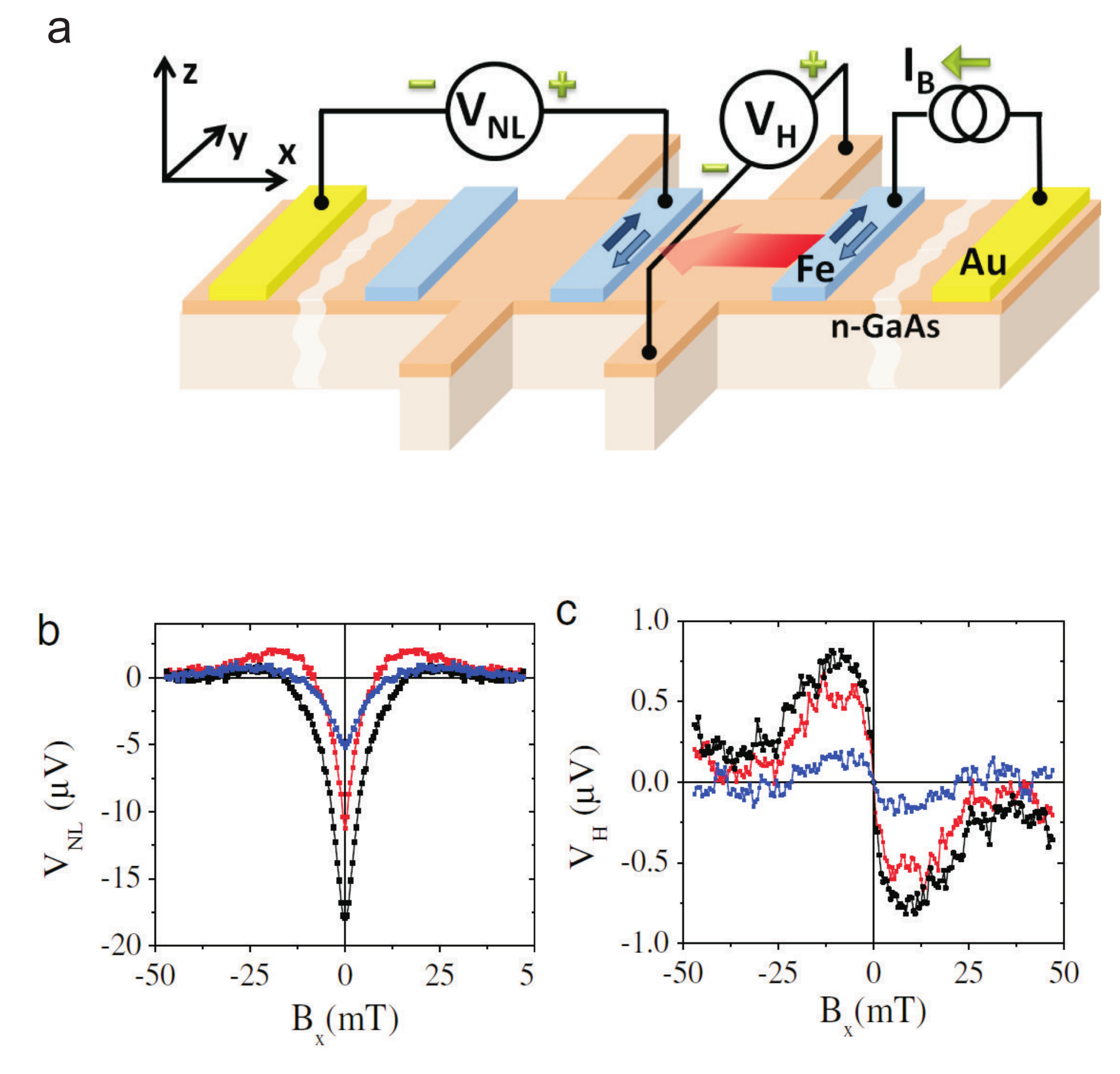}}
  \caption{(a) Schematic of the device used to detect the ISHE in $n$-GaAs. Current is injected on the Fe-electrode on the right, the voltage generated by ISHE and by spin accumulation are detected simultaneously with the Hall cross and the Fe-electrode on the left, respectively. The spin transport can be further modified by a drift current applied between the outermost Au electrodes. (b) and (c) show the experimental symmetrized nonlocal spin injection/detection signal and the antisymmetrized ISHE signal in the in-plane hard-axis field for constant spin-injection bias current (300 $\mu$A) and for three different drift currents. From Ref.~\onlinecite{Olejnik2012}}
 \label{FigNL4}		
\end{figure}

The device of \onlinecite{Olejnik2012}, shown in  Fig.~\ref{FigNL4}(a), comprised the $n$-GaAs channel, a Hall-cross, and two Fe electrodes as in Fig.~\ref{FigNL2}(a)  with  Fe Schottky injection contact. The Fe/$n$-GaAs heterostructure was grown epitaxially in a single molecular beam epitaxy chamber without breaking ultrahigh vacuum. The heterostructure contained 250 nm of low Si-doped GaAs ($5 \times 10^{16}$ cm$^{-3}$), 15 nm of GaAs with graded doping, and 15 nm of highly Si-doped GaAs ($5 \times 10^{18}$ cm$^{-3}$). The purpose of the doping profile was to create a narrow tunnel Schottky barrier between GaAs and Fe favorable for spin injection or detection. It was then possible to simultaneously detect the spin-current in $n$-GaAs generated by nonlocal injection from a Fe contact by using the ISHE and the spin accumulation by using the additional Fe contact (Figs.~\ref{FigNL4}(b) and (c)). The spins were manipulated by spin precession with an external magnetic field combined with drift using an external bias \cite{Huang2007b}. In this case, the magnetic field was applied in-plane ($x$-direction) to precess the spin accumulation into the out-of-plane direction, so that it could be detected by the ISHE. The signal first increases at low fields but then is suppressed due to spin dephasing (Fig.~\ref{FigNL4}(c)).

Devices described above required the application of a magnetic field for observing the ISHE. \onlinecite{Seki2008} used a FM (FePt) with an out-of-plane anisotropy, which enabled them to measure the ISHE in Au without magnetic fields. The device was fabricated with the geometry in Fig.~\ref{FigNL2} using ohmic contacts. The measurements presented a rather large background voltage, which is likely due to the flow of charge current at the position of the Hall cross \cite{Mihajlovic2009}. The use of ohmic contacts, as opposed to tunnel barriers, results in inhomogeneous in-plane current injection. Because the width of the Au wire and the distance of the Hall cross were comparable, some current reached the Hall cross contributing to the background. By considering that the voltage was independent of the magnetization of the injector electrodes, Seki {\em et al.} deduced $\alpha_{SH} = 0.113$ for Au at 295 K, which was weakly dependent on temperature. This large $\alpha_{SH}$ was first attributed to resonant scattering in the orbital-dependent Kondo effect of Fe impurities in the Au host metal \cite{Guo2009}. In the follow-up work, \onlinecite{sugai2010} found that $\alpha_{SH}\sim 0.07$ was approximately independent of the Fe concentration.  \onlinecite{Seki2010} further observed a reduction of $\alpha_{SH}$ from 0.1 to 0.03 when the thickness was decreased from 10 nm to 20 nm in Au. Additionally, \onlinecite{gu2010} obtained similar results in Pt-doped Au by co-deposition of Pt and Au with magnetron sputtering (1.4\% Pt). These results in combination with \textit{ab initio} and quantum Monte Carlo calculations for the skew scattering due to a Pt impurity led to the proposal of a much larger $\alpha_{SH}$ in the surface of Au than in the bulk \cite{gu2010}.

\subsubsection{Nonlocal detection of spin Hall effects with vertical spin-current}
\label{nonlocal_abs_sec}

\begin{figure}[t]	
  \centerline{\includegraphics[width=1\columnwidth]
  {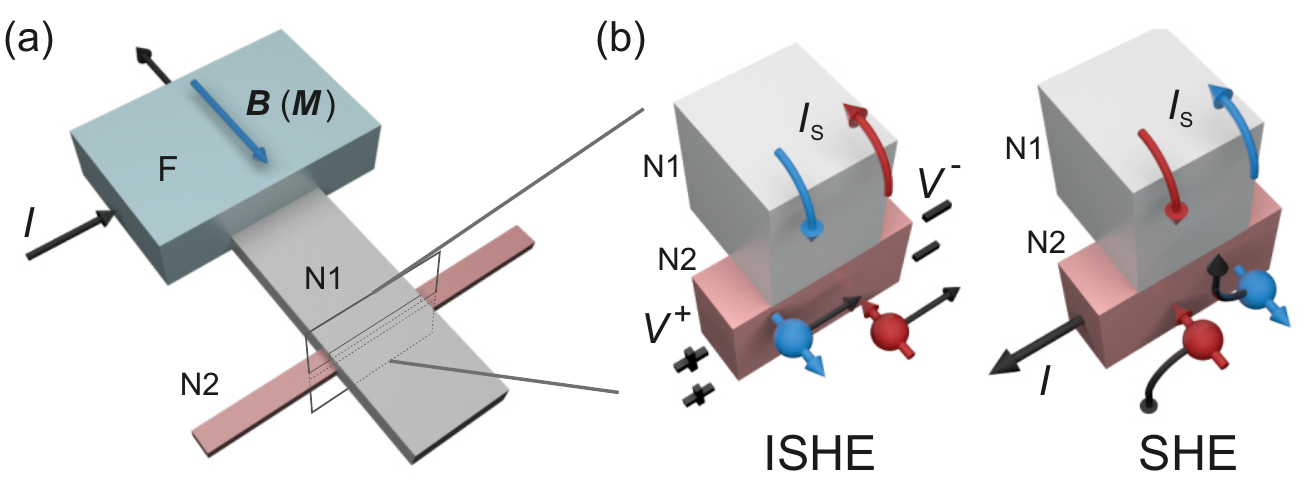}}
  \caption{
(a) Schematic illustration of a nonlocal device to measure the direct and inverse spin Hall effect in materials (N2) with short spin relaxation length $\lambda_{sd}^{N2}$.
(b) Schematic illustration of the charge accumulation process in N2 (left) due to the ISHE when a spin-current is injected from F as in (a). Schematic illustration of the charge to spin-current
conversion due to the SHE when a current is applied to N2. This process generates spin accumulation that is detected by measuring the voltage at which F floats. See \cite{Kimura2007}. Adapted from Ref.~\onlinecite{Valenzuela2012}.
   }	
 \label{FigNL5}		
\end{figure}

The approach described in the previous section enables proper quantification of the spin Hall angle because of the direct measurements of the spin diffusion length. However, it is suitable for materials that have spin diffusion lengths beyond tens of nanometers. For smaller spin diffusion lengths,  \onlinecite{Kimura2007} modified this approach using the device structure shown in Fig.~\ref{FigNL5}(a). The structure comprises a Hall cross where the material of the transverse arm is the large spin-orbit coupling NM N2 with short $\lambda_{sd}$, which acts as a spin-current absorber that induces $V_{SH}$ via the ISHE. The longitudinal arm, on the other hand, is made of a NM N1 with long spin diffusion length that fulfils the purpose of transporting spin information between the FM electrode (F) and N2.

The way the measurements are performed is sketched in Fig.~\ref{FigNL5}(b) (left). A charge current is injected from F into N1 that induces a  spin-current towards N2 polarized in-plane in the direction parallel to the N1 arm. When the distance between F and the cross is smaller than the spin diffusion length in N1, the spin-current is preferably absorbed into the transverse arm N2 because of the strong spin relaxation in N2. The injected vertical spin-current into N2 vanishes in a short distance from the N1/N2 interface because of the short spin diffusion length of N2 and generates a transverse voltage via the ISHE.

This device can be also used to measure the SHE. The bias configuration is modified as shown in Fig.~\ref{FigNL5}(b) (right). Here, N2 acts as a spin-current source, which induces a spin accumulation in N1 that is detected with the FM electrode F, as originally proposed by \onlinecite{Zhang2000}.

\begin{figure}[t]	
  \centerline{\includegraphics[width=1\columnwidth]
  {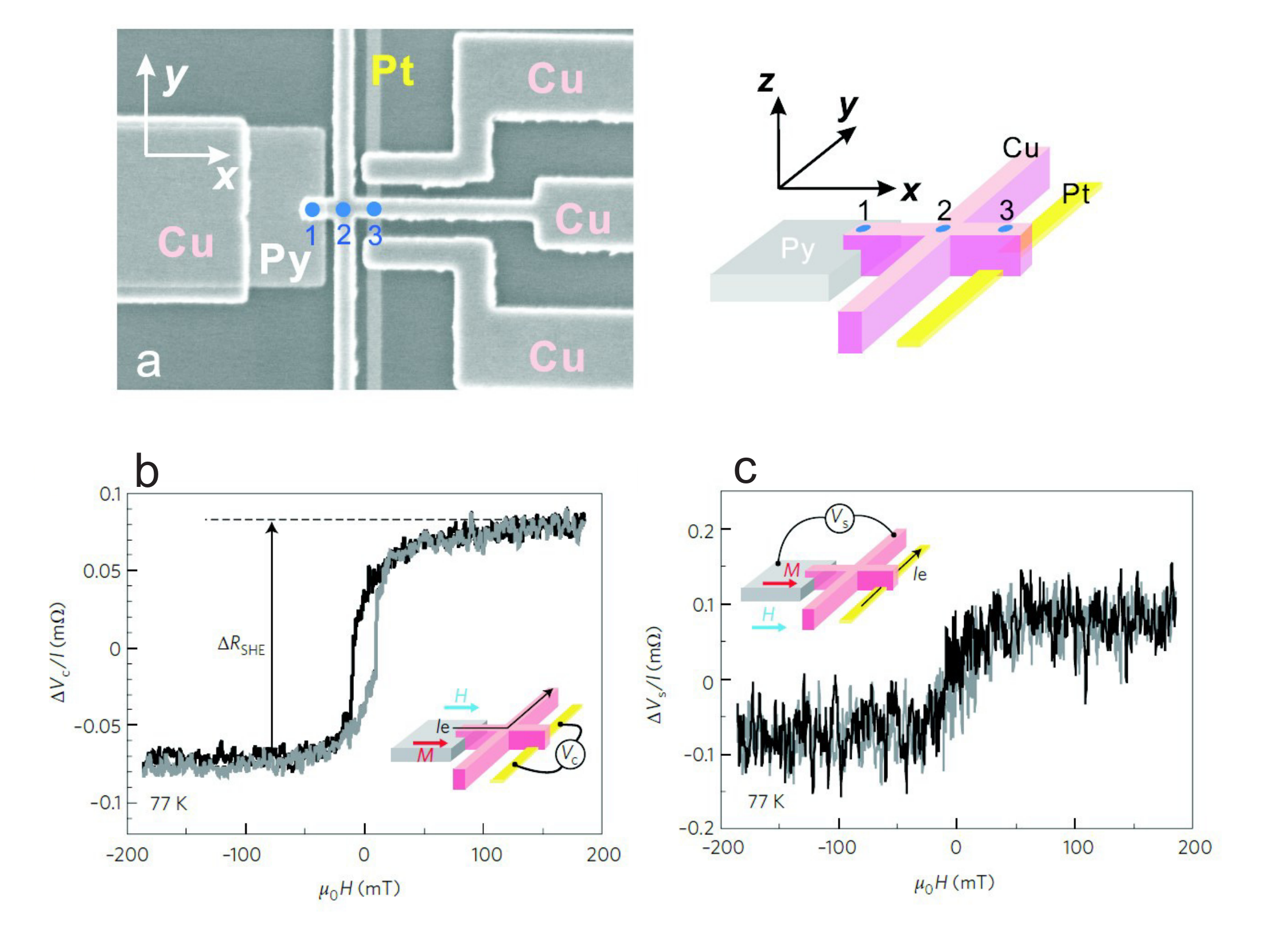}}
  \caption{
(a) Scanning electron microscope (SEM) image of the fabricated spin Hall device to measure the SHE in Pt together with a schematic illustration of the fabricated device.
(b) Signal due to de ISHE at 77 K. The black and grey curves show measurements for the two opposite sweeps of the magnetic field. Spin-accumulation signal generated by SHE at 77 K. Insets: measurement set-up. NiFe, Cu and Pt are in grey, pink and yellow, respectively. From Ref.~\onlinecite{Kimura2007}.
   }	
 \label{FigNL6}		
\end{figure}

\onlinecite{Kimura2007} used permalloy (Py) as the FM source, and Cu and Pt as N1 and N2, respectively (see Fig.~\ref{FigNL6}). The materials were deposited by electron-beam evaporation. The devices were fabricated with transparent interfaces between Py and Cu and between Pt and Cu. Ar ion beam etching was done prior to depositing Cu in order to clean the surfaces of Py and Pt, a method that has been repeated in the other studies described below. The long spin diffusion length of Cu (about 500 nm) assured that the spin-current reached Pt, which was 4 nm thick. The measurements were interpreted with a one-dimensional model by assuming that the induced spin-current at the Cu/Pt interface was completely absorbed by the Pt. The spin relaxation length for Pt was assumed (not measured) to be $\lambda_{sd} = 3$~nm. \onlinecite{Kimura2007} then obtained that $\sigma_{SH} \sim 2.4 \times 10^2$ ($\Omega$cm$)^{-1}$ and $\alpha_{SH} = 3.7 \times 10^{-3}$.

\begin{figure}[t]	
  \centerline{\includegraphics[width=1\columnwidth]
  {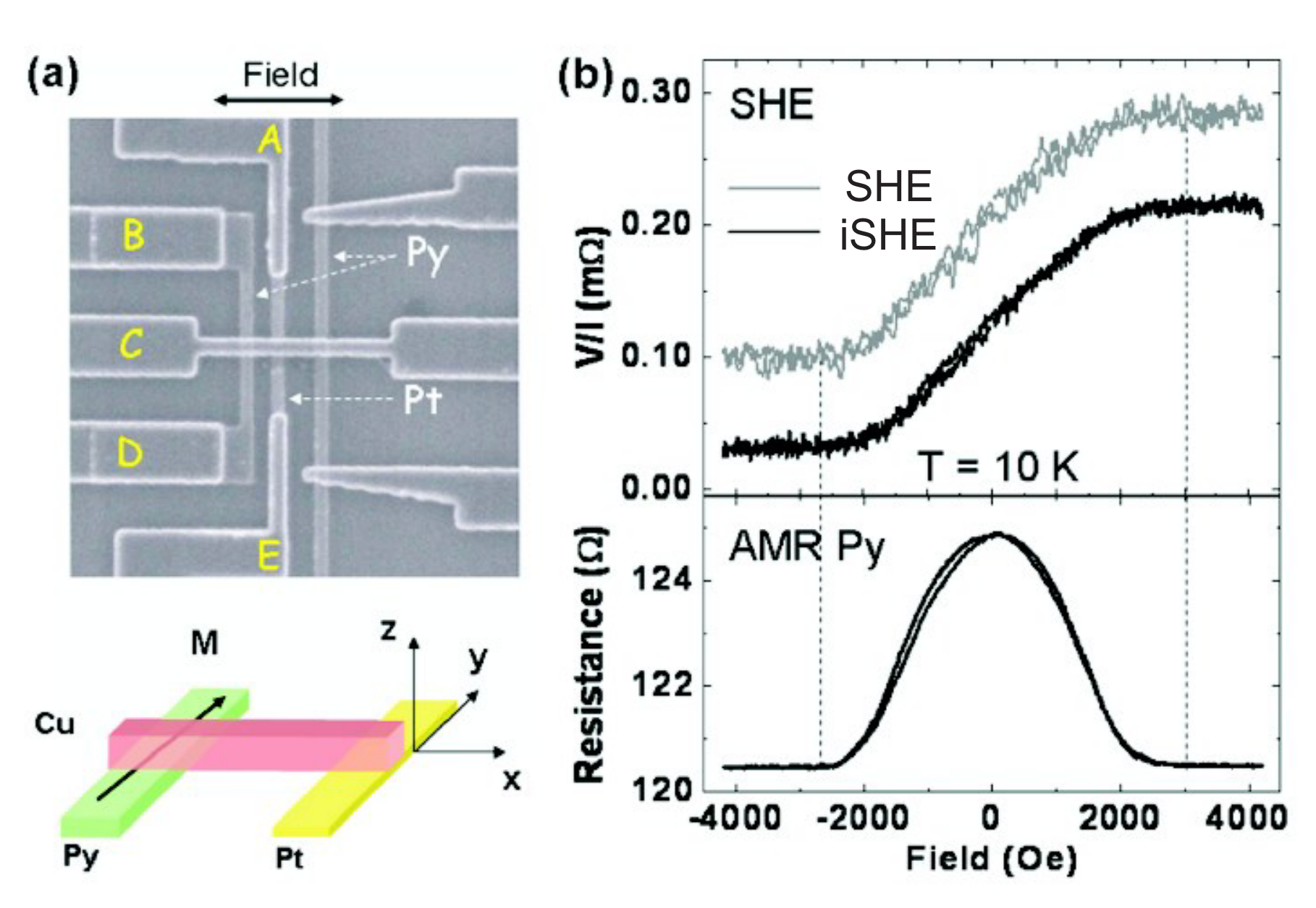}}
  \caption{
(a) SEM image of the typical device for SHE measurements and an illustration of the device. (b) Direct and
inverse SHE (SHE and ISHE) recorded at $T=10$ K using a device with a Pt-thicness of 20 nm, altogether with the AMR from the Py wire measured on the same condition. SHE measurement corresponds to $V_{BC}/I_{AE}$, and ISHE to $V_{EA}/I_{BC}$; with $V$ the voltage, $I$ the applied
current; A, B, C and E are the contact leads as denoted in the SEM image. From Ref.~\onlinecite{Vila2007}.
   }	
 \label{FigNL7}		
\end{figure}

Over the last few years some of the initial simplifications that are mentioned above have been removed, leading to more reliable quantitative interpretations of the experimental results. \onlinecite{Vila2007} noted that the absorption efficiency of the spin-current may depend on the device geometry and temperature. They modified the design of Fig.~\ref{FigNL6} to a conventional nonlocal spin injection/detection structure where a Pt electrode was inserted between the FM Py electrodes (see Fig.~\ref{FigNL7}). This change enabled them to determine explicitly the magnitude of the absorbed spin-current. By comparing with reference devices without the Pt insertion, they observed that the ratio between the spin signal with and without Pt varied from 0.35 at 5 K to 0.2 at room temperature, irrespective of the Pt thickness. They then performed systematic spin absorption studies as a function of the Pt thickness, obtaining that $\lambda_{sd}$ for Pt was 10 nm and 14 nm at room temperature and at 5 K, respectively.

The Pt thickness dependence of the ISHE signal resulted in somewhat lower $\lambda_{sd}$ for Pt of 7 nm and 8 nm at room temperature and at 5 K, respectively, still more than a factor of 2 larger than previously assumed. The obtained value of $\sigma_{SH} \sim 3.5 \times 10^2$ $\Omega$cm$^{-1}$ was larger than that in the \onlinecite{Kimura2007} experiment; this is because the assumption of the complete spin-current absorption into the Pt wire led to underestimating the spin Hall conductivity.

Additionally, \onlinecite{Vila2007} found that the spin Hall conductivity was nearly constant as a function of temperature indicating that the spin Hall resistivity likely evolves in a quadratic form with the Pt resistivity in the analyzed temperature range, which was initially associated to a side jump origin of the SHE. However, this resistivity dependence can also be associated with the intrinsic mechanism \cite{Tanaka2008,Kontani2009}.

\onlinecite{Niimi2011} further included a correction factor $0< x <1$ that accounted for the fact that the transverse charge current induced by the ISHE is partially shunted by the wire N1 above the N1/N2 interface or, conversely, that the charge current that induces the spin-current via the SHE does not only flow through N2 but also leaks into N1 (see also \cite{Liu2011}). In order to determine $x$ experimentally, they measured the voltage drop of two identical N2 nanowires with and without shunting N1 bridges. Within a one-dimensional circuit model, the current flowing into the N2 wire $I_0$ was assumed to divide into two components at the N1/N2 interface: $xI_0$ for the N2 wire and $(1-x)I_0$ for the N1 bridge. With this, $x$ was estimated to be $0.36 \pm 0.08$ for Cu (N1), when using a number of transition metals and alloys as N2 \cite{Niimi2011,Morota2011}, therefore appearing to be rather insensitive to the resistivity of N2. Because of this correction, former reports underestimated $\sigma_{SH}$ by a factor $x^{-1} \sim 2.8$. Such large correction is to be expected given that the N1 wire (usually Cu or Ag) is highly conductive (conductivity $\sim 3-5 \times 10^7$ ($\Omega$m)$^{-1}$) and thick ($\sim 100$ nm), when comparing with N2 ($\sim 10^5-10^7$ ($\Omega$m)$^{-1}$ and $\sim 10$ nm).

In addition, \onlinecite{Niimi2011} and \onlinecite{Morota2011}  pointed out that the spin-currents injected in N2 should dilute when its thickness $t_{N2}$ is larger than the spin diffusion length in N2 leading to smaller spin Hall signals. To correct for this effect, they obtained an aggregate spin-current in N2 by integrating over $t_{N2}$, which was then divided by $t_{N2}$; they also forced the spin-current to be zero at the bottom surface of N2.

\onlinecite{Niimi2011} reported  $\alpha_{SH} = 0.021\pm0.006$ for the skew scattering off Ir in a Cu matrix, which is consistent with experimental work relying on spin polarized currents generated by dilute Mn impurities, for which $\alpha_{SH} = 0.026$ \cite{Fert1981,Fert2011}. The spin Hall angle was extracted with CuIr wires that were prepared with different Ir concentrations (0\%, 1\%, 3\%, 6\%, 9\%, and 12\%) using magnetron sputtering. They measured $\rho^H$ of CuIr as a function of the resistivity induced by the Ir impurities, defined as $\rho_{imp}=\rho_{CuIr}-\rho_{Cu}$, finding a simple linear dependence up to Ir concentration of 12\%. This is presented as a proof that the dominant mechanism of the extrinsic SHE induced by the Ir impurities is the skew scattering, with $\alpha_{SH} = \rho^H/\rho_{imp}$.

\onlinecite{Morota2011} investigated the ISHE and SHE in 4d and 5d transition metals, Nb, Ta, Mo, Pd, and Pt. Nb, Ta, and Mo wires were deposited by magnetron sputtering while Pd and Pt wires were grown by electron-beam evaporation. In particular, for Pt, they obtained a spin Hall angle $\sigma_{SH}= 0.021\pm0.005$ that was roughly 6$\times$ larger than that in \onlinecite{Kimura2007}. Such a difference can be explained with the above corrections. They also found that the sign of the spin Hall conductivity changes systematically depending on the number of $d$ electrons, a tendency that is in good agreement with theoretical calculations based on the intrinsic SHE \cite{Kontani2009}.

More recently, \onlinecite{Niimi2012} studied the ISHE and SHE by introducing a small amount of Bi impurities in Cu. The alloy Cu$_{1-x}$Bi$_{x}$ were deposited by magnetron sputtering from Bi-sintered Cu targets with different Bi concentrations (0\%, 0.3\%, 0.5\%, 1\%, 3\%, and 6\%). The spin Hall resistivity was derived by 1D and 3D calculations as a function of the resistivity induced by the Bi impurities. As for the case with Ir impurities, the experimental results follow the linear variation of the spin Hall resistivity characteristic of skew scattering by dilute impurities but only at the lowest concentrations ($< 1$\%). At larger concentrations, inhomogeneous distribution on Bi results in the departure from the dilute impurity regime. From the slope $\rho^H/\rho_{imp}$ in the linear regime, $\alpha_{SH}$ was estimated with the standard 1D analysis above, and with more accurate 3D calculations, resulting in $\alpha_{SH}(\mathrm{1D})=-(0.12\pm0.04)$ and $\alpha_{SH}(\mathrm{3D})=-(0.24\pm0.09)$ at 10 K.

The 3D calculations yield a larger $\alpha_{SH}$ because spin accumulation is observed to spread at the side edges of the CuBi/Cu junction, which is not taken into account in the 1D model. For the calculations with the 1D model, the spin-current is considered to flow vertically into the CuBi wire, therefore, they cannot take into account the spin escape by lateral spreading. In general, the correction is observed to become important when the spin diffusion length in N2 is longer than $t_{N2}$. For the cases of CuIr or Pt, it produces a small additional error because the spin diffusion length in N2  is usually shorter than $t_{N2}$. For Pt, $\alpha_{SH}$ was estimated to increase from 0.021 (1D model) to 0.024 (3D model).

Nonlocal methods have been used to estimate spin Hall angles in a number of other materials, including IrO$_2$ \cite{Fujiwara2013a}, and Bi \cite{Fan2008}; it was also applied to determine the sign of the spin injection polarization of FMs by using materials with a well established spin Hall angle, which is not possible with standard nonlocal spin injection and detection methods using the same FM material for the two electrodes. This procedure was demonstrated for the Heusler alloy Co$_2$FeSi \cite{Oki2012}. The ISHE in nonlocal geometries was also used as a probe of spin fluctuations in weak FM NiPd alloys \cite{Wei2012}. An anomaly near the Curie temperature was explained by the fluctuation contributions to skew scattering via spin-orbit interactions; the total magnetic moment involved in the experiment was extremely small (less than $10^{-14}$ emu), highlighting the very high sensitivity of the technique.

\subsubsection{Direct detection of the spin Hall induced spin accumulation}
\label{direct_sec}

As discussed in Sec.~\ref{mag}, \onlinecite{Zhang2000} proposed to detect the spin accumulation induced by the SHE via a FM probe directly attached in the side of a thin conductor. The magnetization of the FM points to the direction perpendicular to the plane of the film. The method is based on measuring the voltage at which the FM floats depending on the direction of its magnetization, which gives direct information of the spin accumulation at the edge of the conductor (see section \ref{js}). The implementation of the method took several years because of the local currents that circulate nearby the FM, which result in spurious signals that are avoided by the nonlocal methods, as described above.

\begin{figure}[h]	
  \centerline{\includegraphics[width=1\columnwidth]
  {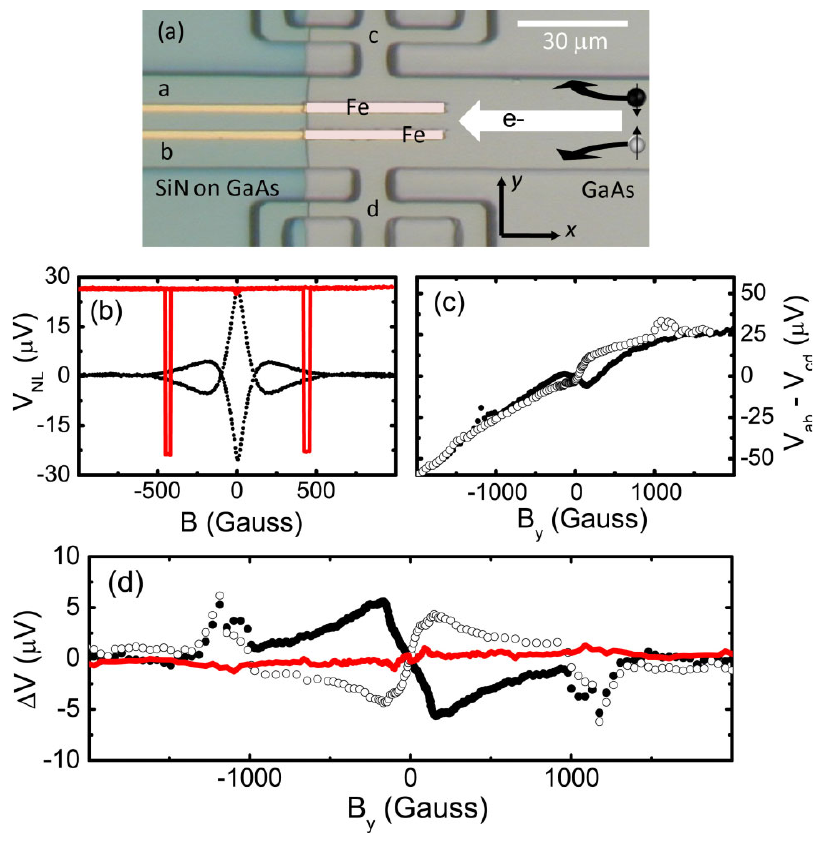}}
  \caption{
(a) Micrograph of a spin Hall device with Fe contacts located 10 $\mu$m from the edges of the GaAs channel. The contact pairs {\em ab} and {\em cd} are used to measure the spin accumulation. (b) Nonlocal spin valve (red lines) and Hanle effect (black dots) data obtained on a GaAs device at T = 60~K for injection current $8.2\times10^2$~A/cm$^2$. (c) Measured voltage $V_{ab}-V_{cd}$ for a GaAs device with Fe contacts 2~$\mu$m from the edges at T = 30~K for a channel current $5.7\times10^3$~A/cm$^2$. An offset voltage of 13.2~mV has been subtracted from the data. In (b) and (c) data is shown for both parallel and antiparallel states of injector and detector. (d) spin Hall signal for both positive (full circle) and negative (open circle) currents, after removing background and extracting antisymmetric signal. The spin Hall signal in the antiparallel state is shown as the solid red line. From Ref.~\onlinecite{Garlid2010}.
   }	
 \label{FigNL8}		
\end{figure}

\onlinecite{Garlid2010} implemented a similar device based on epitaxial Fe/In$_x$Ga$_{1-x}$As heterostructures (Fig.~\ref{FigNL8}). The active layers consisted of a 2.5 $\mu$m thick Si-doped ($3-5 \times 10^{16}$ cm$^{-3}$) channel, a highly doped Schottky tunnel barrier ($5 \times 10^{18}$ cm$^{-3}$), and a 5 nm thick Fe layer. Heterostructures with In concentrations 0, 0.03, 0.05, and 0.06 were processed using lithographic and etching techniques into devices with 30 $\mu$m-wide channels oriented along the [110] direction, which is the $x$ direction in Fig.~\ref{FigNL8}.

It is technically difficult to fabricate a thin film with a FM attached at its edge with the  magnetization orientation proposed by \onlinecite{Zhang2000}. To circumvent this obstacle,  \onlinecite{Garlid2010} patterned pairs of Fe electrodes so that the centers of the contacts in each pair are 2, 6, or 10 $\mu$m from the edges of the channel. However, since the contacts are magnetized along $x$, and the spin polarization generated by the SHE is oriented along $z$, a magnetic field along $y$ was applied to precess the spin accumulation into the $x$ direction so that it could be detected. The spin accumulation is identified through the observation of the Hanle effect in the voltage measured between the pairs of FM contacts. The voltage first increases at low fields but then is suppressed due to spin dephasing in large fields.

The local character of the measurement causes a large background signal due to imperfect cancelation of the background HE voltage induced by the applied magnetic field, of local HEs due to  fringe fields generated by the FM contacts, and voltages due to the small fraction of the channel current that is shunted through the Fe contacts. The HE voltages were eliminated by using the expected symmetries of the signal, while the shunting effect was reduced by subtracting the voltages for the two current directions.

The results showed that the magnitude of the spin Hall conductivity was in agreement with models of the extrinsic SHE due to ionized impurity scattering. The bias and temperature dependences of the SHE indicated that both skew and side-jump scattering contribute to the total spin Hall conductivity. By analyzing the dependence of the SHE on channel conductivity, which was modified with the In content, \onlinecite{Garlid2010} determined the relative magnitudes of the skew and side-jump contributions to the total spin Hall conductivity.

\onlinecite{Ehlert2012,Ehlert2014} reported measurements of the SHE using a similar structure based on $n$-GaAs layers with relatively low carrier concentration ($5 \times 10^{16}$ cm$^{-3}$) and corresponding low conductivity. The FM voltage probes were implemented with (Ga,Mn)As/GaAs Esaki diode structures. The heterostructures were grown by molecular-beam epitaxy and consisted of a 1 $\mu$m-thick $n$-type transport channel, a 15-nm thick $n\rightarrow n^+$ GaAs transition layer ($5 \times 10^{18}$ cm$^{-3}$), a 2.2-nm Al$_{0.36}$Ga$_{0.64}$As diffusion barrier, and a 15-nm-thick layer of Ga$_{0.95}$Mn$_{0.05}$As. The highly doped (Ga,Mn)As/GaAs $p-n$ junction forms an Esaki diode. This structure was covered on the top  by 2 nm of Fe and 4 nm of Au. The purpose of Fe was to make the contacts harder magnetically, which helped to keep the magnetization aligned along their long axes during Hanle measurements. The values of spin Hall conductivities that were extracted are consistent with those calculated by \onlinecite{Engel2005} but smaller than those observed by \onlinecite{Garlid2010}.  \onlinecite{Ehlert2012,Ehlert2014} observe that the combined results of these two experiments show that both the skew and side-jump contributions to the spin Hall conductivity cannot be treated as fully independent of the conductivity of the channel.

\subsubsection{Spin Hall injection and detection without ferromagnets}
\label{nometals_sec}

\begin{figure}[h!]
\hspace*{-0cm}\epsfig{width=1\columnwidth,angle=0,file=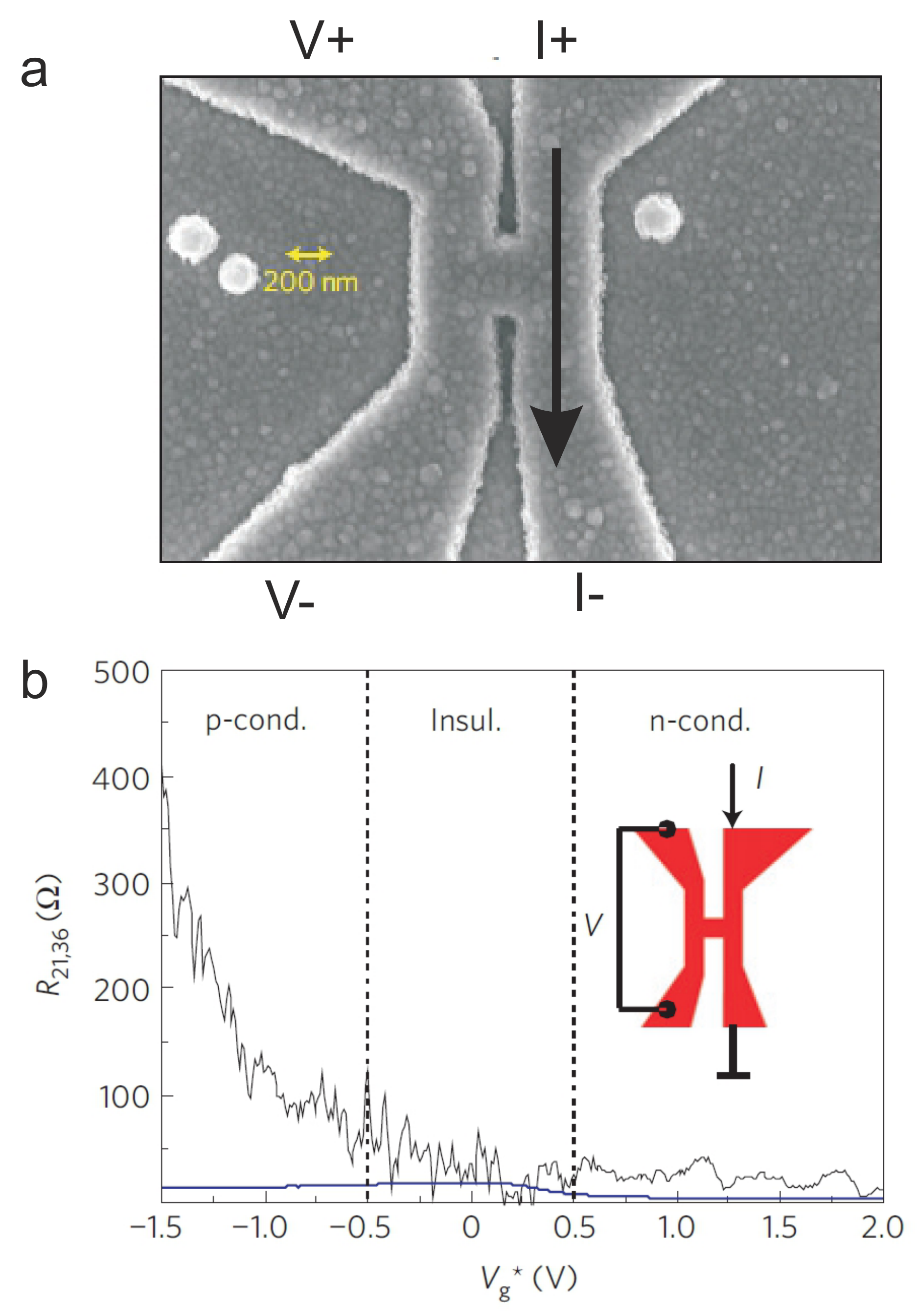}
\caption{(a) Scanning electron micrograph of  a H-shape device and probe configuration for spin injection via SHE and spin detection via ISHE. (b) The inset indicates the
measurement configuration for current injection (arrows) and voltage probes. The black curve in the main panel shows the non-local ISHE resistance signal. The blue solid curve indicates the residual voltage owing to current spreading. From Ref.~\onlinecite{Brune2010}.}
\label{fig_SHE/ISHE}
\end{figure}

Spin injection by the  SHE combined with spin detection by the  ISHE in one device \cite{Hirsch1999} was implemented by \onlinecite{Brune2010} using a device geometry proposed by \onlinecite{Hankiewicz2004b}. The original \onlinecite{Hirsch1999} proposal required a transverse strip connecting the edges of a slab on which spin accumulation was generated due to the SHE. A spin-current would circulate in the transverse strip which would then generate a measurable voltage transverse to it (see also Sec.~\ref{over} and Fig.~\ref{fig_exp_MHH}(c)). The fabrication of such structure is challenging, albeit not impossible. \onlinecite{Hankiewicz2004b} considered the same concept but on a planar structure shaped as an H, which is much simpler to fabricate. The device and measurement principle is shown in Fig.~\ref{fig_SHE/ISHE} (see also Sec.~\ref{over} and Fig.~\ref{fig_exp_MHH}(b)). An electric current is applied in one of the legs of the H-shaped structure and generates a transverse spin-current owing to the SHE. The spin-current propagates towards the other leg through the connecting part and produces a measurable voltage via the ISHE. This non-local voltage in the second leg dominates local contributions if the separation between the legs is large enough.

\onlinecite{Brune2010} used devices based on high-mobility HgTe/(Hg, Cd)Te quantum wells with a top gate electrode. The H-structures consisted of legs 1~$\mu$m long and 200~nm wide, with the connecting part being 200~nm wide and 200~nm long. The estimated mean free path in the system was $\ge2.5$~$\mu$m, i.e., the samples are well within the quasi-ballistic regime. Sweeping the gate voltage in the sample allowed to vary the strength of the Rashba spin-orbit coupling by a variation of both the electrical field across the quantum well and the Fermi level in the quantum well. In the sample it was possible to electrically tune the carrier density  from strongly $n$-type, through insulating, down to a $p$-type regime. This resulted in a strong modulation of the ISHE voltage, as shown in Fig.~\ref{fig_SHE/ISHE}. In the $p$-regime, where the spin-orbit coupling is strong, the signal is at least one order of magnitude larger than in the weakly spin-orbit coupled $n$-regime. Detailed numerical calculations  confirmed that the observed spin Hall signals had the ballistic intrinsic origin \cite{Brune2010}.

An H-shaped structure was also used in graphene devices \cite{Abanin2011a}. Here, a large Hall response was observed near the graphene neutrality point in the presence of an external magnetic field. The results were ascribed to spin-currents that resulted from the imbalance of the Hall resistivity for the spin-up and spin-down carriers induced by the Zeeman interaction; a process that does not involve a spin-orbit interaction, i.e. is not of the SHE origin, and that is largest in the cleanest graphene samples \cite{Abanin2011b}. More recently, the controlled addition of small amounts of covalently bonded hydrogen atoms has been reported to induce an enhancement of the spin-orbit interaction by three orders of magnitude in graphene \cite{Balakrishnan2013,CastroNeto2009}. Such large enhancement was estimated from nonlocal signals of up to 100 $\Omega$, which are observed at zero external magnetic fields and at room temperature. From the magnetic field and the length dependence of the non-local signal, a spin orbit strength of 2.5 meV was extracted for samples with 0.05\% hydrogenation.

\subsubsection{Spin Hall magnetoresistance}
\label{SHE-MR-sec}
In bilayer FM/NM systems a new type of magnetoresistance has been recently discovered which is directly associated with the SHE \cite{Huang2012,Weiler2012}. 
The observed magnetoresistance is  given by 
\begin{equation}
\rho=\rho_0+\rho_1(\hat{m}\cdot(\hat{j}\times\hat{z}))^2,
\label{SHMR}
\end{equation}
where  $\rho_0$ is the normal resistance, $\rho_1$ is the anisotropic resistance amplitude, and $\hat{j}$, $\hat{m}$, and $\hat{z}$ are the directional vectors of the current, the magnetization, and  the normal to the interface. This means that the magnetoresistance depends on the in-plane component of the magnetization perpendicular to the current.
In contrast, the conventional non-crystalline AMR  \cite{McGuire1975} has the form of 
\begin{equation}
\rho=\rho_0+\rho_1(\hat{j}\cdot\hat{m})^2, 
\label{AMR}
\end{equation}
with $\hat{j}\cdot\hat{m}=\cos(\theta_{j-m})$, where $\theta_{j-m}$ is the angle between the current and the magnetization.  

\begin{figure}[h!]
\hspace*{-0cm}\epsfig{width=1\columnwidth,angle=0,file=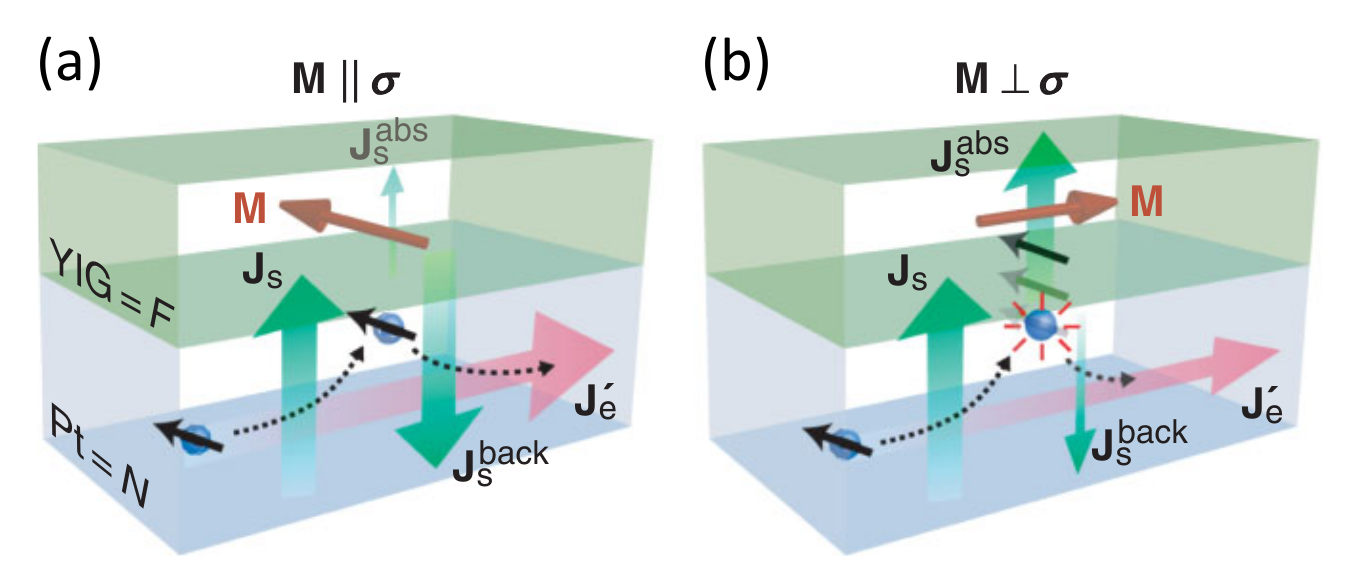}
\caption{Illustration of the SHE magnetoresistance. (a) When the magnetization aligns with the polarization of the SHE spin-current,  its back flow reflection generates an ISHE current that contributes to the longitudinal current. (b) When the magnetization is perpendicular to polarization of the SHE spin-current, the spin-current is absorbed and no ISHE current affects the longitudinal current.  From Ref.~\onlinecite{Nakayama2013}.}
\label{SHE-MR-schematics}
\end{figure}
This phenomenon has been termed the spin Hall magnetoreristance (SHMR) \cite{Hahn2013,Nakayama2013,Weiler2012,Vlietstra2013,Chen2013}. Its origin is  illustrated in Fig.~\ref{SHE-MR-schematics}. 
When a current flows parallel to the FM/NM interface a SHE spin-current is generated in the NM directed to the interface. If the magnetization is parallel to the polarization of the  spin-current generated by the SHE, it gets reflected at the interface and a spin-current back flows, as sketched in Fig.~\ref{SHE-MR-schematics}(a). This back flow spin-current then gets transformed into a charge current via the ISHE in the direction of the longitudinal current. If the magnetization is instead perpendicular to the polarization of the spin-current generated by the SHE, it can enter the FM and dephase, as shown in Fig.~\ref{SHE-MR-schematics}(b). In this case there is no spin-current back flow and no contribution via the ISHE to the longitudinal current in the NM. 

\begin{figure}[h!]
\hspace*{-0cm}\epsfig{width=1\columnwidth,angle=0,file=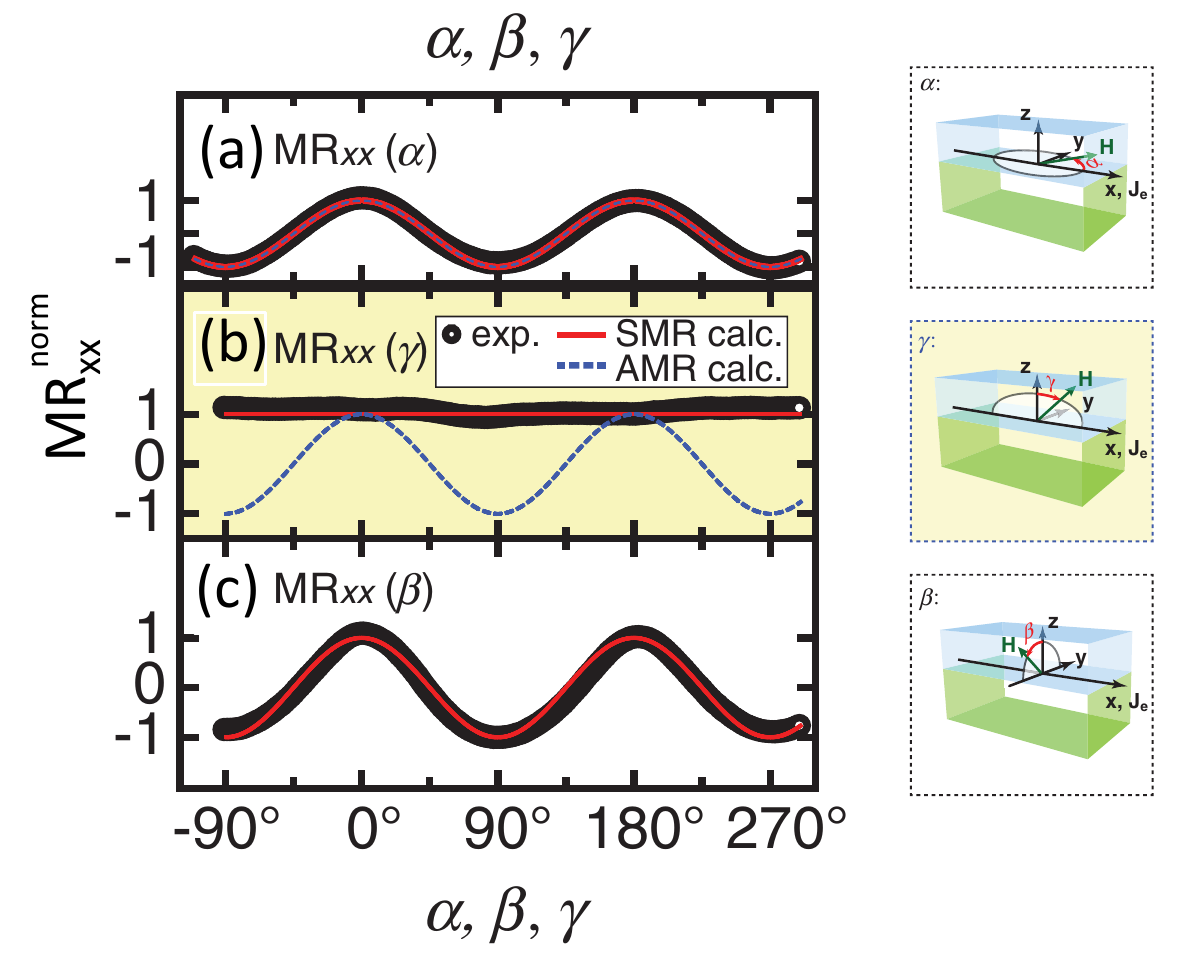}
\caption{Magnetoresistance curves as a function of the angles (a) $\alpha$, (b) $\gamma$, and (c) $\beta$, illustrated in the right panel.  
The key contrast to conventional AMR is the trace in (b), where no dependence is observed, while conventional AMR would give the sinusoidal form illustrated in the dashed-blue line.
From Ref.~\onlinecite{Nakayama2013}.}
\label{SHE-MR-exp}
\end{figure}

The typical experimental results are illustrated in Fig.~\ref{SHE-MR-exp}, where the bilayer system was YIG/Pt. The magnetoresistance traces are measured as a function of the magnetization angle in the x-y plane parallel to the interface, and in the z-y and z-x planes that are perpendicular to the interface. The measured angular dependencies are consistent with the SHMR phenomenology described by Eq.~(\ref{SHMR}) and are inconsistent with the AMR expression (\ref{AMR}).  
The  theory of the effect was derived by \onlinecite{Chen2013} based on the scattering formalism and the spin-charge drift-diffusion equations.

\subsection{Spin Hall effect coupled to magnetization dynamics}
\label{she-mag}

When the SHE is studied by coupling to magnetization dynamics three 
different FMR-based techniques can be found: (i)	Ferromagnetic resonance -- spin pumping (FMR-SP), (ii) modulation of damping (MOD) experiments, and (iii) spin Hall effect -- spin transfer torque (SHE-STT). The general underlying principle for the three methods is similar. In a bilayer NM/FM structure, the FM is used to inject or absorb a dynamic spin-current into or from the NM. (Note that  these studies have been also extended to replacing  the SHE/ISHE generating NM with another FM or antiferromagnet \cite{Miao2013,Azevedo2014,Freimuth2010,Mendes2014}.)

In FMR-SP, a spin-current is injected from the FM into the NM. The injected spin-current is a pure AC spin-current which is not accompanied by a charge current but which nevertheless can be detected electrically since it is converted into a charge current by means of the ISHE in the NM \cite{Saitoh2006}. The efficiency of the conversion process can be quantified by the spin Hall angle. Since in the process of spin injection angular momentum is lost in the FM, the FMR-SP leads to a broadening of the FMR line \cite{Mizukami2001,Urban2001,Heinrich2003}. 

In MOD experiments, the direct SHE induced in the NM by a DC electrical current is used to modify the damping in the FM which is concomitantly driven into FMR by the application of an RF magnetic field. In this approach, the DC spin-current generated by the SHE and injected across the NM/FM interface leads to a damping or antidamping-like torque acting on the precessing magnetization of the FM. Modulation of the damping is observed as a function of the applied DC charge current and a detailed line-width analysis allows extraction of the spin Hall angle \cite{Saitoh2012}. Note that the pure DC spin-current is generated in the bulk of the NM and that in order to quantitatively determine the spin Hall angle it is important to know the transmissibility of the NM/FM interface for the pure spin-current.

In the SHE-STT, a spin-current is used to transfer spin angular momentum and thus to exert a torque on the magnetic moments.  In these experiments an AC current sent along the NM/FM interface can create a RF excitation of the magnetization of the FM via the SHE-STT. In conventional STT junctions, an electrical current is sent perpendicular to a  stack with two FM electrodes to transfer angular momentum from one FM to the other FM \cite{Ralph2008}. SHE-STT experiments, on the other hand,  exploit the use of a perpendicular pure spin-current generated by an in-plane electrical current  in the attached NM via the SHE.

In both the MOD experiments and the SHE-STT, the  torques in the FM that are generated by the SHE in the NM would be in addition to the ISGE-related SOTs present at the inversion asymmetric FM/NM interface \cite{Kurebayashi2014,Garello2013,Freimuth2013}. Hence, in these experiments the spin Hall angle is in reality a parametrization of the total  torques generated by the currents and therefore it should be considered instead as the effective spin Hall angle for the specific bilayer system.  

In the rest of the section we expand on the details and recent results of each of these FMR-based techniques. FMR-SP is the more widely used technique to measure the effective spin Hall angle thus we detail this technique more extensively.

\subsubsection{Ferromagnetic resonance spin pumping}
\label{spin-pumping-exp}
\begin{figure}
\centering
\includegraphics[width=8cm]{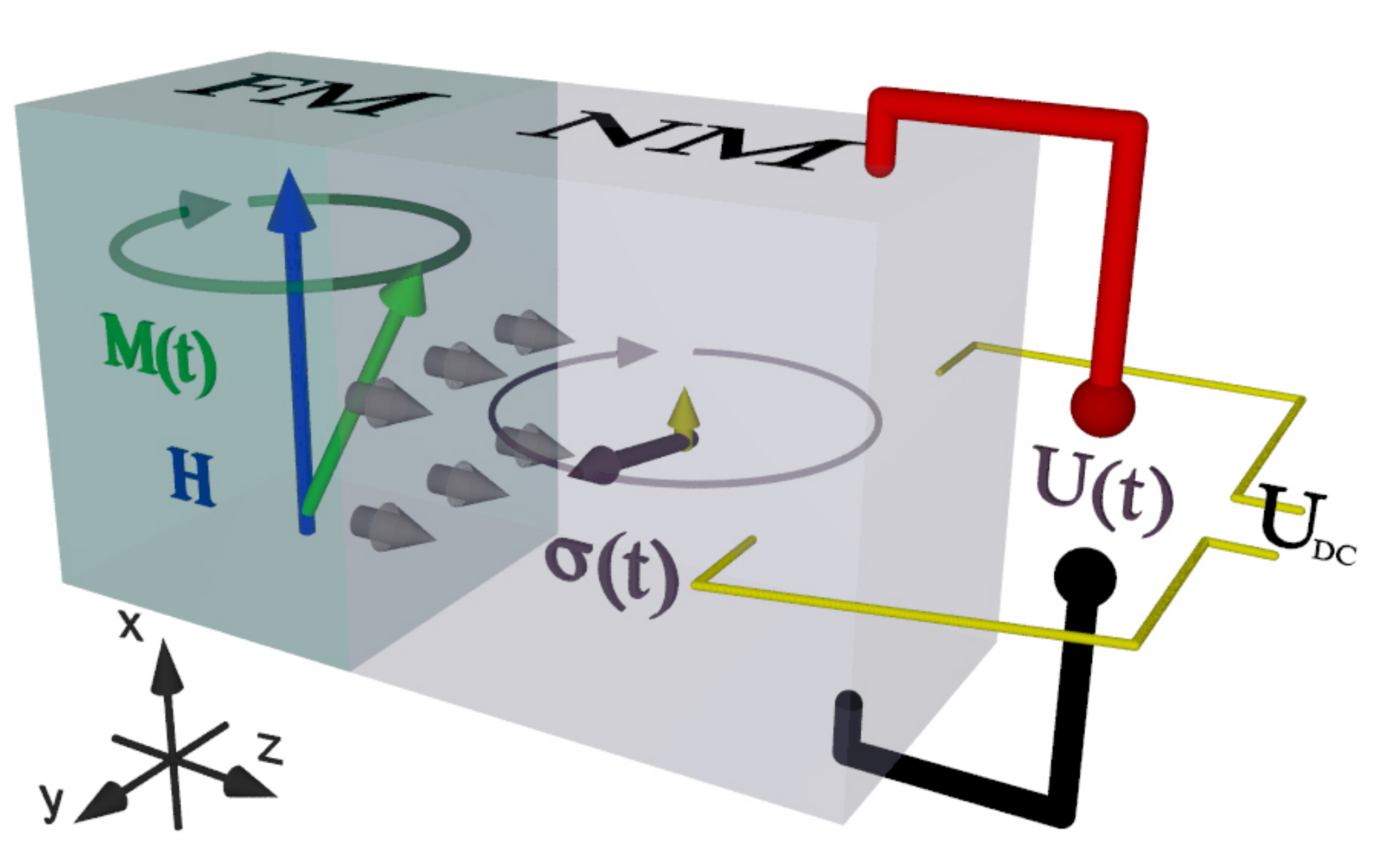}
\caption{A spin-current is generated by SP at the FM/NM interface (grey arrows). The time dependent spin polarization of this current (indicated as a dark grey arrow) rotates almost entirely in the $y-z$ plane. The small time averaged DC component (yellow arrow) appears along the $x$ axis. Both components lead to charge currents in NM and can be converted into AC and DC voltages by placing probes along the $x$ and $y$ direction, respectively. From Ref.~\onlinecite{Wei2014}.}
\label{schematics}
\end{figure}

As described in the theory section (Sec.~\ref{spin-pumping-theory}), \onlinecite{Tserkovnyak2002,Tserkovnyak2005} have shown that the precessing magnetization in a FM generates a spin-current strictly at the FM/NM interface, as sketched in Fig.~\ref{schematics}.
The spin-current generated at the interface propagates into the NM and consequently decays on a length scale connected to the effective spin diffusion length $\lambda_{sd}$ of the NM. As mentioned in the theory section, we note that the term effective is used here, since the determination of the spin diffusion length for a NM interfaced with a FM may also be connected to spin memory loss and proximity polarization at the interface. 
In the case of Pt and Pd  in contact with a FM metal, proximity effects are well known from x-ray magnetic circular dichroism experiments. 

The direction of the injected pure spin-current points from the FM to the NM and its polarization is time-dependent. Its projection onto the static magnetization direction of the FM leads to a small DC component of the injected spin-current into the NM. Performing time averaging one obtains a net DC spin-current given by Eq.~(\ref{ispump}) from Sec.~\ref{spin-pumping-theory}:
\begin{equation}
\label{ispump}
j_\mathrm{s,dc}  = \frac{\hbar  \omega}{4 \pi} \tilde{A}_\mathrm{r} sin^2\Theta, \nonumber
\end{equation}
where $\omega$ is the driving RF and $\Theta$ is the cone  angle of precession. Here $\tilde{A}_\mathrm{r}$ is the effective SP conductance.  If the thickness of the NM is smaller than the spin-diffusion length, the build-up of spin accumulation will yield a back-flow spin-current which will reduce the total spin-current into the NM.  
The SP conductance, $\tilde{A}_\mathrm{r}$, is proportional to the real part of the mixing conductance, discussed in Sec.~\ref{spin-pumping-theory}, and is reduced  by this back flow. 
The reduction depends on the ratio 
 $\tau_{tr}/\tau_{sf}$, the reduction being strongest as this ratio increases. Hence, the effective spin mixing conductance may become small even though a pure spin-current is efficiently transferred across the FM/NM interface. 
 Recently, spin flip scattering near the FM/NM interface has been divided up into a spin memory loss occuring directly at the interface (interface scattering) and the decay of the spin polarization as described above \cite{Rojas-Sanchez2014}. 

\begin{figure}[h!]
\hspace*{-0cm}\epsfig{width=1\columnwidth,angle=0,file=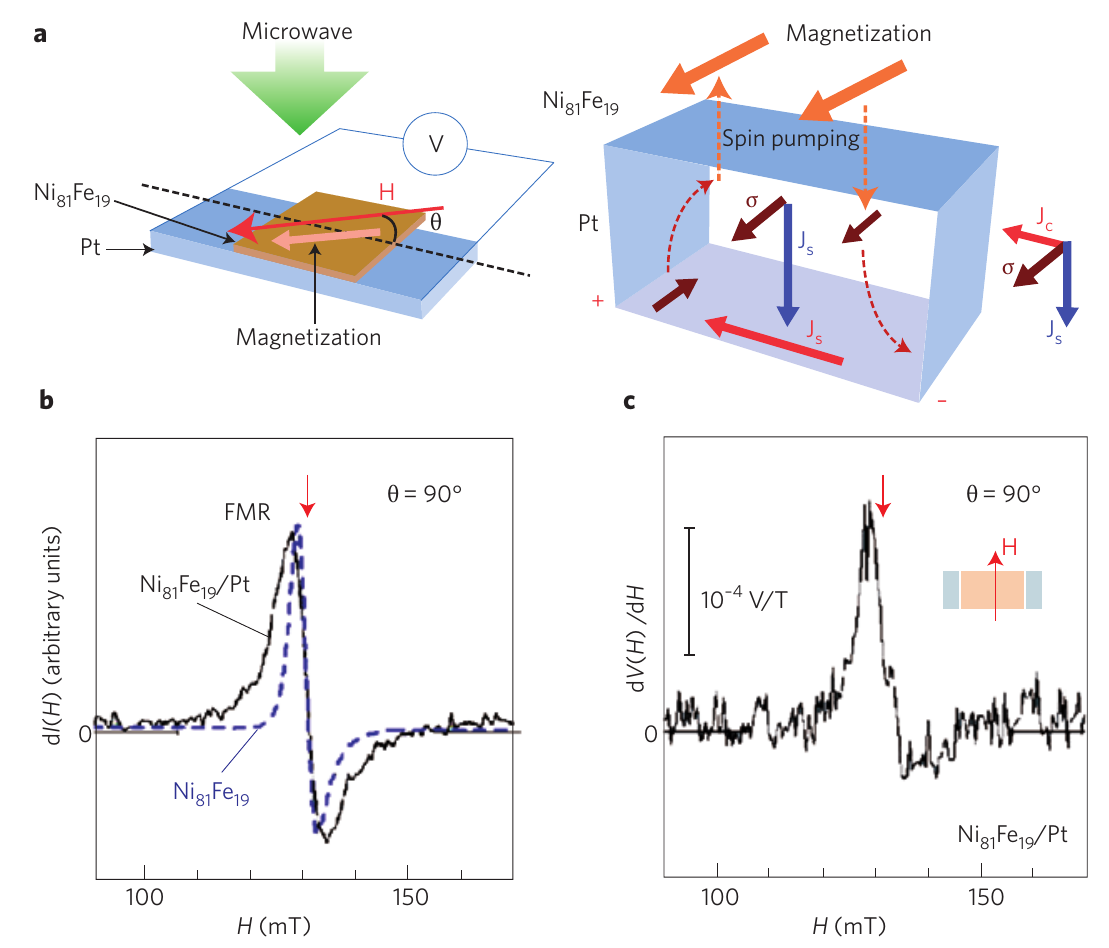}
\caption{Observation of the ISHE  in a metal device with spin injection from a FM by FMR-SP.  (a) Schematic illustration of the NiFe/Pt sample system used in the study and of the SP effect and the ISHE. (b) Magnetic field dependence of the FMR signal for the NiFe/Pt bilayer film and a bare NiFe film. $I$ denotes the microwave absorption intensity. (c) Magnetic field dependence of  $dV(H)/dH$ for the NiFe/Pt sample. $V$ denotes the electric-potential difference between the electrodes on the Pt layer. From Ref.~\onlinecite{Saitoh2006}.}
\label{fig_8_spin_pumping}
\end{figure}

The ISHE is used 
to electrically detect pure spin-currents generated by the SP \cite{Saitoh2006}, as shown Fig~\ref{fig_8_spin_pumping}. In spin-orbit coupled NMs like Pt or Pd,  the ISHE converts the pure spin-current into a detectable charge current
given by Eq.~(\ref{jcacsp}) Sec.~\ref{spin-pumping-theory}:
\begin{equation}
\vec{j}_\mathrm{c} = \alpha_\mathrm{SH} \; \frac{2e}{\hbar} \; \vec{j}_\mathrm{s} \times \vec{\sigma}(t).\nonumber
\end{equation}
Here, the vector of the spin-current density $\vec{j}_\mathrm{s}$ points perpendicular to the NM/FM interface into the NM. Note that the vector of the spin-current polarization $\vec{\sigma}(t)$ is a time varying quantity, which we do not average here, since it has now been demonstrated that the AC component is also measurable \cite{Wei2014,Weiler2014}.  In Fig.~\ref{fig_8_spin_pumping}, only the DC component of the spin-current polarization is depicted.

To measure the effect of the injected spin-current via ISHE, i.e. to measure the generated charge current, contact electrodes have to be attached to the sample. If the coordinate system of Fig.~\ref{schematics} is considered, placing electrodes along the y-direction allows detecting the small DC component of the SP-induced ISHE. In contrast, if the contact electrodes are attached along the x-direction, the much larger AC component in the GHz frequency range can be detected when high frequency lines are used. 

In case of DC detection the time averaged DC component of the injected spin-current pointing along the x-direction (yellow arrow in Fig.~\ref{schematics}) leads to a charge current which is converted to a potential drop across the resistance of the NM and can be measured as a voltage signal.
When performing FMR-SP experiments not only voltages due to ISHE are generated, but also due to, e.g., the AMR  or the AHE. Thus, great care has to be taken to disentangle these contributions. 

In the geometry sketched in Fig.~\ref{schematics} the propagation direction of the spin-current is along $z$ and its polarization is along $x$-direction. 
Equation~(\ref{jcsp}) from Sec.~\ref{spin-pumping-theory} is then used to convert between this spin-current and the measured voltage.


In the original experiments by \onlinecite{Saitoh2006}, the bilayer is placed in a FMR cavity  in which the magnetic-field component of the microwave mode with frequency 9.45~GHz is maximized while the electric-field component is minimized.  The voltage probes are placed on the sides of the millimeter-sized sample (see Fig.~\ref{fig_8_spin_pumping}). A similar set-up was used by \onlinecite{Azevedo2011}. Here, the sample is rotatable in the cavity and the cavity (i.e. the direction of the RF excitation field) is kept fixed with respect to the DC external magnetic field. This experimental geometry has advantages and disadvantages. The main advantage is that it is possible  to find an in-plane angle between excitation RF field and angular position of the voltage probes where the AMR contribution to the signal vanishes exactly while ISHE is detectable. Second, in the in-plane excitation geometry typically used, the sensitivity is large due to the large in-plane susceptibility at FMR. A major disadvantage is that it is not easy to perform frequency dependent measurements and that due to the use of a cavity the exact amplitude of the excitation field, and thus the cone angle of precession which enters Eq.~(\ref{ispump}) in Sec.~\ref{spin-pumping-theory}, is usually not well known.  Finally, since typically large, millimeter sized samples are used in the experiments, spurious RF electric fields may lead to additional contributions due to the AHE. It is therefore not straightforward to obtain an exact quantitative value of the spin Hall angle from cavity FMR-type measurements.

In experiments shown in Fig.~\ref{fig_8_spin_pumping}, the measured FMR spectrum of the NiFe/Pt sample is compared to a reference NiFe sample (see Fig.~\ref{fig_8_spin_pumping}(b)). The FMR line width of the NiFe/Pt sample is larger than that of the reference NiFe film which demonstrates the presence of the SP effect in the NiFe/Pt. The induced voltage signal measured simultaneously across the sample along an axis parallel to the NiFe/Pt  interface is shown in Fig.~\ref{fig_8_spin_pumping}(c). \onlinecite{Saitoh2006} and \onlinecite{Ando2008b} demonstrated that the signal is present only when the spin polarization vector of the injected spin-current has a component perpendicular to the measured electric field across the sample, consistent with the ISHE.  

\onlinecite{Ando2009} reported electrical detection of a spin wave resonance in nanostructured NiFe/Pt samples. 
Electrical tuning of the spin signal in a semiconductor has been recently demonstrated also by  \onlinecite{Ando2011}. In the experiment, spins were  injected from NiFe into GaAs through a Schottky contact using the  FMR-SP.  Tuning of the SP efficiency was achieved by applying a bias voltage across the NiFe/GaAs Schottky barrier and interpreted as a consequence of a suppressed or enhanced spin coupling across the interface. The FM/semiconductor SP experiments in Ref.~\onlinecite{Ando2011} were performed also on samples with an ohmic contact between NiFe and GaAs. The measurements indicate that the resistance mismatch problem in ohmic metal/semiconductor spin-injection devices can be circumvented by using the FMR-SP technique.
Similar experiments have recently been performed also for spin injection into Si \cite{Ando2010}, Ge \cite{Jain2012} and organic semiconductors \cite{Watanabe2014}. 

\begin{figure}[h!]
\hspace*{-0cm}\epsfig{width=1\columnwidth,angle=0,file=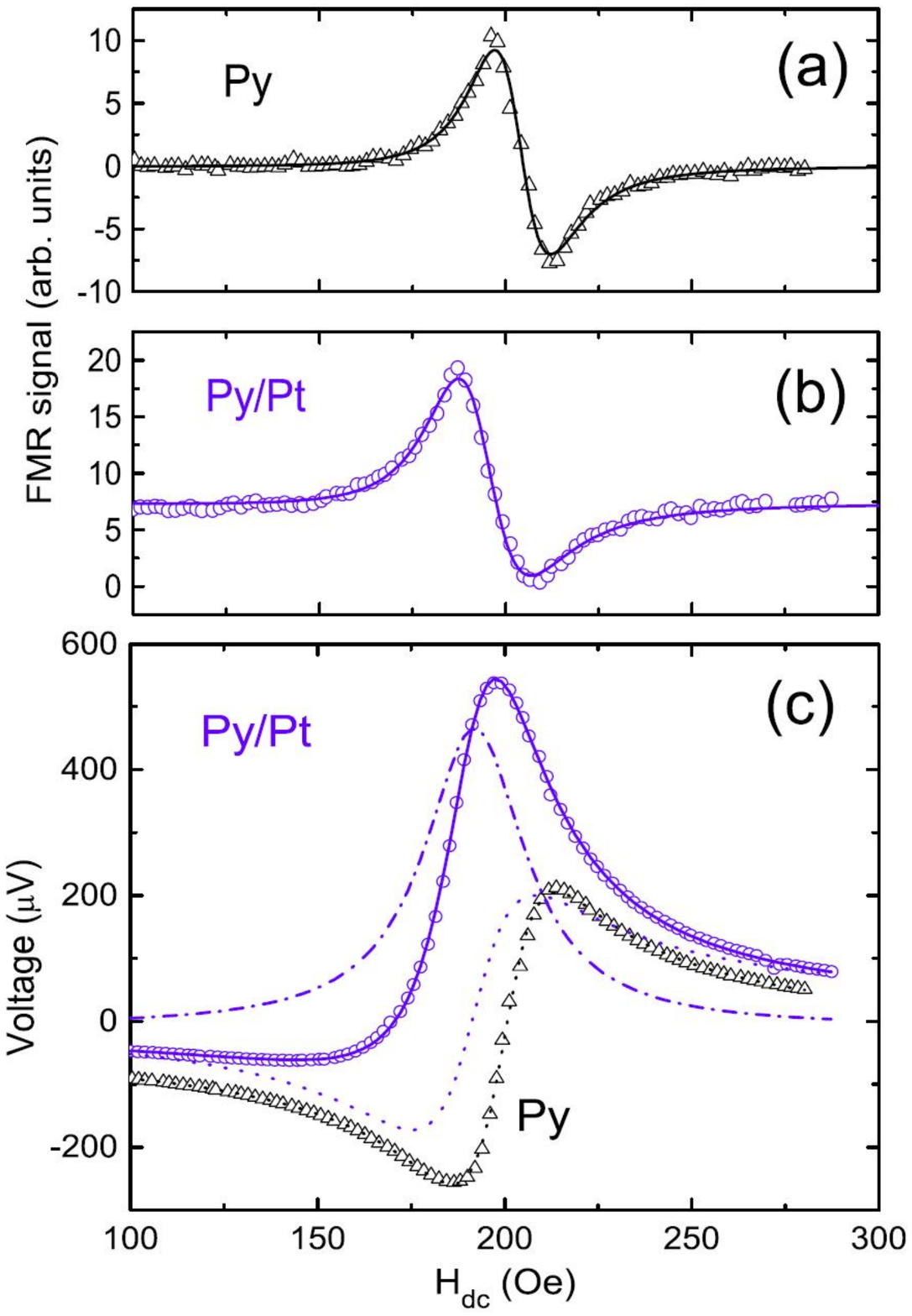}

\caption{(a),(b) Derivative of FMR spectra for Py/Pt (blue open circles) and Py (black triangles). The solid lines are fits to a Lorentzian FMR absorption function. (c) Voltage along the
samples vs. field DC magnetic field (Py/Pt: blue open circles; Py: black triangles). Dotted and dashed lines show the decomposition of the spectrum into a symmetric (ISHE) and antisymmetric (AMR) contribution. The solid line shows the combined fit for the Py/Pt sample. From Ref.~\onlinecite{Mosendz2010}.}
\label{resonance_trace}
\end{figure}

\begin{figure}[h!]
\hspace*{-0cm}\epsfig{width=1\columnwidth,angle=0,file=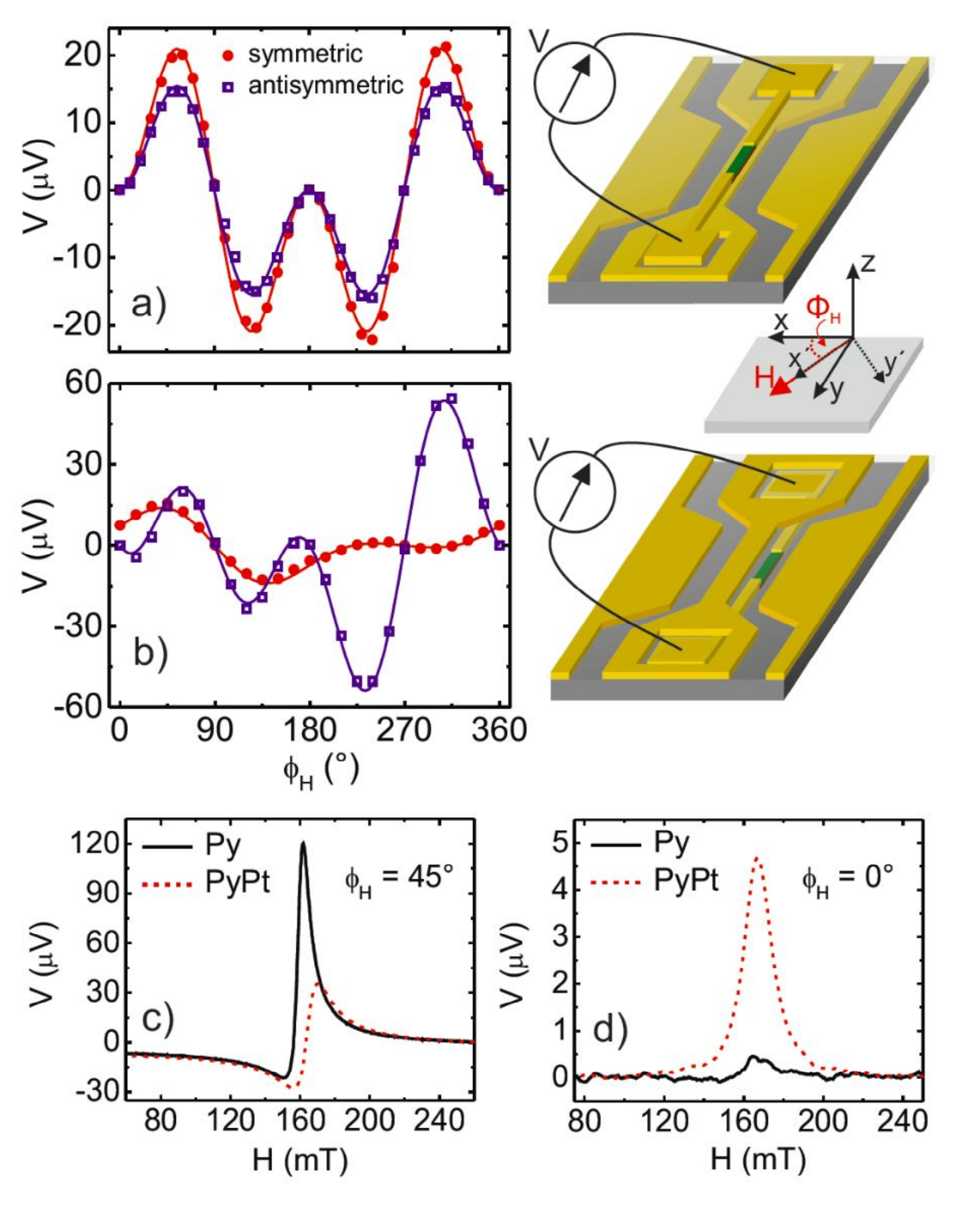}

\caption{Symmetric (red dots) and antisymmetric (blue open squares) voltage signals amplitudes at FMR (at 12 GHz) for a Py/Pt bilayer as a function of angle $\phi_\text{H}$. In (a) the magnetic excitation field is in-plane placing a Py/NM bilayer on top of the signal line of a CPW. Both symmetric and antisymmetric amplitudes obey a $\sin (\phi_\text{H}) \sin (2\phi_\text{H})$ behavior. (b) The magnetic excitation field generated by the CPW is out-of-plane with respect to the Py/Pt layers. The amplitudes of the antisymmetric part follow a $(a\sin (\phi_H)+b)\sin (2\phi_\text{H})$ behavior. The symmetric part obeys $(c \sin(\phi_\text{H}) + d)\sin(2\phi_\text{H}) + e \cos(\phi_\text{H})$, which reflects the fact, that the symmetric part is due to AMR and ISHE. (c) Voltage at FMR for $\phi_\text{H}=45^\circ$, and (d) $\phi_\text{H}=0^\circ$ for a single Py layer and a PyPt bilayer. From Ref.~\onlinecite{Obstbaum2014}.}
\label{setup_Obstbaum}
\end{figure}

A second possibility to quantify the spin Hall angle has been pioneered by \onlinecite{Mosendz2010}. They use a microstructured co-planar wave guide (CPW) with integrated bilayer structure on top of the center wave guide. This geometry allows excitation of FMR in the FM layer over a wide frequency range while the driving RF field is in the plane of the bilayer at 90$^\circ$ to the long axis of the several hundred micrometer long device (see Fig.~\ref{setup_Obstbaum}). The use of a wave guide structure allows precise knowledge of the amplitude of the RF fields and thus the cone angle of the precessing magnetization. Voltage pick-up at the ends of the wire are used, perpendicular to the direction of the RF driving field. \onlinecite{Mosendz2010} applied the external magnetic bias field at an angle of 45$^\circ$ to the long axis of the wave guide. In this experimental geometry both ISHE and AMR signals are detected at the voltage probes as can be seen directly in the recorded voltage traces (see Fig.~\ref{resonance_trace}). 

AMR leads to a parasitic DC voltage signal at FMR due to the mixing of the time dependent resistivity (AMR and precessing magnetization) with a capacitively or inductively coupled microwave current $I(t)$ in the bilayer.  The AMR of the bilayer can be taken into account by considering the orientation of the magnetization with respect to the current direction: $R_A=R_{\parallel}-R_{\perp}$. The general formula describing the parasitic voltage pick-up due to the AMR is given by $\langle V(t)\rangle = \langle I(t)R_\text{A}\alpha_\text{ip}(t) \sin(2\phi_\text{H})\rangle$, \cite{Mecking2007,Obstbaum2014} and it follows that this time-averaged DC voltage is to first order proportional to the in-plane dynamic cone angle of the magnetization $\alpha_\text{ip}(t)$. The cone angle of precession can easily be calculated from the simultaneously measured susceptibility at FMR in the exactly known geometry of the CPW structure. The angle $\phi_\text{H}$ is defined in Fig.~\ref{setup_Obstbaum}.
Note that according to  \onlinecite{Bai2013}, spurious effects due to the AMR can be excluded by carefully analyzing the high frequency characteristics of the CPWs used in the experiments with in-plane excitation, leading to a quantitative determination of the spin Hall angles.

Another possibility is to place the bilayer in the gap of the CPW (see Fig.~\ref{setup_Obstbaum}). Now the in-plane dynamic cone angle relevant for the AMR is given by $\alpha_\text{ip}(t) = \chi_\text{y'y'} h_\text{x}(t) \sin(\phi_\text{H}) + \chi_\text{y'z} h_\text{z}(t)$. The formula contains both in-plane and out-of-plane magnetic fields, together with the corresponding tensor elements of the susceptibility ($\chi_{ij}$). 
Since the out-of-plane field produced by the CPW is about three orders of magnitude larger than its in-plane component, one is tempted to simply neglect the terms arising form the in-plane field.
This approach is justified as long as only a single layer is studied. 
However, as soon as a FM/NM bilayer with a highly conductive NM  is used, the inductively or capacitively coupled microwave current largely flows in the NM and therefore generates an in-plane Oersted field of the same frequency and phase and with an amplitude comparable to the RF field generated by the CPW. Hence the RF current distribution in the bilayer has a significant effect on the magnetization dynamics in the FM layer and can even be the dominating source of DC voltage generation by the AMR \cite{Obstbaum2014}. Using standard electro-magnetic wave simulation codes, the RF magnetic field contribution can be calculated rather accurately.

When performing angular dependent measurements, the symmetric and antisymmetric contributions due to the ISHE and the AMR can be traced (see Fig.~\ref{setup_Obstbaum}(a) and (b)). While for in-plane excitation the signal shows the same angular dependence, for the out-of-plane excitation case the antisymmetric contribution can be suppressed completely at an angle of $\phi_\text{H}=0$ (see Fig.~\ref{setup_Obstbaum}(d)). The voltage contribution at this angle is thought to arise from ISHE exclusively and allows quantitative determination of the spin Hall angle. Note that in these measurements both symmetric and antisymmetric contributions can be observed in a bare FM layer when the angle is set to $\phi_\text{H}=45^\circ$ (see Fig.~\ref{setup_Obstbaum}(c)).


\subsubsection{Spin Hall effect modulation of magnetization damping}

A MOD experiment that is the inverse of the FMR-SP was 
proposed by \onlinecite{Ando2008b}. In the MOD described in Fig.~\ref{setup_Ando}, a FM/NM bilayer (in this case Py/Pt) is placed in a microwave cavity (frequency 9.4 GHz) and subjected to an RF driving field. By adjusting the external field, the bilayer can be brought into FMR. A typical FMR trace $dI(H)/dH$ is shown in Fig.~\ref{setup_Ando}(b). The direction of the external magnetic field encloses an angle $\theta$ with the direction of current flow. Since the mm-sized sample consists of 10 nm NiFe and 10 nm Pt, the effect of SP which contributes to the relaxation of the precessing magnetization can be observed as a line width broadening when comparing to the data obtained for a plain NiFe film. Fig.~\ref{setup_Ando}(c) illustrates the effect of a DC current sent through the bilayer sample due to the combined action of the SHE and STT. Due to the SHE a spin-current is generated in the Pt layer and enters the NiFe film. Its flow direction is perpendicular to the interface and its polarization direction $\vec{\sigma}$ depends on the direction of current flow. The spin-current exerts a torque on the precessing magnetization which either adds to the damping torque or opposes it. The effect is maximized when the external magnetic field points perpendicular to the direction of current flow. 
For the situation sketched here, the spin-current density can be written as
$\vec{j}_s=\alpha_\text{SH}\frac{\hbar}{2 e}\hat{n}\times\vec{j}_\text{c}=\alpha_\text{SH}\frac{\hbar}{2 e}\abs{\vec{j}_\text{c}}\hat{\sigma}$.
The effect of the injected spin-current on the precessing magnetization can be modelled in terms of an additional STT contribution to the Landau-Lifshitz Gilbert equation \cite{Ando2008b,Liu2011} that has to be added on top of the SP contribution:

\begin{equation}
\vec{\tau}_\text{STT}=-\mu_0\gamma\alpha_\text{SH}\eta\frac{\hbar}{2 {e}}\frac{{j_\text{c}}}{\mu_0\Ms^2d_\text{Py}}\vec{M}\times\left(\vec{M}\times\hat{\sigma}\right)
\end{equation}

Here, $d_\text{Py}$ is the thickness of the Py layer. For the sake of simplicity, the factor 

\begin{equation}
\kappa=\alpha_\text{SH}\eta\frac{\hbar}{2{e}}\frac{{j_\text{c}}}{\mu_0\Ms^2d_\text{Py}}
\end{equation}
is introduced. Note that this factor is dimensionless and $\tstt<0$ for $j_c>0$ due to the negative electron charge. 
The parameter $\eta$ defines the so called injection efficiency and contains the effects of spin-current losses near the interface. There is no consensus on the exact ingredients for this parameter, so it could be useful to use $\eta\cdot\alpha_\text{SH}$ as an effective quantity parametrizing the STT efficiency. 

\begin{figure}[h!]
\hspace*{-0cm}\epsfig{width=1\columnwidth,angle=0,file=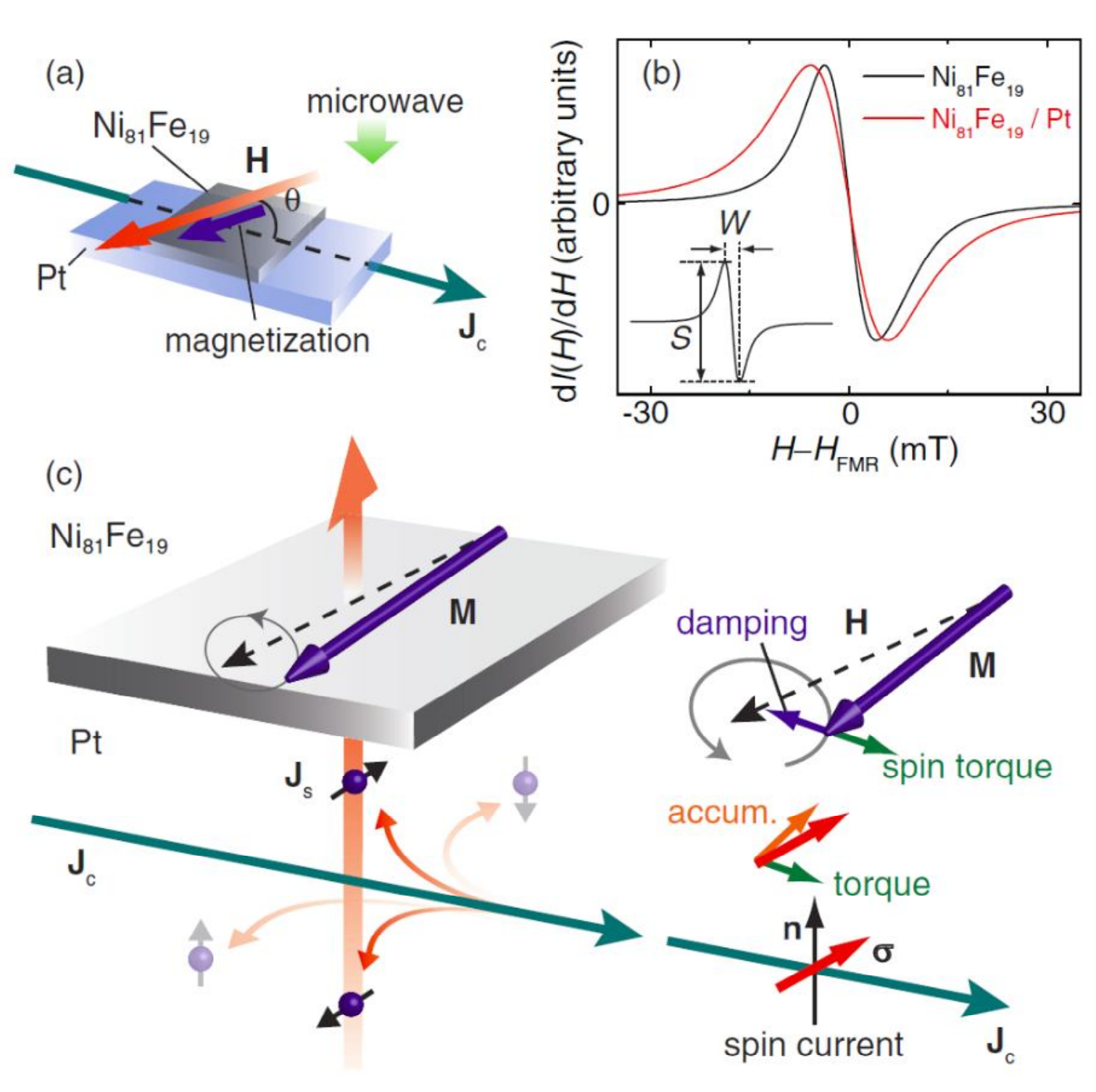}

\caption{ (a) A schematic illustration of the MOD experiment to determine the spin Hall angle. $H$ is the external magnetic field, and $J_c$ represents the applied electric current density. (b) Magnetic field dependence
of the FMR signal for a NiFe/Pt bilayer film (red) and a pure NiFe film (black). Note the linewidth broadening for NiFe/Pt due to SP. (c) Schematic illustration of the spin
Hall and the spin-torque effects. $\vec{M}$, $\vec{J_s}$, and $\vec{\sigma}$ denote the magnetization, the flow direction of the spin-current density, and the spin-polarization vector of the spin-current, respectively. From Ref.~\onlinecite{Ando2008b}.}
\label{setup_Ando}
\end{figure}

Figure~\ref{setup_Ando} shows the MOD experimental findings. When a current flows through the FM/NM bilayer, the STT generated by the spin-current traversing the NM/FM interface due to the SHE alters the FMR line width when the current flow direction and the external magnetic field direction enclose an angle of 90$^\circ$ while no effect is observed for collinear orientation, consistent with the theoretical expectation.

\begin{figure}[h!]
\hspace*{-0cm}\epsfig{width=1\columnwidth,angle=0,file=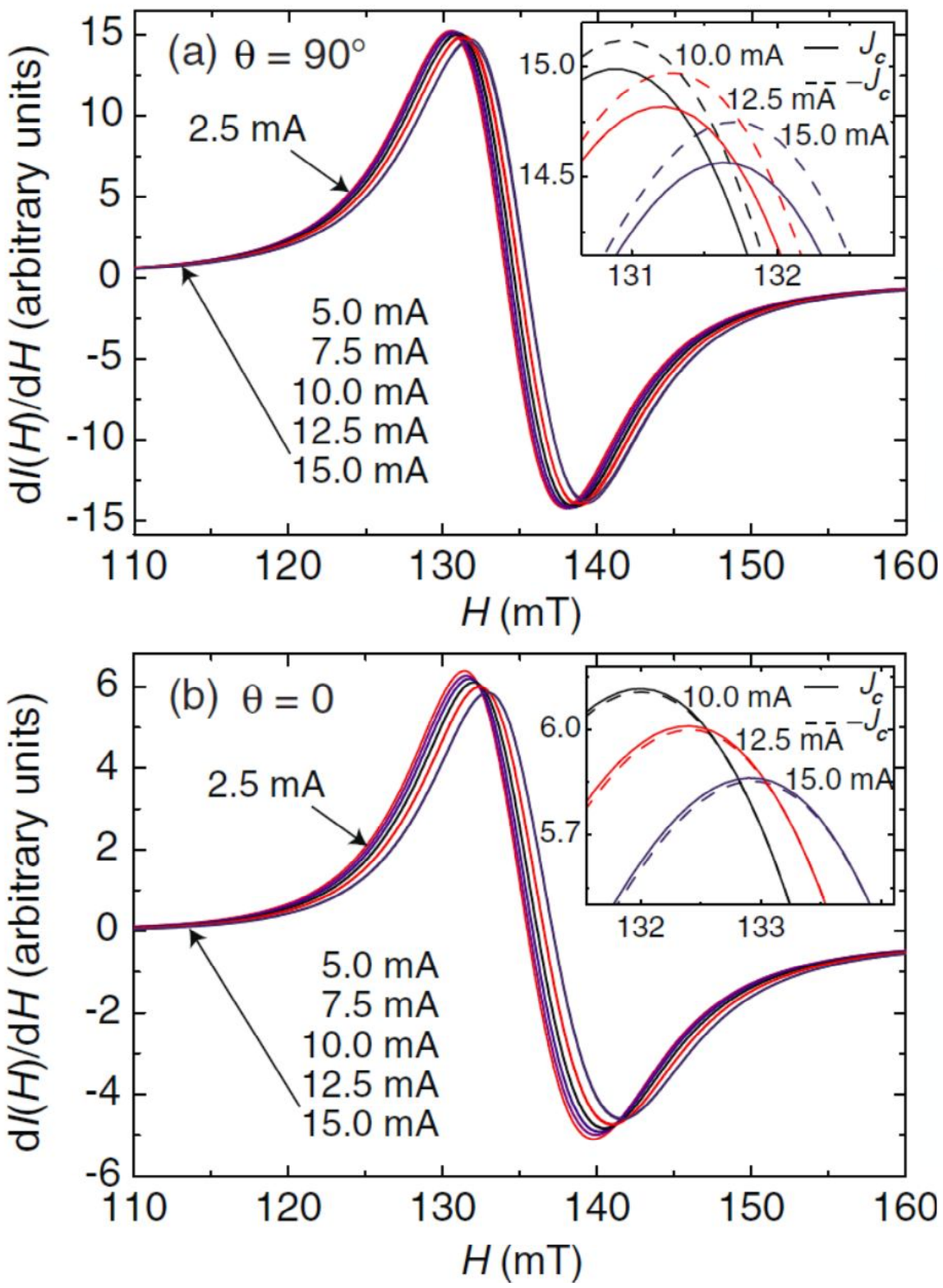}
\caption{FMR spectra for the NiFe/Pt bilayer measured at various electric current density values $J_c$ when the
magnetic field direction is (a) 90$^\circ$ and (b) 0$^\circ$. The inset shows magnified views around the peaks of the spectra, where the solid and dashed curves are the FMR spectra measured with
electric current densities $J_c$ and $-J_c$, respectively. From Ref.~\cite{Ando2008b}.}
\label{setup_Ando}
\end{figure}
 
Similar experiments have been performed by \onlinecite{demidov2011a,demidov2011b} using Brillouin Light scattering methods. The key finding in these experiments is the control of the FMR line width of the FM film by employing the SHE which generates a pure spin-current in the adjacent NM. 
Ultimately, in suitable nanostructured materials, the application of a large enough charge current density should lead to the generation of coherent auto-oscillations in the FM nano object due to a DC charge current \cite{demidov2012}.

\subsubsection{Spin Hall effect - spin transfer torque}
\label{she-stt}

\begin{figure}[h!]
\hspace*{-0cm}\epsfig{width=1\columnwidth,angle=0,file=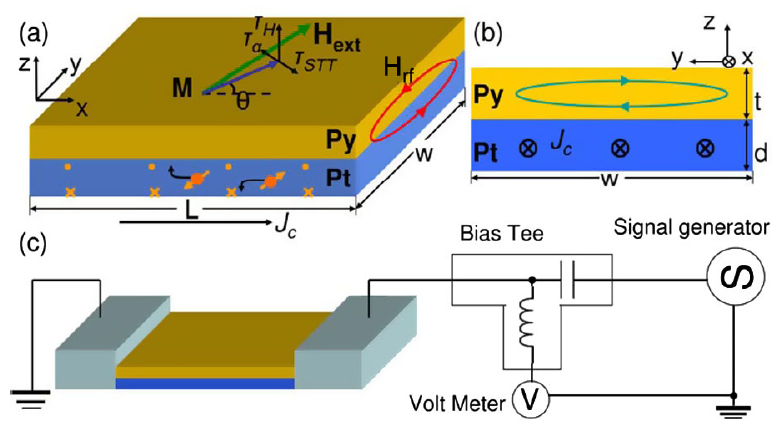}
\caption{a) Schematic of Pt/Py bilayer thin film illustrating the STT induced by the SHE rising form the RF current through NM as well as the damping torque and the torque due to the Oersted field when the magnetization of FM is aligned in an external magnetic field.  b) Shows the dimensions of the sample and the Oersted field due to a current flowing through FM. c) Depicts the electrical measurement scheme. Figure from Ref.~\onlinecite{Liu2011}.}
\label{fig_Liu2008}
\end{figure}

Finally, a third FMR technique has been employed that allows accessing the spin Hall angle experimentally.
\onlinecite{Liu2011} applied a microwave frequency charge current in the plane of a NiFe/Pt sample and observed the FMR in NiFe. 
Due to the action of the SHE a transverse spin-current is generated in the NM, in this case Pt, which is injected into the FM layer. Consequently, an oscillatory STT acts on the magnetic moments in the FM, inducing precession of the magnetization (see Fig.~\ref{fig_Liu2008}). The oscillatory magnetiziation in the FM leads to an oscillatory AMR which in turn leads to an oscillatory resistance. This high frequency resistance mixes with the RF current and leads to a detectable DC voltage across the device which can be picked up using a bias tee (Fig.~\ref{fig_Liu2008}(c)). 

In these experiments the external magnetic field is typically fixed at an angle of 45$^\circ$ and swept in the plane of the films to achieve the FMR condition. 
In the set-up, different torques act on the magnetization of the FM which is aligned along the magnetic field direction as depicted in  Fig.~\ref{fig_Liu2008}(a). The torques include all the STTs due to the SHE in the NM, the torque induced by the Oersted field due to the RF current through the device, and  the torque already modified by SP. We also emphasize  that the  torques  generated by the SHE in the NM would be in addition to the ISGE-related SOTs present in the FM near the interface \cite{Kurebayashi2014,Garello2013,Freimuth2013}.

Landau-Lifshitz-Gilbert equations including all relevant torques can be used to model the DC voltage response of the bilayer device and the result shows that the mixing voltage contains the contributions of symmetric and anti-symmetric Lorentzian lines \cite{Liu2011}. 
According to \onlinecite{Liu2011}, the detailed analysis of the resonance properties of this voltage enables a quantitative measure of the spin-current absorbed by the FM and of the spin Hall angle.
\onlinecite{Liu2011} shows that the ratio of the symmetric and antisymmetric components of the resonance curve, when scaled properly by material parameters like the saturation magnetization, thickness and width of the FM, and the external magnetic field,  is linked to the ratio of spin and charge currents and thus to the spin Hall angle. The authors emphasize that the measurement method is (in a reasonable thickness regime of the FM and the NM) self calibrating since the strength of the torque from the spin current is measured
relative to the torque from the RF magnetic field, which can be calculated  from the geometry of the sample.
The same method has been applied to various combinations of FMs and NMs \cite{Liu2011,Liu2012,Pai2012}.

\begin{figure}[h!]
\hspace*{-0cm}\epsfig{width=1\columnwidth,angle=0,file=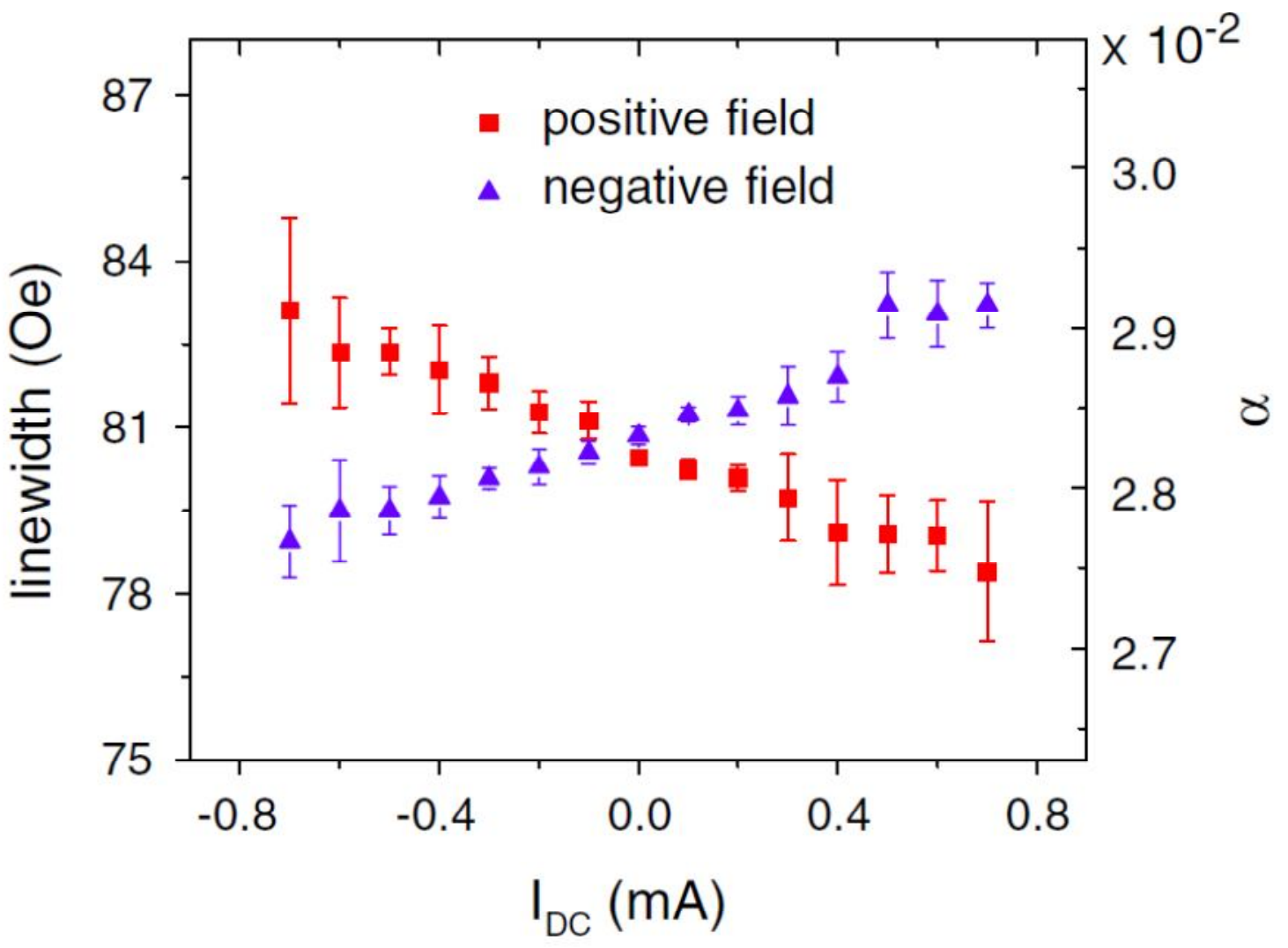}
\caption{Effective damping as a function of current density through the Pt layer in a Py(4~nm)/Pt(6~nm) bilayer. From Ref.~\onlinecite{Liu2011}.}
\label{Liu_Fig3}
\end{figure}

Also in these types of experiments the tunability of the effective damping parameter has been demonstrated by \onlinecite{Liu2011,Kasai2014}. An example is illustrated in Fig.~\ref{Liu_Fig3} for the case of Py/Pt where the effective damping parameter is shown to be tunable as a function of the current direction and amplitude \cite{Liu2011}.

\subsubsection{Spin Hall effect induced switching of the magnetization}
\label{SHE_switching}

For sufficiently large current densities pushed through the NM and large spin Hall angles, it is possible to even reverse the magnetization in a FM nanoelement placed on top of the NM current carrying line, as has been demonstrated by \onlinecite{Liu2012} (see Fig~\ref{she-stt}). In these experiments it is important to use a NM/FM combination where, when placing the NM in contact with the FM, the induced damping due to SP remains negligible. This is the case for CoFeB/Ta. On one hand, $\beta$-Ta shows a giant spin Hall angle \cite{Liu2012}), on the other hand, enhancement of damping due to SP is not observed in the CoFeB layer. Furthermore, due to the large resistivity of the CoFeB layer a large portion of the applied current is pushed through the Ta layer where it produces the pure spin-current due to the SHE. Another important feature is that the bilayer is capped with MgO to induce a large perpendicular anisotropy in CoFeB. The thin layer of MgO (1.6~nm) is used as a tunnel barrier between the thin CoFeB free layer (1.6~nm) and the thicker CoFeB reference layer (3.8~nm) so that the tunnelling magnetoresistance (TMR) effect can be used to determine the relative orientation of their magnetization. 
The results of these experiments are summarized in Fig~\ref{she-stt} and may be viewed as a paradigm change in the mechanism for switching magnetic nanoelements in spintronic devices since here switching is driven be a purely in-plane electrical current and not via a current perpendicular to the layer stack. Similar results have been obtained for W/CoFeB layers \cite{Pai2012}. One should note that while the exact value for the spin Hall angle extracted from these experiments is still under debate, the fact that switching can be achieved for these devices underpins not only the technological relevance, but also that a sizeable SHE (possibly in combination with other ISGE-related SOTs) must be generated in these structures.

\begin{figure}
\centering
\includegraphics[width=8cm]{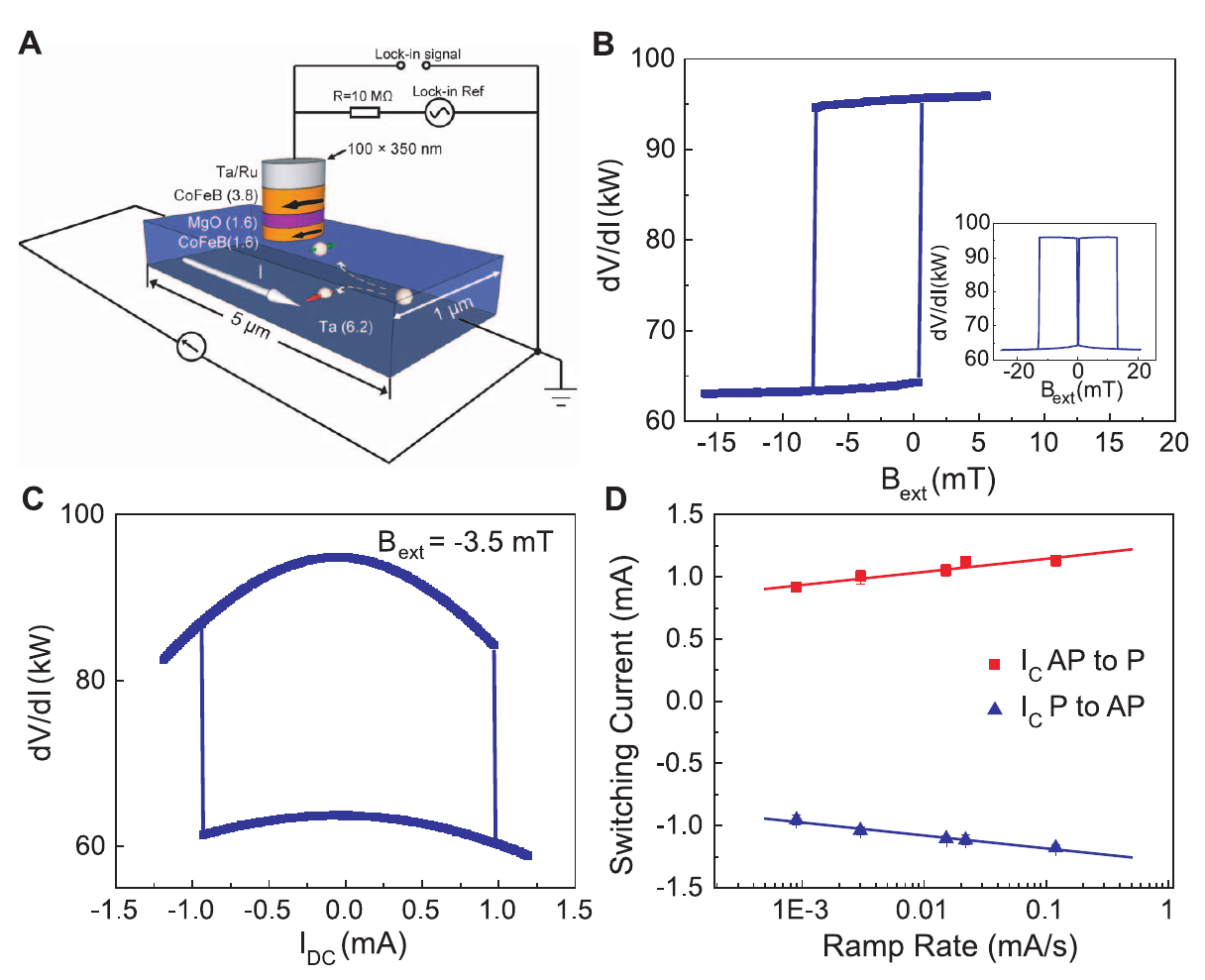}
\caption{SHE induced switching for an in-plane magnetized nanomagnet at room temperature. (A) Schematic of the three-terminal SHE devices and the circuit for measurements.  (B) TMR minor loop of the magnetic tunnel junction as a function of the external applied field $B_{ext}$ applied in-plane along the long axis of the sample. (Inset) TMR major loop of the device. (C) TMR of the device as a function of applied dc current $I_{DC}$. An in-plane external field of -3.5~mT is applied to set the device at the center of the minor loop. (D) Switching currents as a function of the ramp rate for sweeping current. Red squares indicate switching from antiparallel (AP) to parallel (P) magnetizations; blue triangles indicate switching from P to AP. Solid lines represent linear fits of switching current versus log(ramp rate). Error bars are smaller than the symbol size. From Ref.~\onlinecite{Liu2012}.}
\label{she-stt}
\end{figure}

\begin{figure}
\centering
\includegraphics[width=8cm]{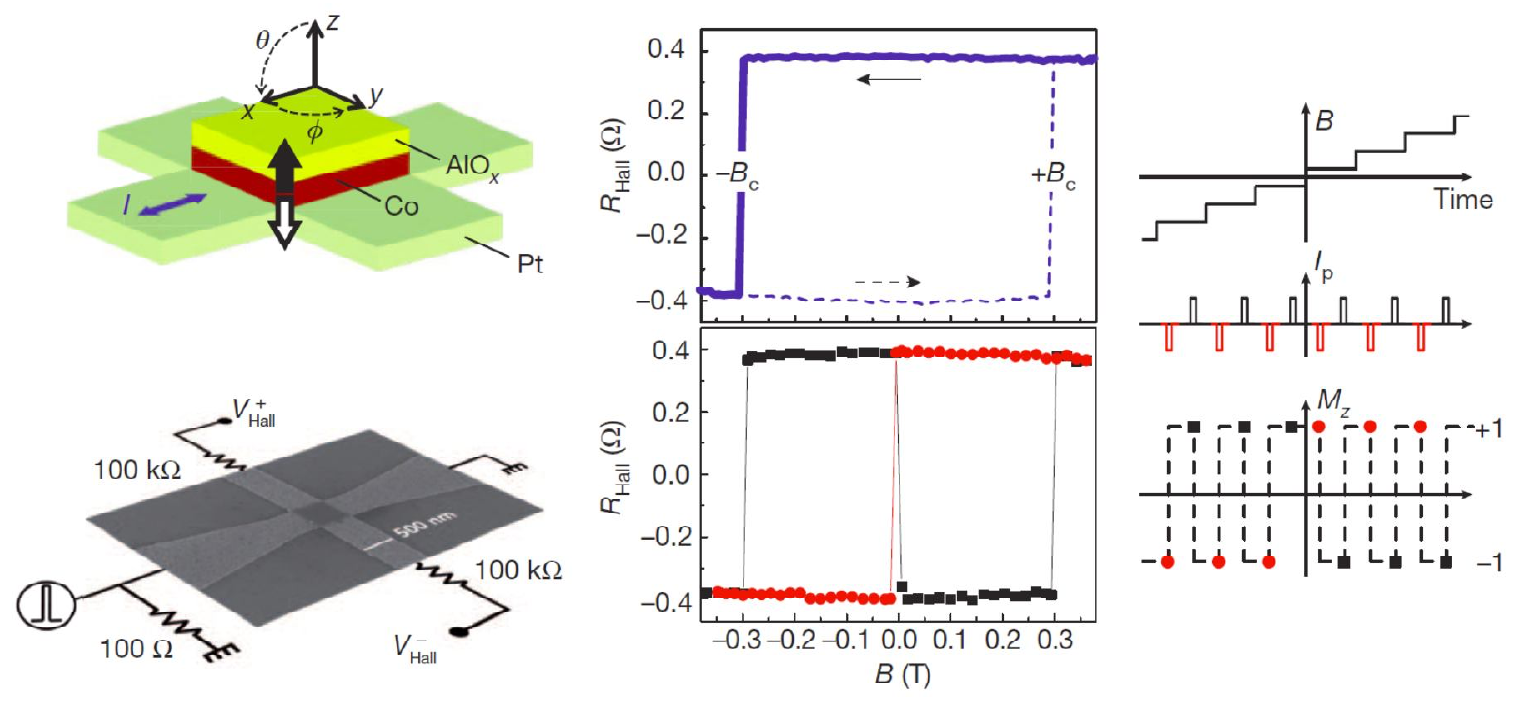}
\caption{Top left: Device schematic and current-induced switching. Hall cross geometry. Black and white arrows indicate the up€™ and €˜down equilibrium magnetization states of the Co layer, respectively. Bottom left: Scanning electron micrograph of the sample and electric circuitry used in the measurements. Shown are the terminals for the Hall voltage measurements as well as the current line where a pulsed current is applied for the switching experiments.
Middle: The state of the perpendicular magnetization is measured via the anomalous Hall resistance as a function of applied field, $B$. After injection of positive (black squares) and negative (red circles) current pulses of amplitude $I_p=52.58$~mA the Hall resistance is measured. The data are reported during a single sweep of $B$. Right: The measurement schematics and pulse sequence. From Ref.~\onlinecite{Miron2011}.}
\label{Miron2011_fig1}
\end{figure}

\onlinecite{Miron2010,Miron2011,Haazen2013} have demonstrated similar results in earlier experiment using ultrathin FM layers. Their devices are based on the Pt/Co/AlOx system with ultrathin Co layers of a thickness  of only 0.6~nm sandwiched between Pt (3~nm) and AlOx (1.6~nm). The use of ultra thin Co in contact with Pt leads to a strong perpendicular anisotropy. When a current is driven through the Pt layer, switching of the Co magnetization can be observed by monitoring the AHE of the device (see Fig.~\ref{Miron2011_fig1}).
While in the original interpretation the driving force for the observed switching was thought to arise mostly from the Rashba symmetry ISGE  due to the broken inversion symmetry along the growth direction of the layer stack, detailed analysis in later three-dimensional vector measurement \cite{Garello2013} point towards significant contributions from the SHE. 
Similarly, the interpretation of the results of current driven domain wall motion experiments in the same type of layer stacks has to be revisited \cite{Miron2010}. Current and even field induced domain wall motion experiments in layer stacks where ISGE, SHE,  and proximity polarization of the NM can contribute are complicated for interpretation, and disentangling the relative strength of these contributions is not straightforward. Experimentally, however, it has been observed that the inclusion of relativistic torques, of either the SHE or ISGE origin, leads to a large increase of domain wall velocities for optimally tuned materials which is potentially of great technological interest \onlinecite{Emori2013,Ryu2013}.

Spin-orbit coupling together with broken inversion symmetry introduces yet another important aspect into the physics of these systems. To fully understand the underlying mechanisms in these experiments one needs to take into account also the fact that these domain walls are chiral due to the Dzyaloshinski-Moriya interaction present at the FM/NM interface. This opens a new field connecting spintronics with the skyrmion physics.  

We conclude by discussing in more detail  that  in the NM/FM bilayer systems the relativistic torques inducing magnetization dynamics are, in general, not only due to the SHE but  the ISGE-induced SOTs may also contribute \cite{Manchon2009,Chernyshov2009,Miron2010,Fang2011}. The ISGE originate from spin-orbit coupling which, combined with broken inversion symmetry in the crystal, can produce spin-polarization when electrical current is driven through a NM. In combination with FMs, the ISGE and the SHE can drive magnetization dynamics in devices with similar geometries. Disentangling these contributions in NM/FM bilayer systems and engineering them for maximal effect is at present a highly active field in spintronics. 

However, the discrimination of the SHE and ISGE based microscopic mechanisms between the field-like and the antidamping-like torque components is difficult to achieve for several conceptual reasons. The original theoretical proposals \cite{Aronov1989,Edelstein1990,Malshukov2002} and experimental observations \cite{Silov2004,Kato2004b,Ganichev2004b,Wunderlich2004,Wunderlich2005} of the ISGE were made in NMs with no FM component in the structure. The corresponding non-equilibrium spin-density, generated in the ISGE by inversion-asymmetry terms in the relativistic Hamiltonian, has naturally no dependence on magnetization. Hence, in the context of magnetic semiconductors \cite{Bernevig2005c,Chernyshov2009,Endo2010,Fang2011} or FM/NM structures \cite{Manchon2008b,Miron2010,Pi2010,Suzuki2011,Miron2011b}, the ISGE may be expected to yield only the field-like component of the torque $\sim{\bf \hat{m}}\times\hat{\boldsymbol\zeta}$, where  the vector $\hat{\boldsymbol\zeta}$ is independent of the magnetization vector ${\bf \hat{m}}$. However, when carriers experience both  the spin-orbit coupling  and magnetic exchange coupling,  the inversion asymmetry can generate a non-equilibrium spin-density component of extrinsic, scattering-related \cite{Pesin2012,Wang2012b} or intrinsic, Berry-curvature \cite{Garate2009,Kurebayashi2014,Freimuth2013} origin which is magnetization dependent and yields an antidamping-like torque $\sim{\bf \hat{m}}\times({\bf \hat{m}}\times\hat{\boldsymbol\zeta})$. Experiments in (Ga,Mn)As confirmed the presence of the ISGE-based mechanism \cite{Chernyshov2009,Endo2010,Fang2011} and demonstrated that the field-like and the Berry-curvature antidamping-like SOT components can have comparable magnitudes \cite{Kurebayashi2014}.

The STT is dominated by the antidamping-like component \cite{Ralph2008} in weakly spin-orbit coupled FMs with $\tau_{ex}\ll\tau_s$, where $\tau_{ex}$ is the precession time of the carrier spins in the exchange field of the FM and $\tau_s$ is the spin life-time in the FM. This, in principle, applies also to the case when the spin-current is injected to the FM from a NM via the SHE. However, at finite $\tau_s$, the STT also acquires a field-like component \cite{Ralph2008}. Experiments in W/Hf/CoFeB structures confirmed the presence of the SHE-based mechanism in the observed torques and showed that the SHE-STT can have both antidamping-like and field-like components of comparable magnitudes \cite{Pai2014}.

In the commonly studied polycrystalline  transition-metal FM/NM samples, the dependence of the torques on the angle of the driving in-plane current also does not provide the direct means to disentangle the two microscopic origins. The lowest order inversion-asymmetry spin-orbit terms in the Hamiltonian have the Rashba form for which  the vector $\hat{\boldsymbol\zeta}$ is in the plane parallel to the interface and perpendicular to the current, independent of the current direction. The same applies to the spin-polarization of the SHE spin-current propagating from the NM to the FM. The ${\bf \hat{m}}$ and $\hat{\boldsymbol\zeta}$ functional form of the field-like and antidamping-like SHE-STTs is the same as of the corresponding SOT components. In the observed lowest order torque terms in Pt/Co and Ta/CoFeB structures \cite{Garello2013} the ISGE-based and the SHE-based mechanism remained, therefore, indistinguishable. The simultaneous observation of  higher order torque terms in these samples pointed to SOTs due to structural inversion-asymmetry terms beyond the basic Rashba model. From the Ta thickness dependence measurements in the Ta/CoFeB structure it was concluded that  in these samples both the ISGE-based and the SHE-based mechanisms contributed to both the field-like and the antidamping-like torques \cite{Kim2013}.

\subsection{Spin Hall angles}
\label{she_angles}
In this subsection we collect  within Table~\ref{SHEangle-table} experimental measurements of the SHE in different materials. The list, in such an active and evolving field, is by no means exhaustive.  As discussed in this experimental section, more things are learned about the techniques and systematic errors are better understood and corrected, 
the measurements begin to  converge for several materials, particularly for the transition metals.

In this table we show the material, the temperature the measurement was taken at, the spin diffusion length either measured or used in the analysis, the conductivity, the spin Hall angle,  the reference of the work and the type of technique as well as relevant comments.

\begin{table*}
\begin{scriptsize}
\begin{tabular}{lrrrrll}
\hline
      & T[K] &  $\lambda_{sd}$~     &    $\sigma_{NM}$~[$10^{6} $S/m]                 & $\alpha_{SH}$~(\%)  &  ~~~Comment      & Ref. \\
\hline \hline

Al    & 4.2&  $455 \pm 15$    & 10.5     &                   $0.032 \pm 0.006$    &~~~NL                                 & (Valenzuela {\it et al} 2006,2007) \\ 
      & 4.2&  $705 \pm 30$    & 17       &                     $0.016 \pm 0.004$    &~~~NL                                 &(Valenzuela {\it et al} 2006,2007) \\ 
\hline
Au    & 295&  $86 \pm 10$     & 37       &                        11.3                &~~~NL (10-nm thick films)                               &(Seki {\it et al} 2008,2010) \\ 
      & 295&  $83$            & 37       &                            3                   &~~~NL (20-nm thick films)                            & \cite{Seki2010}    \\
      & 4.5&  $65^*$            & 48.3       &                            $<2.3$                   &~~~NL (SHE/iSHE)                            & \cite{Mihajlovic2009}    \\
      & 295&  $36^*$            & 25.7       &                            $<2.7$                   &~~~NL (SHE/iSHE)                            & \cite{Mihajlovic2009}    \\
      & 295&  $35 \pm 4$            & 28       &                            $7.0\pm 0.1$        &~~~NL                                &\cite{sugai2010}    \\
      & 295&  $27 \pm 3$            & 14       &                            $7.0\pm 0.3$        &~~~NL (0.95 at.\% Fe)                 & \cite{sugai2010}    \\
      & 295&  $25 \pm 3$            & 14.5       &                            $12\pm 4   $        &~~~NL (1.4 at.\% Pt, 10-nm thick films)                 & \cite{gu2010}    \\
      & 295&  $50 \pm 8$            & 16.7       &                            $0.8\pm 0.2   $        &~~~NL (1.4 at.\% Pt, 20-nm thick films)                 & \cite{gu2010}    \\
       & $<10$&  $40\pm16$     & 25      &                        $1.4\pm0.4$          &~~~NL                            & \cite{Niimi2014}   \\
     & 295&  $35 \pm 3^*$    & 25.2     &                        $0.35\pm 0.03$      &~~~SP                             & \cite{Mosendz2010}    \\
      & 295&  $35$    & 20     &                        $0.25\pm 0.1$      &~~~SP                             & \cite{Vlaminck2013}   \\
      & 295&  $35 \pm 3^*$    & 5.25     &                        $1.6\pm 0.1$      &~~~SP                             & \cite{Hung2013}   \\
      & 295&  $35 \pm 3^*$    & 7     &                        $0.335\pm 0.006$      &~~~SP                             & \cite{Hung2013}   \\
      & 295&  $35^*$          &       &                        $1.1\pm0.3$          &~~~SP                            & \cite{Obstbaum2014}   \\
       & 295&  $60$             &    20.4   &                        $8.4\pm 0.7$          &~~~SP                            & \cite{Wang2014}   \\
\hline
Ag  & 295&  $700$             &    15   &                        $0.7\pm0.1$          &~~~SP                            & \cite{Wang2014}   \\
\hline
Bi    & 3 &  $0.3\pm0.1$          & -         &                            $>0.3$        &~~~local, ~ signal decreases with $\rho_N$ 
																									 & \cite{Fan2008}   \\
      & 295 &  -          & $2.4\pm0.3$(I)          &                           $-(7.1\pm0.8)$(I)       &~~~SP as a function of Bi thickness  & \cite{hou2012}   \\
      &   &             & $50\pm12$(V)          &                           $1.9\pm0.2$(V)        &~~~volume (V) and interfacial (I) param.   &    \\
\hline
Cu  & 295&  $500$             &    16   &                        $0.32\pm0.03$          &~~~SP                            & \cite{Wang2014}   \\
\hline
CuIr  & 10 &  $5-30$          &          &                            $2.1\pm 0.6$        &~~~NL (Ir concentrations from 0 to 12 \%)                                 & \cite{Niimi2011}    \\
CuMn$_x$T$_y$  & - &          &          &                            0.7(Ta);2.6(Ir)       &~~~T = Lu, Ta, Ir, Au, Sb ($y\sim 1-20\times 10^{-4}$)                                & \cite{Fert1981}    \\
      &   &             &           &                           1.35(Au);1.15(Sb)         &~~~Mn ($x\sim 1-2\times 10^{-4}$) creates $I_s$    &    \\
      &   &             &           &                                 -1.2(Lu)                    &~~~Note factor 2 in the definition of $\alpha^{skew}_{SH}$   & Ref. 15 in  \cite{Fert2011}    \\
\hline
CuBi  & 10 &  $\sim 100$;  $\sim 30$                &                          &  $-11$        &~~~NL (Bi = 0.3\%; 0.5\%),  $\alpha^{skew}_{SH}= -(24\pm 9)$ on Bi                               & \cite{Niimi2012}    \\
      &    &            $\sim 10$;$\sim 7$       &          &                                                             &~~~similar  in AgBi \cite{Niimi2014}                        & \\
\hline
$n$-GaAs& 4.2 &  $2200$          &       0.0056   &                            0.15                &~~~NL, $n\approx 10^{17}$cm$^{-3}$                                  & \cite{Olejnik2012}    \\
      & 4.2 &  $8500$          & 0.00137         &                            0.08                &~~~LSP, $n\approx 10^{16}$cm$^{-3}$                                 & \cite{Ehlert2012}    \\
     & 30 &  $ $          & 0.0036         &                            0.08                &~~~LSP, $n\approx 3-5\times10^{16}$cm$^{-3}$    & \cite{Garlid2010}\\
\hline
Ge    & 2  &                  &          &                           $\approx -0.001$     &~~~MR, $\alpha_{SH}$ T-dependent,~sign change at $\approx$ 10 K        & \cite{Chazalviel1975} \\
    & 295  &                  &    0.027      &                           $ 0.001;\, 0.00044$     &~~~SP, not annealed and annealed values        & \cite{Rojas-Sanchez2013} \\
\hline
$n$-InGaAs & 30 &   $\sim 3000$   & $\sim 0.002$                 &  $\approx 0.02$       &~~~KRM, $x=0.07$,$n\approx 3\times10^{16}$cm$^{-3}$ & \cite{Kato2004}    \\
{\tiny (Si-doped) }    & 30 &  $  $          & 0.003-0.005         &                            $\approx0.1$;$\approx0.25$;$\approx0.38$     &~~~LSP, $x=0.03,0.05,0.06$,$n\approx 3-5\times10^{16}$cm$^{-3}$ & \cite{Garlid2010} \\ 
\hline
InSb  & 1.3&                  &          &                           $-0.026\pm 0.005$  &~~~MR, $n\approx 10^{14}$cm$^{-3}$,$\mu \approx 2.2\times 10^{4}$cm$^{2}$/Vs  &(Chazalviel {\it et al} 1972)\\
      & 1.3&                  &          &                           $0.003$              &~~~MR  $n\approx 10^{14}$cm$^{-3}$,$\mu \approx 4 \times 10^{4}$cm$^{2}$/Vs                      &(Chazalviel {\it et al} 1972)\\
\hline
IrO$_2$  & 300&   3.8(P)            &0.5(P);0.18(A)                                   & 4(P);6.5(A)  &~~~NL, polycryst. (P), amorphous (A)      & \cite{Fujiwara2013a} \\ 

\hline
Mo    & 10 &  10              & 3.03     &                       -0.20               &~~~NL                                 & \cite{Morota2009}    \\
      & 10 &  10              & 0.667    &                        -0.075              &~~~NL                                 & \cite{Morota2009}    \\
      & 10 &  $8.6 \pm 1.3$   & 2.8      &                        $-(0.8 \pm 0.18)$   &~~~NL                                 & \cite{Morota2011}    \\
      & 295&  $35 \pm 3^* $   & 4.66     &                       $-(0.05\pm 0.01)$   &~~~SP                             & \cite{Mosendz2010}   \\
\hline
Nb    & 10 &  $5.9 \pm 0.3$   & 1.1      &                        $-(0.87 \pm 0.20)$  &~~~NL                                 & \cite{Morota2011}    \\
\hline
 Pd     & 10 &  $13 \pm 2$      & 2.2      &                        $1.2\pm 0.4$        &~~~NL                                 & \cite{Morota2011}    \\
    & 295&  $9^*$           & 1.97     &                        $1.0$               &~~~SP                             & \cite{Ando2010}   \\
      & 295&  $15 \pm 4^*$    & 4.0      &                        $0.64\pm 0.10$      &~~~SP                             & \cite{Mosendz2010}    \\
      & 295&  $5.5\pm0.5$     & 5     &                        $1.2\pm 0.3$           &~~~SP                             & \cite{Vlaminck2013}   \\
            & 295&  $2.0 \pm 0.1$   & 3.7      &                        $0.8\pm 0.20$      &~~~STT+SHE                             & \cite{Kondou2012}    \\ 

\hline
Pt    & 295&  $3^*$           & 6.41     &                        0.37                &~~~NL                                 & \cite{Kimura2007}    \\
      & 5  &  8              & 8.0      &                         0.44                &~~~NL ($\lambda_N=14$ nm from spin-absorption)                                & \cite{Vila2007}      \\
      & 295&  7              & 5.56     &                        0.9                 &~~~NL ($\lambda_N=10$ nm from spin-absorption)                                 & \cite{Vila2007}      \\
      & 10 &  $11 \pm 2$      & 8.1      &                        $2.1\pm 0.5$        &~~~NL                                 & \cite{Morota2011}    \\
      & 10 &  $\sim 10$       & 8.1      &                        $2.4$               &~~~NL (3D corrected \cite{Morota2011})                                & \cite{Niimi2012}    \\
      & 295&  $7^*$           & 6.4      &                       $8.0$               &~~~SP                             & \cite{Ando2008d}   \\
      & 295&  $10 \pm 2^*$    & 2.4      &                       $1.3\pm 0.2$        &~~~SP                             & \cite{Mosendz2010}    \\
      & 295&  $10^*$           & 2      &                       $4.0$               &~~~SP                             & \cite{Ando2011}   \\
      & 295&  $3.7\pm0.2$      & 2.42      &                       $8\pm 1$               &~~~SP                             & \cite{Azevedo2011}   \\ 
      & 295&  $8.3\pm0.9$    & $4.3\pm0.2$     &                       $1.2\pm 0.2$               &~~~SP                             & \cite{Feng2012}   \\ 
      & 295&  $7.7\pm 0.7$    &$1.3\pm 0.1$     &                  $1.3\pm 0.1$      &~~~SP                             & \cite{Nakayama2012}   \\
       & 295&  $1.5-10^*$    &$2.45\pm 0.1$     &                  $3^{+4}_{-1.5}$      &~~~SP, spin Hall MR                             & \cite{Hahn2013}   \\
      & 295&  $4$             & 4     &                        $2.7\pm 0.5$          &~~~SP                             & \cite{Vlaminck2013}   \\
      & 295&  $8\pm 1^*$  &1.02     &                        $2.012\pm 0.003$      &~~~SP                             & \cite{Hung2013}   \\
      & 295&  $1.3^*$             &  2.4     &                        $2.1\pm1.5$          &~~~SP                            & \cite{Bai2013}   \\
      & 295&  $1.4^*$             &       &                        $12\pm4$          &~~~SP                            & \cite{Obstbaum2014}   \\
      & 295&  $7.3$             &  2.1     &                        $10pm1$          &~~~SP                            & \cite{Wang2014}   \\
       & 295&  $1.2 \pm 0.1$   & 3.6      &                        $2.2\pm 0.4$      &~~~STT+SHE                             & \cite{Kondou2012}    \\
     & 295&  $3 (<6)$           & 5.0      &                   $7.6^{+8.4}_{-2.0}$ &~~~STT+SHE                              & \cite{Liu2011}  \\
      & 295&  $2.1\pm0.2$             & 3.6     &                        $2.2\pm 0.8$          &~~~STT+SHE                            & \cite{Ganguly2014}   \\
      & 295&  $2.1\pm0.2$             & 3.6     &                        $8.5\pm 0.9$          &~~~STT+SHE, modulation of damping                            & \cite{Ganguly2014}   \\  
       & 295&  $2.4^*$    &$1.2$     &                  $\sim4$      &~~~spin Hall MR                             & \cite{Nakayama2013}   \\
        & 295&  $1.5\pm 0.5$  &0.5-3     &                        $11\pm 8$      &~~~spin Hall MR (variable Pt thickness)                            & \cite{Althammer2013}   \\    
\hline
$p$-Si   & 295  &                  &          &                           $\approx 0.01$     &~~~SP, $\tau_s\sim 10$ ps  $n\approx 2\times 10^{19}$cm$^{-3}$     & \cite{Ando2012} \\
\hline
Ta    & 10 &  $2.7 \pm 0.4$   & 0.3      &                        $-(0.37 \pm 0.11)$  &~~~NL                                 & \cite{Morota2011}    \\
     & 295&  $1.9$             &    0.34   &                        $-7.1\pm0.6$          &~~~SP                            & \cite{Wang2014}   \\
      & 295&  $1.8\pm 0.7$    &$0.08-0.75$                     &  $-(2^{+0.8}_{-1.5})$      &~~~SP, spin Hall MR     ~(variable Ta thickness)                         & \cite{Hahn2013}   \\
          & 295&  $ $           & 0.53      &                   $-(12\pm4)$              &~~~STT+SHE ($\beta$-Ta)                              & \cite{Liu2012}  \\ 
\hline
W & 295&  $2.1$             &    0.55   &                        $-14pm1$          &~~~SP                            & \cite{Wang2014}   \\
  & 295 &  $ $   & $0.38 \pm 0.06$     &                        $-(33 \pm 6)$  &~~~STT+SHE ($\beta$-W, lower in $\alpha$-W $\alpha_{SH}$)                                 & \cite{Pai2012}    \\ 
\hline
\end{tabular}
\end{scriptsize}
\caption{Experimental spin Hall angles and related parameters. SP=spin pumping, NL=nonlocal, STT+SHE=spin transfer torque combined with spin Hall effect, MR=magnetoresistance,
LSA=local spin accumulation, MR=Magnetic Resonance, KRM=Kerr rotation microscopy. Values marked with $*$ are taken not measured but taken from the literature.}
\label{SHEangle-table}
\end{table*}

\section{Future directions and remaining challenges}


We conclude this review with our own personal view of possible interesting directions and remaining challenges. As such, this is not entirely scientific and it reflects merely our own preference and intuition and should only be taken as such. We apologize for any omissions of the many interesting possibilities that others may consider. We only know for certain that such future outlook is bound to always fail in a field that continues to bring unexpected surprises. 

Transition metals have traditionally played a dominant role in spintronics both in basic research and, in particular, in applications. It is therefore not surprising that the SHE field has gained new momentum when bringing non-magnetic transition metals in the game. And they have played their role particularly well. When brought out of equilibrium by an applied electric field, the SHE in  some non-magnetic transition metals can generate sufficient flux of spin angular momentum to reorient magnetization in an adjacent  transition metal FM. Entirely new concepts for writing information in magnetic tunnel junctions or domain-wall based spintronic devices have emerged  from this discovered surprising strength of the SHE in the common and technologically relevant family of materials. 

Ta, W, Ir, or Pt are examples among the non-magnetic transition metals with large SHE. The strength of the effect is derived from the large spin-orbit coupling in these heavy elements. Apart from the new opportunities for applications, this brings also new challenges for the basic research of the SHE in transition mentals. We have mentioned in the review the pitfalls in attempting to microscopically describe the SHE in structures comprising heavy transition metals from theories of spin transport in weakly spin-orbit coupled systems. The proper description and microscopic understanding of the SHE structures in the strong spin-orbit coupling regime is among the key remaining challenges in the SHE field. 

The flurry of recent SHE studies in transition metals may give an impression that the field is forgetting its semiconductor roots. Robust FMs are typically dense moment systems and their switching requires comparably large electrical current densities generating the SHE spin-current. Highly conductive transition metals are clearly favorable from this perspective when compared to semiconductors. Moreover, the reported spin Hall angles in semiconductors do not reach the record values in transition metals. 

We nevertheless foresee semiconductors playing vital role in future SHE research; in particular when considering spintronics concepts without FMs.  In the transition metal context the SHE is used as an efficient spin-current generator or detector but these studies rarely consider spin manipulation in the non-magnetic SHE system. Especially in the strongly spin-orbit couple heavy metals, the spin diffusion length is of the order of nanometers, too short for implementing any spin manipulation tools along the non-magnetic transport channel. 
For semiconductors, on the other hand, we have mentioned in the review several examples of electrical manipulation of the output SHE signal. A gate electrode can be used to control coherent spin precession along the channel, additional drift current was shown to modify the spin-current profile along the channel, or non-linear inter-valley transport can strongly enhance the spin Hall angles bringing the values close to their heavy metal counterparts. 

The physics is, however, no different in principle between metals and semiconductors. Large SHE requires large spin-orbit coupling which, on the other hand, tends to suppress spin coherence/diffusion length. Semiconductors with their simpler electronic structure and model spin-orbit fields offer unique ways how to get around this problem. As has been already demonstrated, a proper tuning of the Rashba and Dresssehaus spin-orbit fields can significantly enhance spin coherence in the presence of strong spin-orbit coupling. 
Experiments outside the SHE field have recently made major progress in controlling these two canonical spin-orbit fields in common semiconductor structures and we envisage new developments in semiconductor SHE devices utilizing the coherent spin-manipulation techniques.  

Combining optical selection rules with SHE makes semiconductors also favorable materials for exploring new concepts in opto-spintronics. These may include optical spin-torque structures, electrical polarimeters, spin-photovoltaic cells, switches, invertors and interconnects. The opto-spintronic subfield of the SHE research is still at its infancy and we expect growing activity in this direction in the future. 

The fascinating feature of the SHE is that it can generate a large spin-current, and a resulting large spin-accumulation by bringing weakly out of equilibrium a non-magnetic system. It is, therefore, natural that non-magnetic materials have been traditionally in the center of interest of the SHE research. However, limiting ourselves to paramagnetic or diamagnetic materials, whether metallic or semiconducting, is not necessary when considering the spin Hall phenomena.  Recently, several transition metal FMs and antiferromagnets were demonstrated to act as efficient ISHE spin-current detectors which opens a new broad area of future materials research in the SHE. 

It also brings us back to the opening paragraphs of this section where we mentioned SHE-induced spin torques in NM/FM heterostructures. Since in strongly spin-orbit coupled systems these torques are limited to a few atomic layers around the NM/FM interface, and considering the likely material intermixing at the interface, it is not meaningful to speak strictly about a non-magnetic layer SHE in these structures. The difference than becomes blurred whether including magnetism via intermixing or proximity polarization at hetero-interfaces, or directly considering the SHE in bulk FM or antiferromagnetic materials. Within this notion, an important challenge arises not only for the normal-metal/magnet interfaces but also for monolayer magnets to identify the microscopic origin of the observed spin torques. It remains an open question whether the current induced torques in the magnet are better linked to a SHE-induced spin-current origin or to one of the variants of the ISGE-induced non-equilibrium spin-polarization. Resolving these contributions is an important academic exercise with potentially large implications for the utility of these spin-orbit coupling phenomena in spintronic information technologies.

We conclude by emphasizing that the field of SHE does not live in a vacuum. Its interconnects to other emerging fields, e.g. topological insulators and spin-caloritronics, makes its growth and possibilities very difficult to predict since many things that we have discussed here and that have emerged from its link to these fields were not known or expected a few years ago. It is a rapid evolving field that produces discoveries at a neck breaking speed and we all look forward to its exciting future.

\section*{List of symbols and abbreviations}

\noindent{\bf 2DEG}: Two-dimensional electron gas 

\noindent{\bf 2DHG}: Two-dimensional hole gas 

\noindent{\bf AC}: Alternating current

\noindent {\bf AHE}: Anomalous Hall effect

\noindent {\bf AMR}: Anistropic magnetoresistance

\noindent {\bf CPW}: Coplanar wave guide

\noindent{\bf DC}: Direct current

\noindent{\bf FM}: Ferromagnet

\noindent {\bf FMR}: Ferromagnetic resonance

\noindent {\bf HE}: Hall effect

\noindent {\bf ISGE}: Inverse spin galvanic effect

\noindent {\bf MOD}: Modulation of damping

\noindent {\bf MRAM} Magnetic random access memory

\noindent{\bf NM}: Non-magnetic material

\noindent {\bf QHE}: Quantum Hall effect

\noindent {\bf QSHE}: Quantum spin Hall effect

\noindent {\bf RF}:  Radio frquency

\noindent {\bf SGE}: Spin galvanic effect

\noindent {\bf SHE}: Spin Hall effect

\noindent {\bf SOT}: Spin orbit torque

\noindent {\bf SP}:  Spin pumping

\noindent {\bf STT}: Spin transfer torque

\noindent {\bf TMR}: Tunnelling magnetoresistance

\section*{Acknowledgements}
We thank all colleagues who have given us the permission
to show their results in this review. 
We also thank all our colleagues within the spintronics community that engaged us in
 many fruitful interactions and spirited discussions. 
J. S. acknowledges partial support by the Alexander von Humboldt Foundation and the  European Research Council (ERC)  Synergy Grant No. 610115.
S. O. V. acknowledges partial support from the European Research Council under grant agreement 308023 SPINBOUND and from the Spanish Ministry of Economy and Competitiveness (MINECO) under contracts MAT2013-46785-P and SEV-2013-0295.
J. W. acknowledges partial support from EMRP JRP IND08 MetMags and the ERC Synergy Grant No. 610115.
C. B. acknowledges partial support from the German research foundation
(DFG) through programs SFB 689 and SPP 1538 and from the European Research
Council (ERC) through starting grant no. 280048 ECOMAGICS.
T. J. acknowledges partial support from the ERC Advanced Grant No. 268066, the Ministry of Education of the Czech Republic Grant No. LM2011026, the Grant Agency of the Czech Republic Grant No. 14-37427G, the Academy of Sciences of the Czech Republic Praemium Academiae


\end{document}